\documentclass[preprint,12pt]{elsarticle}

\usepackage{amssymb}

\usepackage{amsmath,amsthm,amsfonts,amscd,amsmath,amsfonts} 
\usepackage{upgreek}
\usepackage{mdframed}
\usepackage{multicol}
\usepackage{graphicx}
\usepackage{nomencl}
\usepackage[nottoc]{tocbibind}
\usepackage{setspace}
\usepackage[caption=false]{subfig}  
\usepackage{verbatim}   	
\usepackage{makeidx}    	
\usepackage{color}
\usepackage{xcolor}
\usepackage{enumerate}
\usepackage{url}	
\usepackage{tcolorbox}
\usepackage{multirow}
\usepackage[linesnumbered,ruled,vlined]{algorithm2e}
\usepackage{ulem}
\usepackage[pdftex,bookmarks]{hyperref}

\usepackage[percent]{overpic}
\usepackage{multirow}
\usepackage{arydshln}

\usepackage[margin=1.2in]{geometry} 

\usepackage{etoolbox} 
\makeatletter
\patchcmd{\@startsection}
  {\@afterindenttrue}
  {\@afterindentfalse}
  {}{}
\makeatother

\journal{Journal of Risk and Uncertainty in Engineering Systems -- }

\begin{document}

\begin{frontmatter}



\title{
Chance-Constrained Optimal Design of Porous Thermal Insulation Systems Under Spatially Correlated Uncertainty
}


\author[inst1]{Pratyush Kumar Singh}

\affiliation[inst1]{organization={Department of Mechanical and Aerospace Engineering, University at Buffalo},
            city={Buffalo},
            state={NY},
            country={USA}}

\author[inst1]{Danial Faghihi\corref{cor1}}

\cortext[cor1]{Corresponding Author, \texttt{danialfa@buffalo.edu} (D. Faghihi)}


\begin{abstract}

This paper presents a computationally efficient method for the optimal design of silica aerogel porous material systems, balancing thermal insulation performance with mechanical stability under stress concentrations. The proposed approach explicitly accounts for additive manufacturing uncertainties by modeling material porosity as a spatially correlated stochastic field within a multiphase finite element formulation. A risk-averse objective function, incorporating statistical moments of the design objective, is employed in conjunction with chance constraints that enforce mechanical stability by restricting the probability of exceeding critical stress thresholds. To mitigate the prohibitively high computational cost associated with the large-dimensional uncertainty space and Monte Carlo estimations of the objective function’s statistical moments, a second-order Taylor expansion is utilized as a control variate. Furthermore, a continuation-based smoothing strategy is introduced to address the non-differentiability of the chance constraints, ensuring compatibility with gradient-based optimization. The resulting framework achieves computational scalability, remaining agnostic to the dimensionality of the stochastic design space. The effectiveness of the method is demonstrated through numerical experiments on two- and three-dimensional thermal break systems for building insulation. The results highlight the framework’s capability to solve large-scale, chance-constrained optimal design problems governed by finite element models with uncertain design parameter spaces reaching dimensions in the hundreds of thousands.

\end{abstract}

\begin{keyword}
Thermal insulation \sep
Optimal design \sep  
chance constraint optimization \sep
high-dimensional parameters \sep
uncertainty quantification
\end{keyword}

\end{frontmatter}




\section{Introduction}

The development of high-performance insulation materials is critical for improving energy efficiency and reducing the environmental impact of buildings. Silica aerogels, renowned for their ultra-low thermal conductivity \cite{an2021wearable, zhou, an2023flexible}, offer exceptional insulating properties but are highly susceptible to mechanical failure due to stress-concentration. Advances in additive manufacturing have facilitated precise control over the spatial distribution of material properties, potentially enabling mechanically robust and lightweight insulation systems. However, the inherent uncertainties in the thermal and mechanical performance of these engineered materials pose significant challenges to their reliable integration into construction applications, necessitating robust design strategies.


While finite element-based topology optimization techniques have advanced considerably \cite{woischwill2018multimaterial, guilleminot2019topology, keshavarzzadeh2020stress, keshavarzzadeh2017topology, ng2014multifidelity, zheng2019level, da2021three, dos2018reliability, ding2024multi}, design of additively manufactured aerogel components presents two primary challenges.
The first challenge arises from uncertainties in spatially varying design parameters, particularly those induced by the layer-by-layer fabrication process. This manufacturing variability introduces spatially correlated stochastic fluctuations along the deposition direction, leading to a high-dimensional design space when discretized using finite element methods. Consequently, the computational cost becomes prohibitive due to the necessity of solving multiple nonlinear finite element solution of the porous material response.
The second challenge involves incorporating stress constraints into the optimization framework. These constraints introduce significant nonlinearity and discontinuities, particularly when addressing localized stress concentrations. Global stress constraints, often formulated using aggregated measures such as von Mises stress, require carefully chosen aggregation functions to maintain both computational efficiency and sensitivity to critical stress regions. Furthermore, the presence of uncertainty necessitates a stochastic formulation of stress constraints, integrating probabilistic constrained into the optimal design formulation.


To address the first challenge and ensure robustness in optimal design under uncertainty, risk-averse optimization methods incorporates statistical moments, such as mean and variance, into the objective function \cite{alexanderian2017mean, chen2019taylor, chen2021optimal, kruger2023efficient, maute2014topology, ng2014multifidelity, kouri2016risk, luo2023efficient, liao2020heuristic}. However, directly integrating these moments into finite element-based optimization with high-dimensional parameters imposes significant computational costs. Various approximation techniques have been developed to alleviate this burden. Notably, Ghattas et al. \cite{chen2021taylor, chen2019taylor, alexanderian2017mean} demonstrated that Taylor series expansions of statistical moments provide scalable and computationally efficient solutions for optimization problems governed by finite element models. Kriegesmann et al. \cite{kruger2023efficient, kranz2023generalized, kriegesmann2019robust} introduced first-order second-moment methods for efficient gradient-based variance estimation. More recently, Tan et al. \cite{tan2024scalable} extended Taylor-based approximations to multi-objective optimization, incorporating phase-field regularization to enhance spatial sparsity control.
To address the second challenge, several studies have explored the integration of stress constraints into optimization frameworks across various materials and manufacturing processes \cite{wang2023multi, wang2021stress, da2019stress, li2022structural, de2015stress, dos2018reliability, zhao2022stress, duysinx1998topology, yang2018stress, le2010stress, banh2024novel, guo2024topology, herrero2024adaptive, hoang2022robust, wu2024robust, deng2020topology}. Efforts have also focused on incorporating uncertainty into stress-constrained optimization \cite{cheng2024uncertainty}, addressing factors such as variable loading and heat sources. For example, Silva et al. \cite{da2019stress} accounted for uncertainties in boundary conditions during manufacturing, while Oh et al. \cite{oh2022stress} incorporated uncertainties in loading positions in topology optimization. Mart{'\i}nez and Ortigosa \cite{martinez2021risk} developed a risk-averse approach for fail-safe structures, treating collapse as an uncertainty source rather than a deterministic event. Liu et al. \cite{liu2022uncertain} introduced interval reliability-based topology optimization, considering multi-source uncertainties in dynamic responses.

We present a novel computational framework for the optimal design of aerogel-based insulation systems under spatially varying stochastic design parameters, incorporating chance constraints to ensure mechanical stability. The thermomechanical property of the porous silica aerogels is characterized by a multiphase finite element model, where material porosity is represented as a spatially correlated stochastic field. The objective is to optimize the spatial distribution of material porosity to maximize thermal insulation while mitigating stress concentrations of the insulation component. 
To alleviate the computational burden associated with Monte Carlo-based moment estimation in the mean-variance cost functional, we employ second-order Taylor approximations to efficiently approximate both the design objective and chance constraints.
Building on our previous work in \cite{tan2024scalable}, this study introduces the following key contributions:
(1) Explicit integration of probabilistic stress constraints to directly regulate stress concentrations within the optimization framework, ensuring mechanical robustness. 
(2) Rather than directly employing the Taylor expansion of the moments of the design objective, this approximation is leveraged as a control variate in Monte Carlo estimation, substantially enhancing accuracy in high-variance uncertainty scenarios.
(3) Implementation of an inexact Newton-Conjugate Gradient algorithm, enabling scalable optimization for large-scale problems while maintaining convergence guarantees.
The effectiveness and computational efficiency of the proposed approach are demonstrated through two- and three-dimensional numerical experiments of a beam-thermal break component within a building envelope.


The remainder of the paper involves presenting the thermomechanical model for silica aerogels, incorporating spatially correlated uncertainties and the mean-variance cost functional for optimal design in Section \ref{sec:duu}. Section \ref{sec:taylor} presents the Taylor approximations and Monte Carlo estimators with control variates. Section \ref{sec:optimizor} details the integration of chance constraints within the gradient-based optimization framework. Numerical experiments demonstrating the method’s effectiveness are provided in Section \ref{sec:results}, followed by conclusions and directions for future research in Section \ref{sec:conclusions}.

\section{Problem formulation: Design of aerogel thermal break under uncertainty}\label{sec:duu}

\begin{figure}
    \centering
    \includegraphics[width=0.35\linewidth]{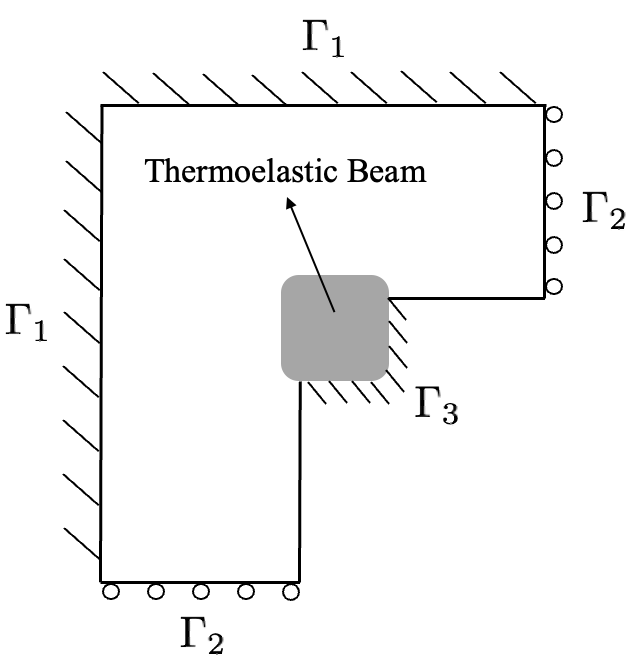}~
    \includegraphics[width=0.35\linewidth]{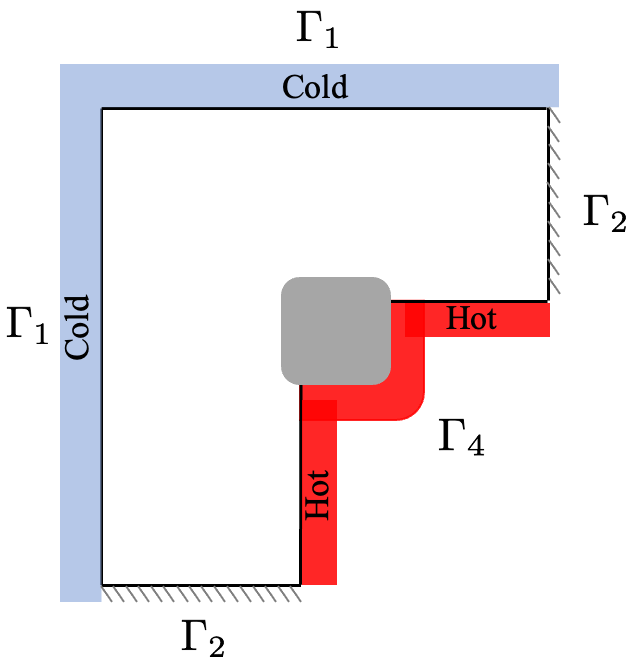}
    \caption{The domain of a beam-insulator system used for numerical experiments, indicating both the mechanical (left) and thermal (right) boundary conditions.
    The solid displacement are fixed at the boundaries $\Gamma_1$ and $\Gamma_3$ and roller boundary conditions are implemented at $\Gamma_2$. 
    At $\Gamma_1$ and $\Gamma_4$, we consider convective heat transfer cold ambient temperature in the building exterior and hot ambient temperature in the building interior. no heat flux boundary is implemented at $\Gamma_2$.
    }
    \label{fig:domain}
\end{figure}

Figure \ref{fig:domain} illustrates a building insulation system involving a concrete beam with an integrated silica aerogel thermal break. Concrete’s high thermal conductivity promotes heat transfer through building envelopes, forming thermal bridges, while thermal breaks mitigate this effect. However, thermal expansion of the concrete beam can generate stress concentrations in the aerogel, risking mechanical failure and reducing insulation effectiveness. This section formulates the optimal design problem under uncertainty, incorporating thermomechanical models governed by partial differential equations (PDEs) and detailing the modeling of uncertainty in design parameters.
%

\subsection{Thermomechanical finite element models and design objectives}\label{sec:duufwd}
For silica aerogel materials, we adopt a multiphase model based on continuum mixture theory \cite{tan2021predictive}, capturing the interactions between an incompressible solid aerogel phase and a compressible fluid (gaseous) phase.
The governing equations consist of a system of PDEs that describe heat transfer in both phases, solid-phase deformation, and pore pressure evolution, over the domain $\Omega$ and boundaries $\Gamma_1$, $\Gamma_2$, $\Gamma_3$, and $\Gamma_4$, as 
\vspace{-0.05in}
\begin{equation}
\begin{cases}
    -\nabla \cdot(\phi_s \, \kappa_s \, \nabla \theta_s) + h(\theta_s-\theta_f) = 0 \quad &\text{in} \quad \Omega,
        \\
    -\nabla \cdot(\phi_f \, \kappa_f \, \nabla \theta_f) - h(\theta_s-\theta_f) =0 \quad &\text{in} \quad \Omega,
        \\
    -\phi_s \, \kappa_s \, \nabla \theta_s = h_\mathrm{conv} (\theta_s - \theta_\mathrm{a})  \quad &\text{on} \quad \Gamma_1 \cup \Gamma_4 ,
        \\
    -\phi_f \, \kappa_f \, \nabla \theta_f = h_\mathrm{conv} (\theta_f - \theta_\mathrm{a})  \quad &\text{on} \quad \Gamma_1 \cup \Gamma_4,
        \\
      
\end{cases}
\label{eq:pdes_thermal}
\end{equation}
\begin{equation}
\begin{cases}
    D\,p + (\nabla\cdot\mathbf{u}_s) = 0\quad &\text{in} \quad \Omega,
    \\
    - \nabla\cdot\mathbf{T}^\prime_s  -  (2\,\phi_f - 1)\,\nabla p  = 0 \quad &\text{in} \quad \Omega,
    \\
    \mathbf{u}_s = \mathbf{0} \quad &\text{on} \quad \Gamma_1 \cup \Gamma_3,
    \\
    \nabla \mathbf{u}_s \cdot \mathbf{n} = 0 \quad &\text{on} \quad \Gamma_2
\end{cases}
\label{eq:pdes_mech}
\end{equation}
where $\nabla$ is gradient operator, $\mathbf{n}$ is the outward unit normal vector, and
the effective solid stress is
$\mathbf{T}^\prime_s = 2\,\mu\,\mathbf{E}_s + \lambda\,\mathrm{tr}(\mathbf{E}_s)\,\mathbf{I}
$
with $\mathbf{E}_s(\mathbf{u}_s)$ being the solid strain.
The model parameters are
solid conductivity $\kappa_s$, fluid conductivity $\kappa_f$, fluid compressibility $D$,
the Lam\'{e} constants $\lambda$ and $\mu$, and convective heat coefficient $h_\mathrm{conv}$ with respect to hot or cold ambient temperature $\theta_\mathrm{a}$, and are assumed to be known for silica aerogel from previous studies.
The state variables are $\theta_s$, $\theta_f$, $\mathbf{u}_s$, and $p$ represent the solid and fluid temperatures, solid displacement, and fluid pressure, respectively. 
Additionally, $\phi_s$ and $\phi_f$ are the volume fractions of solid and fluid (porosity) phases.
The concrete beam is assumed to follow a thermoelastic model,
\begin{equation}
\begin{cases}
    -\nabla \cdot( \, \kappa_b \, \nabla \theta_b)  = 0 \quad &\text{in} \quad \Omega_b,
        \\   
    - \kappa_b \, \nabla \theta_b = h_\mathrm{conv} (\theta_b - \theta_\mathrm{a})  \quad &\text{on} \quad \Gamma_3 ,
        \\
        \nabla \cdot \mathbf{T}_b = 0   & \text{in} \quad \Omega_b\\
\end{cases}
\label{eq:pdes_beam}
\end{equation}
where $\theta_b$ represents temperature of beam and Cauchy stress is $\mathbf{T}_b = 2 \mu_b \mathbf{E}_b + \lambda_b \mathrm{tr}(\mathbf{E}_b)\mathbf{I} -\alpha_T (3\lambda_b + 2\mu_b) (\theta_b - \theta_0)$ with $\mathbf{E}_b$ is strain in the beam. 
In \eqref{eq:pdes_beam}, 
$\kappa_b$ is thermal conductivity of beam, $\lambda_b$ and $\mu_b$ are the Lam\'{e} constants for the beam, $\alpha_T$ is coefficient of thermal expansion and $\theta_0$ is the reference temperature.
Finite element solution of the above model requires the weak formulation of the governing equations as:

\textit{Find $\mathbf{u}= (\theta_s, \theta_f, \theta_b, \mathbf{u}_s, \mathbf{u}_b, p)$, such that} 
\begin{eqnarray}
    - \big<  \phi_s\kappa_s \nabla \theta_s, \nabla z_s  \big>
    - \big<  h(\theta_s-\theta_f), z_s  \big>
    - \big<  \phi_s h_\mathrm{conv}(\theta_s-\theta_\mathrm{a}), z_s  \big>
    & = & 0, 
    \nonumber\\
    - \big<  \phi_f\kappa_f \nabla \theta_f, \nabla z_f  \big>
    + \big<  h(\theta_f-\theta_f), z_f  \big>
    - \big<  \phi_f h_\mathrm{conv}(\theta_f-\theta_\mathrm{a}), z_f  \big>
    & = & 0, 
    \nonumber\\
    - \big<  \kappa_b \nabla \theta_b, \nabla z_b  \big>
    - \big<  h_\mathrm{conv}(\theta_b-\theta_\mathrm{a}), z_b  \big>
    & = & 0, 
    \nonumber\\
    - \big<  (\lambda + \frac{1-2\phi_f}{D}) (\nabla \cdot \mathbf{u}_s) \mathbf{I} + \mu\big(  \nabla\mathbf{u}_s + (\nabla\mathbf{u}_s)^T  \big), \boldsymbol{w}_u   \big>
    + \big<  \mathbf{t}, \boldsymbol{w}_u  \big>
    & = & 0, 
    \nonumber\\
    - \big<  (\lambda_b) (\nabla \cdot \mathbf{u}_b) \mathbf{I} + \mu_b \big(  \nabla\mathbf{u}_b + (\nabla\mathbf{u}_b)^T  \big), \boldsymbol{w}_b   \big>
    + \big<  \mathbf{t}, \boldsymbol{w}_b  \big>
    & = & 0, 
    \label{eq:weakform}
\end{eqnarray}
for all test functions
$\mathbf{v} = (z_s, z_f, z_b, \boldsymbol{w}_u, \boldsymbol{w}_b)$.
Here
$\big<\cdot,\cdot\big>$ represents the inner product.

\begin{figure}[h!]
    \centering
    \includegraphics[width=0.22\linewidth]{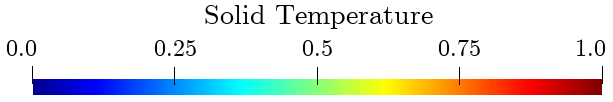}~
    \includegraphics[width=0.22\linewidth]{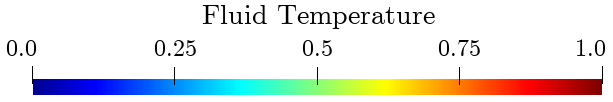}~
    \includegraphics[width=0.22\linewidth]{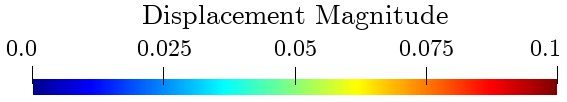}~
    \includegraphics[width=0.22\linewidth]{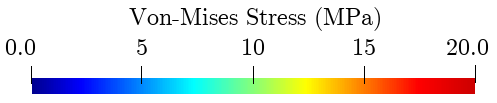}\\
    \includegraphics[trim={5.75in 3.75in 5.75in 3.75in},clip,width=0.22\linewidth]{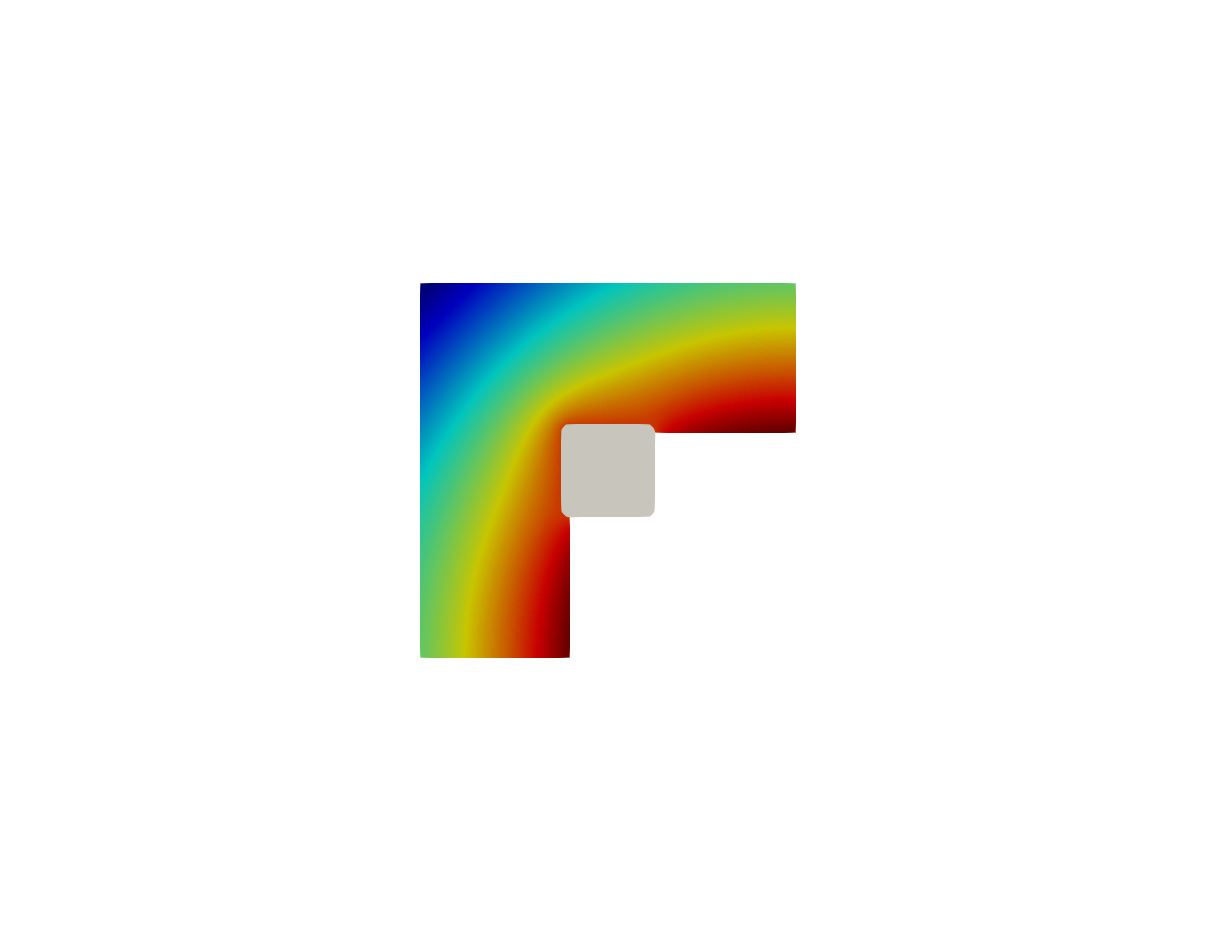}~
    \includegraphics[trim={5.75in 3.75in 5.75in 3.75in},clip,width=0.22\linewidth]{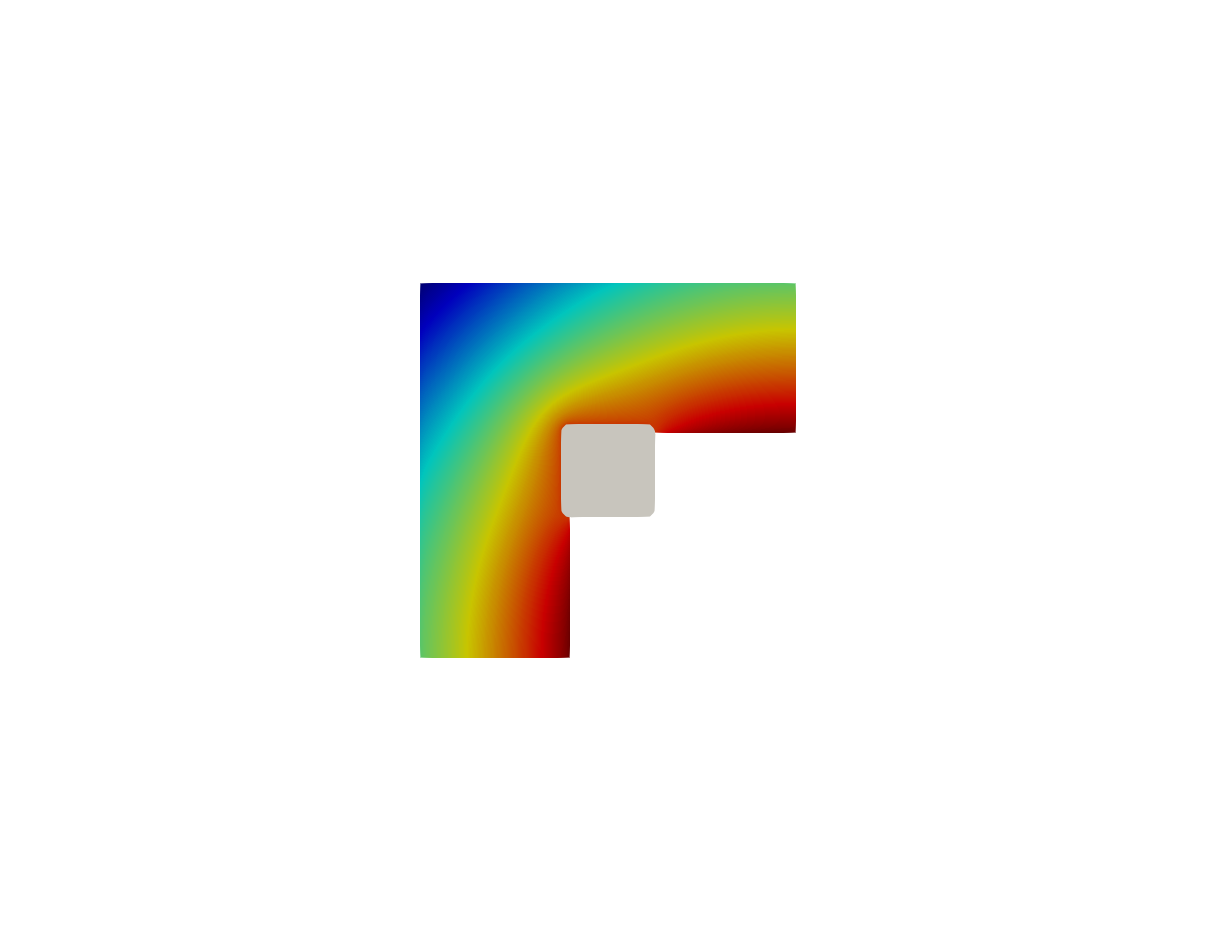}~
     \includegraphics[trim={5.75in 3.75in 5.75in 3.75in},clip,width=0.22\linewidth]{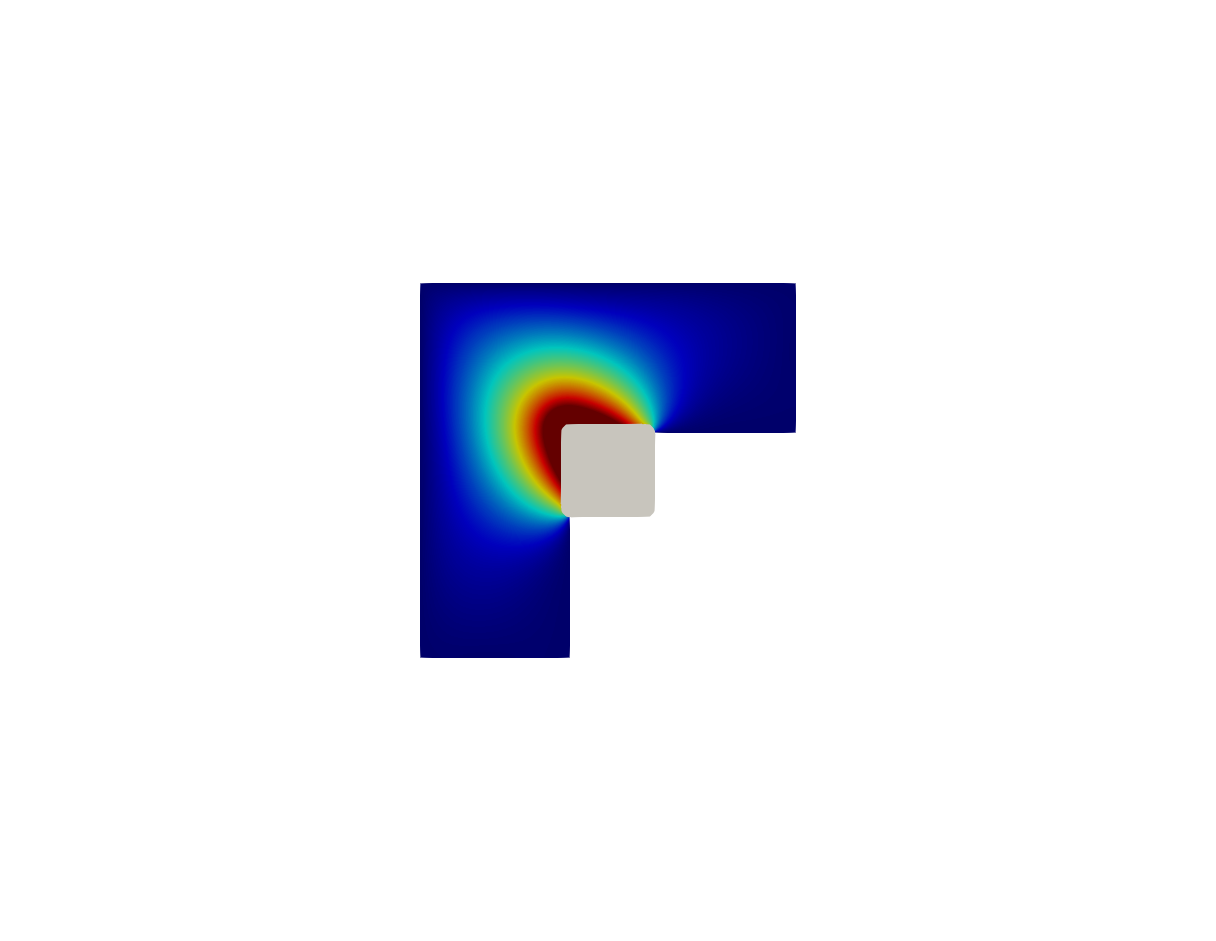}~
    \includegraphics[trim={5.75in 3.75in 5.75in 3.75in},clip,width=0.22\linewidth]{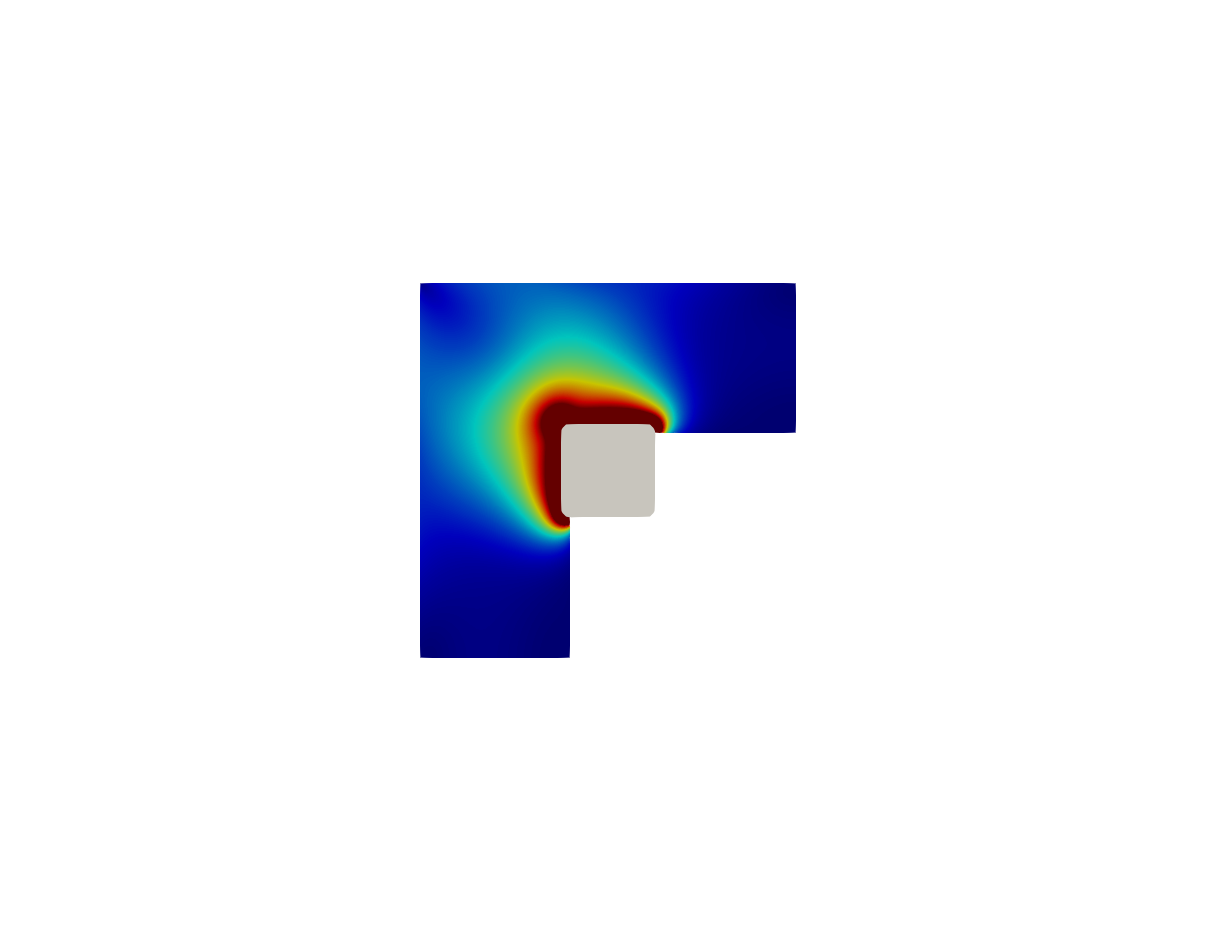}\\
    \caption{Finite element solution of the thermomechanical (forward) model for the beam-insulator system with constant porosity $\phi_f=0.7$, depicting the solid temperature $\theta_s$, fluid temperature $\theta_f$, displacement magnitude $||\mathbf{u}_s||$, and von Mises stress $T_{\text{VM}}$.}
    \label{fig:stress_concentration}
\end{figure}

Figure \ref{fig:stress_concentration} presents the deterministic finite element solution of the forward model for the beam-insulator system under constant porosity $\phi_f=0.7$, utilizing the model parameters listed in Table \ref{tab:parameters}. As shown in this figure, thermal expansion of the thermoelastic concrete beam induces stress concentrations within the aerogel thermal break.
We define the design objective $Q$ as thermal compliance of the thermal break, 
\begin{equation}
    Q = 
    \frac{1}{2}\sum_{i = s,f} \big<\phi_i\,\kappa_i\,\nabla \theta_i, \nabla \theta_i \big> 
    + 
    {\big< \phi_i h_\mathrm{conv}(\theta_i - \theta_\mathrm{a})   ,  \theta_i   \big>},
    \label{eq:qois}
\end{equation}
The design goal is determining the spatial distribution of the fluid volume fraction (porosity), $\phi_f(\mathbf{x})$, to minimize stress concentrations within the thermal break, thereby preventing mechanical failure while preserving its thermal insulation performance.

\subsection{Spatially correlated uncertain design parameter}\label{sec:matern}
\noindent
Due to the inherent challenges in precisely controlling the properties of additively manufactured aerogel, considerable uncertainty arises in the porosity values within the thermal break arising from the layer-by-layer deposition of aerogel ink.
To effectively model such uncertainty, first we define uncertain parameter $m$ as a Gaussian random field characterized by the Mat\'{e}rn covariance kernel.
Specifically, $m$ is distributed as $\mathcal{N}(\Bar{m}, \mathcal{K})$ where $\Bar{m}$ is the mean and $\mathcal{K}$ is the covariance operator.
The sampling of Gaussian random field $m$ is equivalent to solving the elliptic equation with a robin boundary condition, 
\begin{equation}
    \begin{aligned}
        - \gamma \nabla . (\boldsymbol{\Theta} \nabla m) + \delta m = & \: \dot{W} \quad \text{in} \quad \Omega \\
        m + 1.42 \: \nabla m . \: \boldsymbol{n} = & \: 0 \quad \text{on} \quad \Gamma
    \end{aligned}
    \label{eq:matern_pde}
\end{equation}
Here, $W$ represents the spatial White Gaussian noise with a unit variance, and the parameters $\gamma$ and $\delta$ control the variance $\sigma^2= {1}/{4\pi\,\delta\,\gamma}$ and spatial correlation length $L_{CR}=\sqrt{{8\gamma}/{\delta}}$.
The factor ${\sqrt{\delta\,\gamma}}/{1.42}$ is chosen to minimize boundary artifacts \cite{roininen2014}, and $\boldsymbol{\Theta}$ is an anisotropic tensor.
Then we define the design parameter $d$ as a spatial field,  employing a sigmoid mapping function $\phi_f \equiv \text{sigmoid}\left(d + m \right)$, ensuring that porosity values asymptotically remain within the physically valid range, maintaining near-linear behavior within $\phi_f \in [90\%, 10\%]$ \cite{tan2024scalable}.

For simplicity, for the rest of this paper, we express the forward state problems \eqref{eq:pdes_thermal}, \eqref{eq:pdes_mech} and \eqref{eq:pdes_beam} in an abstract form 
$\mathcal{R}(\mathbf{u},m,d) = 0$, 
and the weak form \eqref{eq:weakform} is expressed as $r(\mathbf{u},\mathbf{v},m,d)$,
where $\mathbf{u}$ is the state variables, $\mathbf{v}$ is adjoint variables and $m$ and $d$ are uncertain and design parameters.

\subsection{Chance-constrained PDE-constrained optimal design under uncertainty}
\noindent
We formulate the design problem as minimizing the thermal compliance in \eqref{eq:qois}, which explicitly depends on the uncertain parameter, $Q(d,m)$. The mean-variance cost functional, e.g., \cite{alexanderian2017mean,chen2019taylor, tan2024scalable}, is given by:
\begin{equation}
\mathcal{J}(d) = \mathbb{E}[Q(d,m)] + \beta_V \,\mathbb{V}[Q(d,m)] + \beta_{RG} \, R_p (d).
    \label{eq:cost}
\end{equation}
where $R_p (d)$ is the regularization term, while $\beta_V$ and $\beta_{RG}$ represent the coefficients assigned to the variance and the regularization component, respectively.

%
%
To prevent mechanical failure of the aerogel thermal break due to stress concentration, the von Mises stress $T_{\text{VM}}$ must remain below a critical threshold. 
This requirement is incorporated as a probabilistic (chance) constraint in the optimization problem, as $T_{\text{VM}}$ depends on uncertain parameters.
However, due to the non-differentiable nature of the maximum stress, direct optimization poses significant challenges. This challenge is addressed using stress aggregation function based on p-norm \cite{le2010stress}, which is a smooth approximation of the maximum stress and enables gradient-based optimization techniques.
The p-norm for the von Mises stress over the domain can be computed as
\begin{equation}\label{eq:p_norm}
    T_{\text{pn}} = {\left( \int_{\Omega} T_{\text{VM}}^{p} \,d \Omega \right)}^{\frac{1}{p}}, \quad \text{where} \quad
    T_{\text{VM}}=\sqrt{(3/2) \mathbf{T}_0 : \mathbf{T}_0}, \;
    \mathbf{T}_0 = \mathbf{T} - \frac{1}{3} \mathrm{tr}(\mathbf{T})\mathbf{I}
\end{equation}
The chance constraint function can hence be written as,
\begin{equation}\label{eq:chance_function}
    f = T_{cr} - T_{\text{pn}}.
\end{equation}
In above relation $T_{cr}$ is the limiting critical stress, $\mathbf{T}_0$ is stress deviator tensor, and $\mathrm{tr}(\cdot)$ is the trace.
To avoid stress concentration, we consider a chance constraint:
    \begin{equation} \label{eq:inequality_chance}
        P(f(m,d) \geq 0) = \mathbb{E}[\chi_{[0,\infty)} (f(m,d))] \leq \alpha_{c},
    \end{equation}
    for a critical chance $0 < \alpha_{c} <1$, where $\chi_{[0,\infty)} (f(m,d))$ is an indicator function defined as:
\begin{equation}\label{eq:indicator_function}
 \chi_{[0,\infty)} (f(m,d)) =
\begin{cases}
  1 & \text{if } f(m,d) \geq 0 \\
  0 & \text{if } f(m,d) < 0 \\
\end{cases}
\end{equation}
Thus, the risk-averse optimal design problem under uncertainty can be formulated as a PDE-constrained optimization problem, expressed as:
\begin{equation}\label{eq:problem_formulation}
\begin{aligned}
     d_{\text{opm}} = \min_{{d}}  &\  \mathcal{J}({d}) 
            \\
             \mathrm{s.t.\ } & \mathcal{R}(\boldsymbol{u},m,d) = 0 \\
             & P(f(m,d) \geq 0) \leq \alpha_{c}
\end{aligned}
\end{equation}

%

\section{Scalable framework for chance constrained optimal design }\label{sec:taylor}

A conventional method for calculating the mean and variance components in \eqref{eq:cost} and the probability constrained in \eqref{eq:inequality_chance} is the Monte Carlo estimation for $N_{mc}$ samples, 
\begin{align}
    \mathbb{E}[Q] &\approx \frac{1}{N_{mc}}\sum_{i=1}^{N_{mc}} Q(m^{(i)}),
    \label{eq:mc_mean}
     \\
    \mathbb{V}[Q] &\approx \Big(\frac{1}{N_{mc}}\sum_{i=1}^{N_{mc}} Q^2(m^{(i)}) \Big)   -   \Big(  \frac{1}{N}_{mc}\sum_{i=1}^{N_{mc}} Q(m^{(i)})  \Big)^2,
        \label{eq:mc_var}
    \\
    P(f(.,d)\geq 0) &\approx \frac{1}{N_{mc}} \sum_{i=1}^{N_{mc}} \chi_{[0,\infty)]} (f(m^{(i)},d))
    \label{eq:mc_chance}
\end{align}
where $\{ m^{(i)} \}_{i=1}^{N_{mc}}$ are random samples from $m$.
Achieving an accurate approximation requires a large number of finite element model evaluations due to the slow convergence of the Monte Carlo method, making optimization under uncertainty computationally prohibitive. To address this, we introduce a computationally efficient approach that approximates the mean and variance of the design objective using a quadratic Taylor expansion, leveraging this approximation as a control variate to accelerate Monte Carlo sampling.

\subsection{Quadratic Taylor approximation of the design objective and chance constraint}\label{sec:quad}
The quadratic Taylor approximation of $Q$ at the mean of the uncertain parameter $\Bar{m}$ is given by \cite{tan2024scalable, alexanderian2017mean,chen2019taylor},
    \begin{eqnarray}\label{eq:taylor_objective}
         Q_{\text{QUAD}}(m)  =  \Bar{Q}  +   \big<\Bar{Q}^m, \,  m-\Bar{m}\big>  +  \frac{1}{2} \big< \Bar{Q}^{mm} \, (m-\Bar{m}),  \,  m-\Bar{m}  \big>.
    \end{eqnarray}
Here, the objective $Q$ and its gradient and Hessian with respect to $m$, evaluated at mean $\Bar{m}$, are denoted as $\Bar{Q}$, $\Bar{Q}^m$ and $\Bar{Q}^{mm}$, respectively.
The expectation and variance of $Q_{\text{QUAD}}$ are derived in \cite{tan2024scalable} such that,
\begin{equation}
    \mathbb{E}[Q_{\text{QUAD}}] = \Bar{Q} + \frac{1}{2}
            \mathrm{tr}\big(\Bar{\mathcal{H}}_k\big),
    \quad        
    \mathbb{V}[Q_{\text{QUAD}}(m)]
    =
    \big<  \Bar{Q}^m, \mathcal{K}\,\Bar{Q}^m \big> 
    +
    \frac{1}{2}
    \mathrm{tr}(  (\Bar{\mathcal{H}}_k)^2   ),        
            \label{eq:mean}
\end{equation}
where $\Bar{\mathcal{H}}_k = \mathcal{K}\,\Bar{Q}^{mm}$ represents the preconditioned Hessian for covariance at the mean of the uncertain parameter $m$ and $\mathcal{K}$ is the covariance operator.
The design objective in \eqref{eq:cost} can then be approximated as
\begin{equation}
\mathcal{J}_\mathrm{QUAD}(d) = \Bar{Q} + \frac{1}{2}
    \mathrm{tr}\big(\Bar{\mathcal{H}}_k\big) + 
    \beta_V \, \bigg(
    \big<  \Bar{Q}^m, \mathcal{K}\,\Bar{Q}^m \big> 
    +
    \frac{1}{2}
    \mathrm{tr}(  (\Bar{\mathcal{H}}_k)^2   ) \bigg)+ 
            \beta_{RG} \, R_p (d).
    \label{eq:cost_quad}
\end{equation}

We can also construct a second-order Taylor expansion of the chance constraint function $f$ at $\Bar{m}$ as,
\begin{equation}\label{eq:taylor_chance}
          f_{QUAD}(m)  = \Bar{f}  +   \big<\Bar{f}^m, \,  m-\Bar{m}\big>  +  \frac{1}{2} \big< \Bar{f}^{mm} \, (m-\Bar{m}),  \,  m-\Bar{m}  \big>,
\end{equation}
where
$\Bar{f}$, $\Bar{f}^m$ and $\Bar{f}^{mm}$ are the constraint function $f$ and it gradient and Hessian evaluated at $\Bar{m}$. 
The probability $P(f(.,d) \geq 0)$ by Monte Carlo estimator (\ref{eq:mc_chance}) with the quadratic approximation yields, 
\begin{equation}\label{eq:taylor_constraint}
    P(f(m,d))\geq 0) = \frac{1}{N_{mc}} \sum_{i=1}^{N_{mc}} \chi_{[0,\infty)}(f_{\text{QUAD}}(m_{i},d)).
\end{equation}
In order to compute the second-order Taylor expansion \eqref{eq:taylor_chance}, we need to compute the term $\big< \Bar{f}^{mm} \, (m-\Bar{m}),  \,  m-\Bar{m}  \big>$, which can be approximated through a low-rank approximation as,
\begin{equation}\label{eq:approximation_chance}
    \big< \Bar{f}^{mm} (m - \Bar{m}), m - \Bar{m} \big> = 
    \sum_{i=1}^{N_\mathrm{eig}^f} \lambda_i^f \big| \big< m - \Bar{m}, \mathcal{K}^{-1} \psi_i^f \big> \big|^2,
\end{equation}
where $\lambda_i^f$ and $\psi_i^f$ are the eigenvalues and eigenfunctions corresponding to the solution of a generalized eigenvalue problem. 
The generalized eigenvalue problem corresponding to the second-order Taylor expansion can be formulated as,
\begin{equation}\label{eq:eig_chance}
    \mathcal{K}^{1/2} \: \Bar{f}^{mm} \: \mathcal{K}^{1/2} \: \psi_i^f = \lambda_i^f \psi_i^f, \quad i = 1, 2, \dots, N_\mathrm{eig}^f,
\end{equation}
where $\lambda_{1}^f,\lambda_{2}^f,...,\lambda_{N_\mathrm{eig}^f}^f$ are the $N_\mathrm{eig}^f$ largest eigenvalues.
Computing the approximations of the mean, variance, and constraint function requires a Lagrangian-based formalism to derive the $m$-gradient, $m^f$-gradient, $m$-Hessian and $m^f$-Hessian, which represent the gradients and Hessians of the design objective $Q$ and constraint function $f$ with respect to $m$, as detailed in Section \ref{sec:gH}.

%


Finally, the the mean and variance in \eqref{eq:mean} requires the trace of the covariance-preconditioned Hessian $\Bar{\mathcal{H}}_k$.
Here, we leverage a randomized trace estimator using the dominant eigenvalues $\{\lambda^Q_n\}_{n=1}^{N^Q_\mathrm{eig}}$ of $\bar{\mathcal{H}}_k$, 
\begin{equation}
        \mathrm{tr}\big(\Bar{\mathcal{H}}_k\big) 
        \approx \sum_{n\geq1}^{N^Q_\mathrm{eig}}  \lambda^Q_n
            ,\quad
        \mathrm{tr}\big(\Bar{\mathcal{H}}_k^2\big)  
        \approx \sum_{n\geq1}^{N^Q_\mathrm{eig}}  {(\lambda^Q_n)}^2.
    \label{eq:trace_eig}
\end{equation}
When eigenvalues decay rapidly, $N^Q_\mathrm{eig}$ only depend on the intrinsic low dimensionality of the otherwise high-dimensional parameters.
The eigenvalues $\lambda^Q_n$ are obtained by solving the generalized eigenvalue problem with an input of oversampling factor $N_o$, via double-pass randomized eigenvalue solver (see \cite{tan2024scalable, chen2019taylor} for more details).

\subsection{Monte Carlo with control variate}
The approximation of $\mathbb{E}[Q]$ and $\mathbb{V}[Q]$ in the cost functional (\ref{eq:cost}) using a quadratic Taylor series expansion may lack accuracy for the design problem, particularly for design problems where the variance of the uncertain parameter $m$ is large, as shown in  \cite{tan2024scalable}. To enhance accuracy, the  $Q_{\text{QUAD}}$  is leveraged as a control variate to reduce variance in the sampling estimations of the mean and variance. In particular, A Monte Carlo method is employed to correct the moments of $Q_{\text{QUAD}}$ by computing the mean and variance of the residual between $Q$ and its quadratic approximation $Q_{\text{QUAD}}$.
For mean of control variate approximation
\begin{equation}
\small
\begin{aligned}
&\mathbb{E}[Q_{CV}] \approx \mathbb{E}[Q] = \mathbb{E}[Q_{\text{QUAD}}] + \mathbb{E}[Q - Q_{\text{QUAD}}] := Q(\Bar{m}) + \frac{1}{2}tr(\Bar{\mathcal{H}}_k)\\
 & + \frac{1}{N_{mc}}\sum_{i=1}^{N_{mc}} \left( Q(m^{(i)}) - Q(\Bar{m}) - \langle m^{(i)} - \Bar{m}, Q^m (\Bar{m}) \rangle - \frac{1}{2} \langle m^{(i)} - \Bar{m}, Q^{mm} (\Bar{m}) (m^{(i)} - \Bar{m}) \rangle   \right).
\end{aligned}
\end{equation}
The variance can be written as
\begin{equation}
\begin{aligned}
\mathbb{V}[Q_{CV}] &= \mathbb{E}[{(Q_{\text{QUAD}} -Q(\Bar{m}))}^2] + \mathbb{E}[{(Q - Q(\Bar{m}))}^2 -{(Q_{\text{QUAD}} -Q(\Bar{m}))}^2]\\
&-{(\mathbb{E}[Q_{\text{QUAD}} -Q(\Bar{m})] + \mathbb{E}[(Q - Q(\Bar{m}))-(Q_{\text{QUAD}}-Q(\Bar{m}))])}^2,
\end{aligned}
\end{equation}
which can be approximated by
\begin{equation}
\footnotesize
\begin{aligned}
    &\mathbb{V}[Q_{CV}] := \langle \mathcal{K} Q^{m}(\Bar{m}),Q^{m}(\Bar{m}) \rangle + \frac{1}{4} {(tr({\Bar{\mathcal{H}}_k}))}^2 + \frac{1}{2} tr({\Bar{\mathcal{H}}_k}^2)\\
    &+ \frac{1}{N_{mc}}\sum_{i=1}^{N_{mc}} \left( {(Q(m^{(i)}) - Q(\Bar{m})}^2 - {\left( \langle m^{(i)} - \Bar{m},Q^m(\Bar{m}) \rangle +\frac{1}{2} \langle m^{(i)} - \Bar{m}, Q^{mm}(\Bar{m}) (m^{(i)} - \Bar{m}) \rangle \right)}^2 \right)\\
    &- {\left( \frac{1}{2} tr({\Bar{\mathcal{H}}_k}) + \frac{1}{N_{mc}} \sum_{i=1}^{N_{mc}} \left( Q(m^{(i)}) - \langle m^{(i)} - \Bar{m}, Q^{m}(\Bar{m}) \rangle - \frac{1}{2} \langle m^{(i)} - \Bar{m}, Q^{mm}(\Bar{m})(m^{(i)} - \Bar{m})\rangle \right) \right)}^2.
\end{aligned}
\end{equation}
The number of samples $N_{mc}$ required to achieve the desired accuracy (and hence the computational cost) depends on the variance of the integrand and how closely the quadratic approximation matches the high-fidelity Monte Carlo estimation. Notably, if the quadratic approximation is highly correlated with $Q$, fewer Monte Carlo samples are needed for effective variance reduction and accurate approximation.

\section{Gradient-based optimization}\label{sec:optimizor}
\noindent
To solve the optimization problem \eqref{eq:problem_formulation},
we employ the Inexact Newton Conjugate-Gradient optimization algorithm (INCG), leveraging an approximate Newton approach combined with a backtracking line search to ensure global convergence \cite{chen2021optimal, VillaPetraGhattas21}. In this method, the Hessian is approximated by evaluating it at the mean of the random field, while the gradient is computed precisely. The resulting Newton system is then solved approximately using a preconditioned conjugate gradient method in a matrix-free manner, allowing for efficient computation.
However, applying INCG to solve \eqref{eq:problem_formulation} in a scalable manner requires special treatment of the chance constraints and efficient computation of the gradient of the cost functional with respect to high-dimensional design parameters, as discussed in this section.

\subsection{Continuous approximation of the discontinuous function}
\begin{figure}
    \centering
    \includegraphics[width=0.6\linewidth]{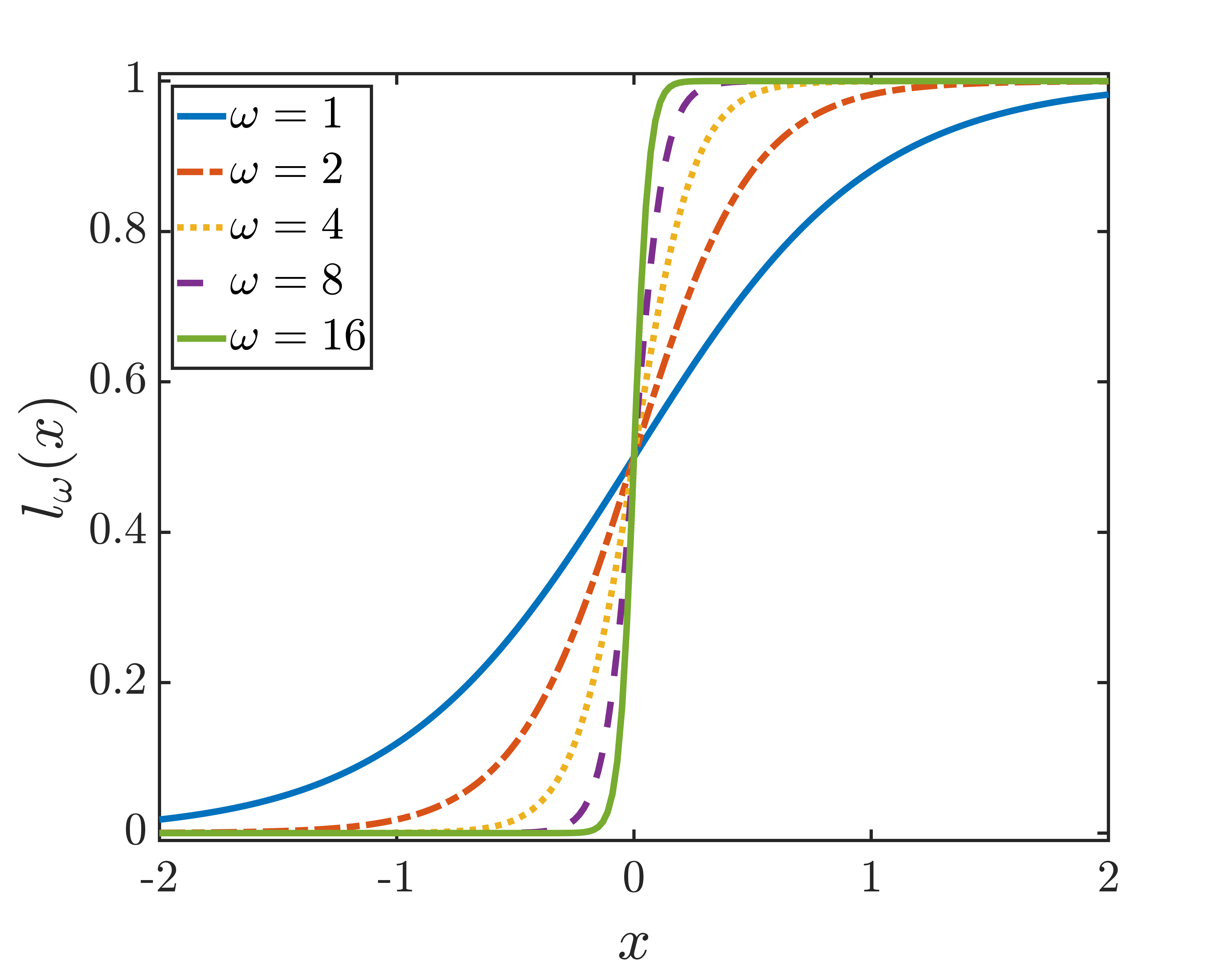}
    \caption{Approximation of the discontinuous indicator function $\chi_{[0,\infty)}(x)$ with a sigmoid function as given in \eqref{eq:smooth_approximation} and its convergence with increasing $\omega$}
    \label{fig:smooth_function}
\end{figure}

The indicator function utilized in the chance constraint is discontinuous at $f(m,d) = 0$. To employ the gradient-based INCG, this discontinuous indicator function must be approximated with a continuous function. A common approach involves using a sigmoid function to approximate the indicator function, as proposed by \cite{chen2021taylor}, given by
\begin{equation}
    \chi_{[0,\infty)}(x) \approx l_{\omega}(x) = \frac{1}{1 + e^{-2 \omega x}},
    \label{eq:smooth_approximation}
\end{equation}
where the parameter $\omega$ controls the sharpness of the transition around $x=0$, as illustrated in Figure \ref{fig:smooth_function}.
A higher value of $\omega$ results in a steeper transition.
 This smooth approximation addresses the non-differentiability of the chance constraint, enabling the cost functional to be differentiable with respect to the design variable.
 The value of $\omega$ is gradually increased through a continuation scheme discussed in Algorithm \ref{algo:adaptive_optimization}. This gradual scaling of $\omega$ helps in avoiding local optima and prevents numerical instabilities.
\subsection{Penalty Method}
\noindent 
 In order to address the inequality constraint in (\ref{eq:inequality_chance}), we convert the original chance-constrained optimization problem into the following unconstrained problem,
\begin{equation}\label{eq: optimization_penalty}
    \underset{d \in [0,1]}{\text{min}} \mathcal{J}(d) + \tau_{\gamma}(\mathbb{E}[l_\omega(f)] - \alpha_c)  .
\end{equation}
Here, $\tau_{\gamma}$ is a quadratic penalty method as outlined in \cite{nocedal2006quadratic}.The quadratic penalty function is given by,
\begin{equation}
        \tau_{\gamma}(x) = \frac{\gamma}{2}{(max{\{0,x\}})}^2,
        \label{eq:quadratic_penalty}
\end{equation}
where $\gamma > 0$ is a constant which determines the penalty weight.
In this manner, the inequality constraint is replaced with a penalization term, that results in the transformation of the constrained problem into an unconstrained form. As $\gamma$ increases as per the continuation scheme in Algorithm \ref{algo:adaptive_optimization}, the penalized solution approaches the feasible region more strictly, making the unconstrained optimization problem effectively mimic the constrained form.
\subsection{Design gradient}
By substituting the trace estimators \eqref{eq:trace_eig} into \eqref{eq:cost_quad} and converting it into unconstrained problem using quadratic penalty \eqref{eq:quadratic_penalty} and continuous approximation \eqref{eq:smooth_approximation}, we have
\begin{equation}
\begin{aligned}
    \mathcal{J}_\mathrm{QUAD}(d)
    &=
    \bigg(
    \Bar{Q} + \frac{1}{2}
            \sum_{j\geq1}^{N_\mathrm{eig}}  \lambda_j^Q
    \bigg)
    +
    \beta_V
    \bigg(
    \big<  \Bar{Q}^m, \mathcal{K}\,\Bar{Q}^m \big> 
    +
    \frac{1}{2}
    \sum_{j\geq1}^{N_\mathrm{eig}}  \left({\lambda_j^Q}\right)^2
    \bigg)
    +
    \beta_{RG} \, R_p (d) \\
    &+ \tau_{\gamma}(\mathbb{E}[l_{\omega}(f_{\text{QUAD}}(m,d))] - \alpha_c)
\end{aligned}
    \label{eq:Jquad}
\end{equation}
Implementing the INCG optimizer requires computing the gradient of the quadratic approximation of the design objective, 
$\mathcal{J}_{\mathrm{QUAD}}(d)$, with respect to the design parameter, referred to as the $d$-gradient. The Lagrangian formalism of the approximated cost, along with variational calculus principles for obtaining the necessary Lagrange multipliers, and an algorithm for computing the $d$-gradient are presented in Section \ref{sec:dCdd}.

The computational cost required to evaluate the quadratic approximation of the cost functional, denoted as $\mathcal{J}_{\mathrm{QUAD}}(d)$ in \eqref{eq:Jquad}, includes solving one potentially nonlinear state PDE, one linear adjoint problem, and $2(N_{\mathrm{eig}}^Q + N_o) + 2(N_{\mathrm{eig}}^f + N_o) + 2$ linear PDEs, as prescribed by the double-pass randomized algorithm (see \cite{tan2024scalable,chen2019taylor}). Furthermore, obtaining the $d$-gradient involves solving $2(N_{\mathrm{eig}}^Q + N_{\mathrm{eig}}^f) + 3$ additional linear PDEs. The design gradient is needed for obtaining the optimal design through a continuation scheme as discussed in Algorithm \ref{algo:adaptive_optimization}.

\section{Numerical Results}\label{sec:results}
\noindent
\begin{table}[hb]
    \centering
    \begin{tabular}{|c|c|}
    \hline
    Parameter & Value\\
    \hline
        $\kappa_s$ & 0.477 $W/m K$  \\
        $\kappa_f$ & 0.085 $W/m K$  \\
        $\kappa_b$ & 5 $W/mK$  \\
        $\alpha_T$ & $10^{-5} \: (1/K)$ \\
        $h$ & 81059 $W/m^2 K$ \\
        D & 0.25 $\times 10^{-8} Pa$ \\
        $\lambda$ & 6.77 $GPa$\\
        $\mu$ & 3.38 $GPa$\\
        $\lambda_b$ & 17.3 $GPa$\\
        $\mu_b$ & 11.5 $GPa$\\
        \hline 
    \end{tabular}
    \caption{List of the forward model parameters and their values used in the design problem.}
    \label{tab:parameters}
\end{table}
This section presents numerical experiments using the proposed framework for design under uncertainty for beam-insulation scenario in Figure \ref{fig:domain} in both 2D and 3D. The objective is to achieve thermal insulation and mechanical stability avoiding stress concentration while mitigating uncertainty in the design process. The analyses presented in this section, investigate the accuracy and convergence of quadratic and Monte Carlo approximation of design objective and explores the effect of different limiting critical stress, critical chance, and uncertain parameter variance and spatial correlation on the optimal design, spatial pattern of $ d_{\text{opt}}$, as well as the probability distributions of the design objectives $Q$. The parameters of the thermomechanical model are summarized in Table \ref{tab:parameters}.
The computational implementation of the framework relies on several open-source libraries like FEniCS \cite{fenics} for the finite element solution of the thermomechanical model, SOUPy \cite{Luo2024} for the quadratic approximation of the design objective and implementation of chance constraint, and hIPPYLib \cite{VillaPetraGhattas21} for the Gaussian random field, trace estimator and INCG optimizer\footnote{The codes and links to the dependent libraries for this work can be found at: \url{https://github.com/pce-lab/InsulationDesignUncertainty}}.

\begin{figure}[h]
    \centering
    \footnotesize
    \includegraphics[width=0.3\linewidth]{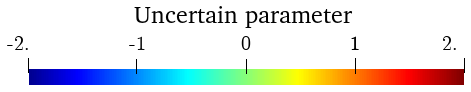}\\
    \includegraphics[trim={2.5in 1.5in 2.5in 1.5in},clip,width=0.23\linewidth]{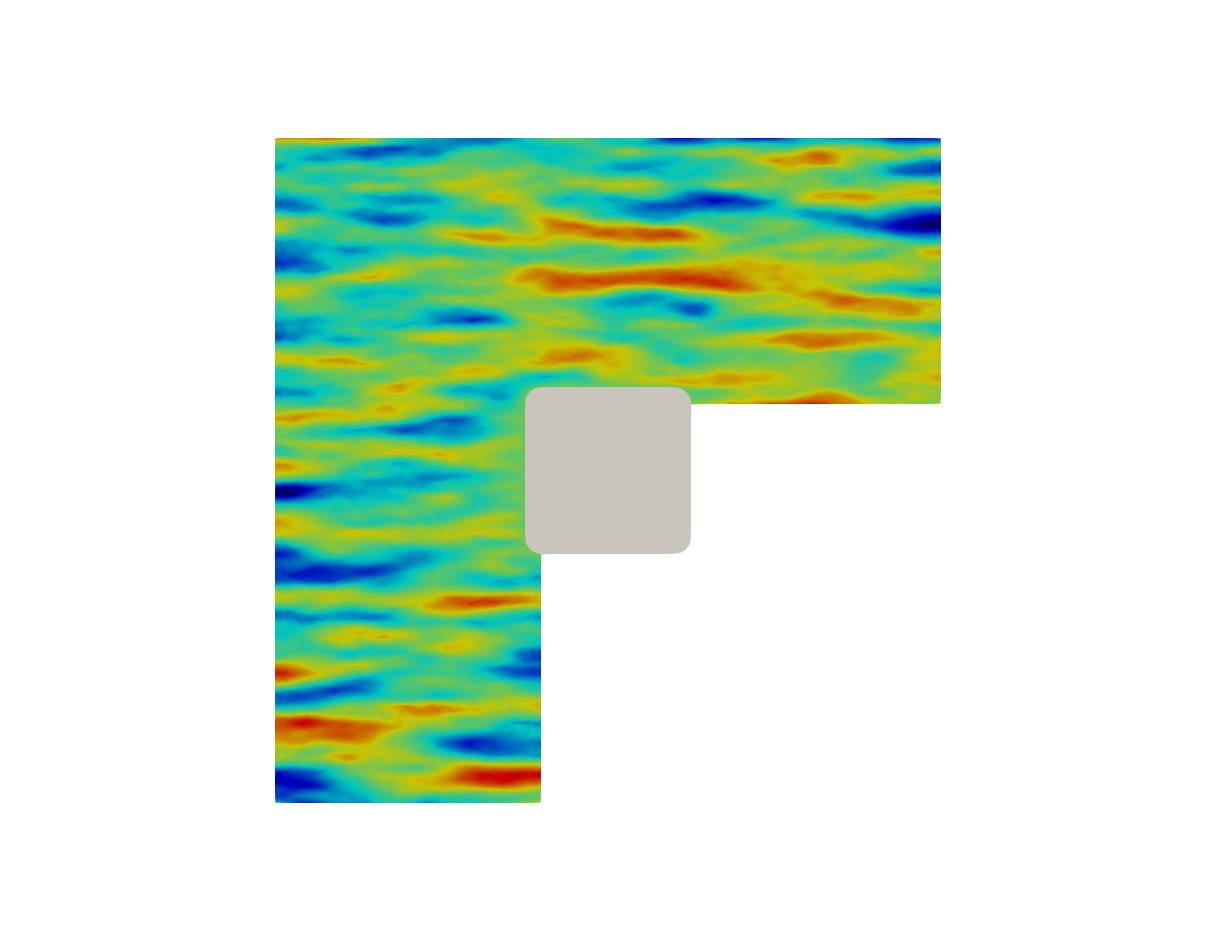}~
    \includegraphics[trim={2.5in 1.5in 2.5in 1.5in},clip,width=0.23\linewidth]{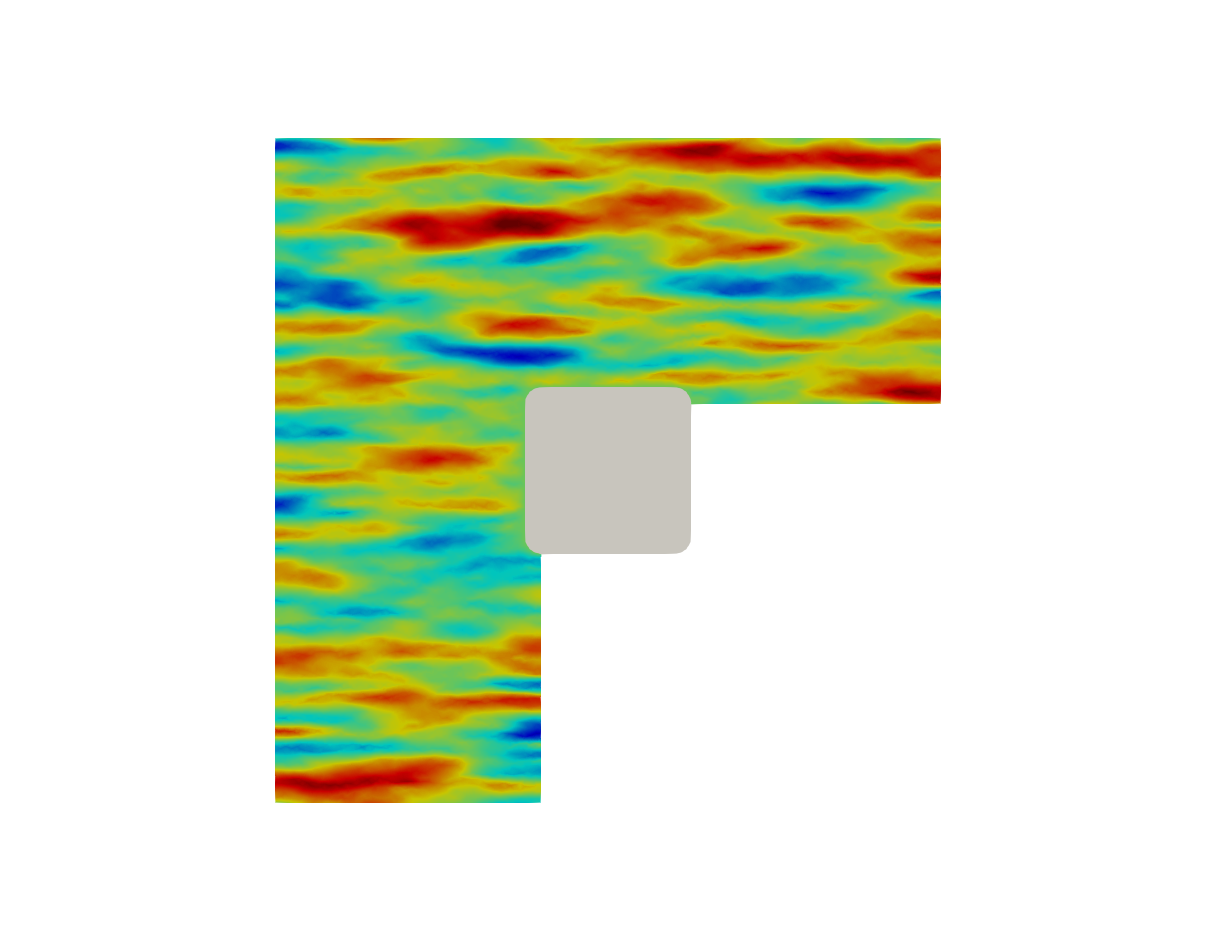}~
    \includegraphics[trim={2.5in 1.5in 2.5in 1.5in},clip,width=0.23\linewidth]{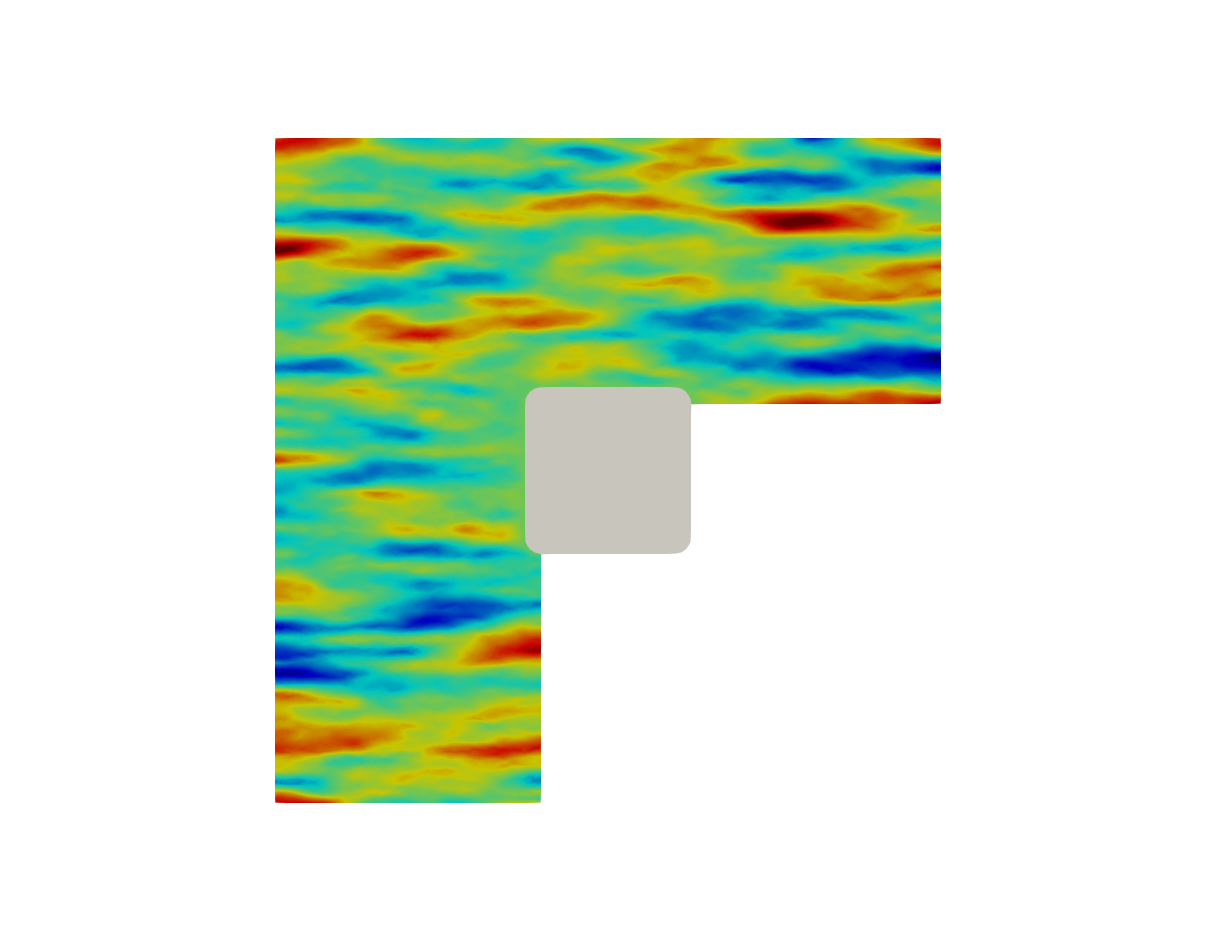}~
    \includegraphics[trim={2.5in 1.5in 2.5in 1.5in},clip,width=0.23\linewidth]{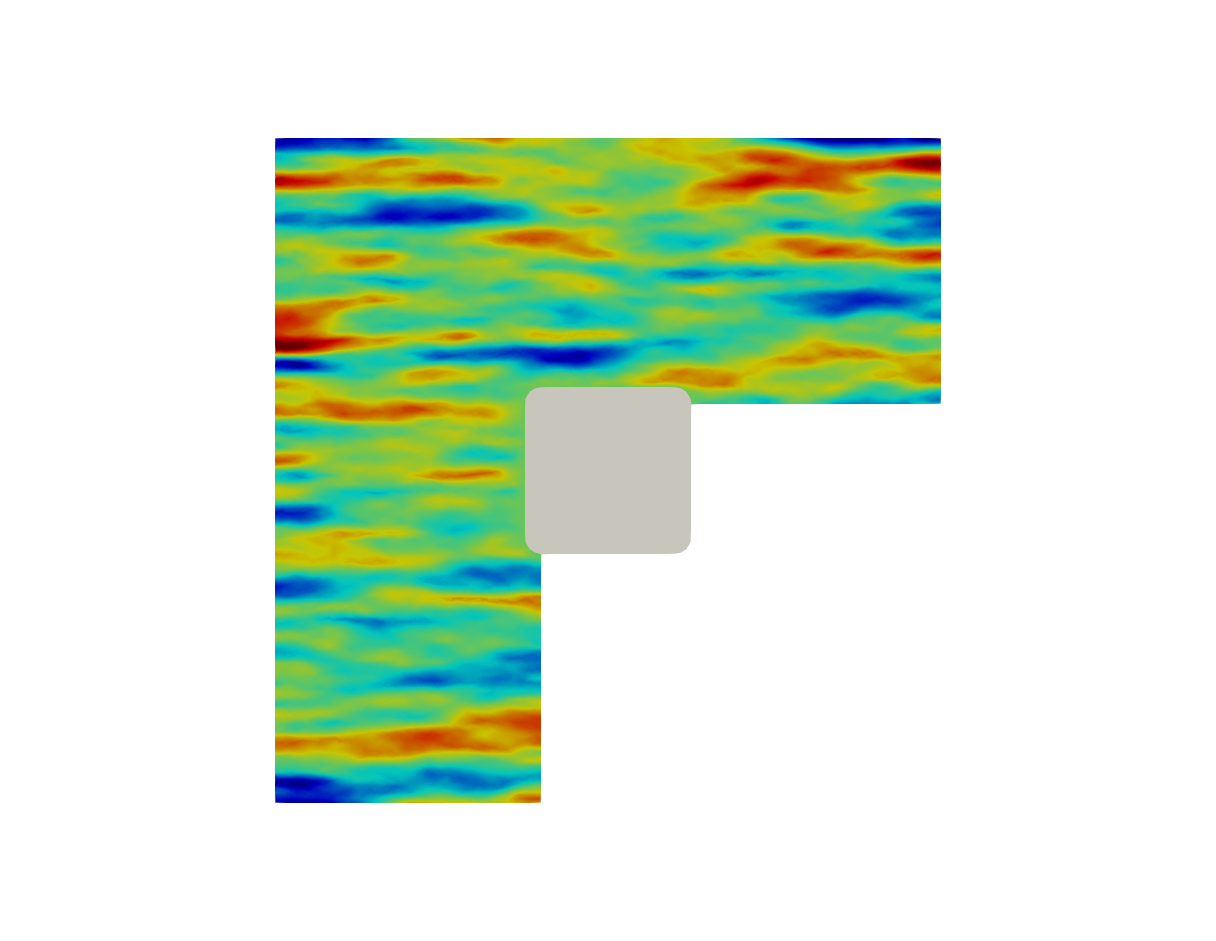}\\
    \includegraphics[width=0.3\linewidth]{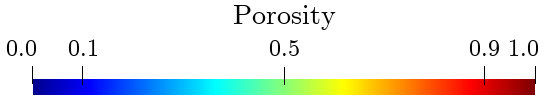}\\
    \includegraphics[trim={2.5in 1.5in 2.5in 1.5in},clip,width=0.23\linewidth]{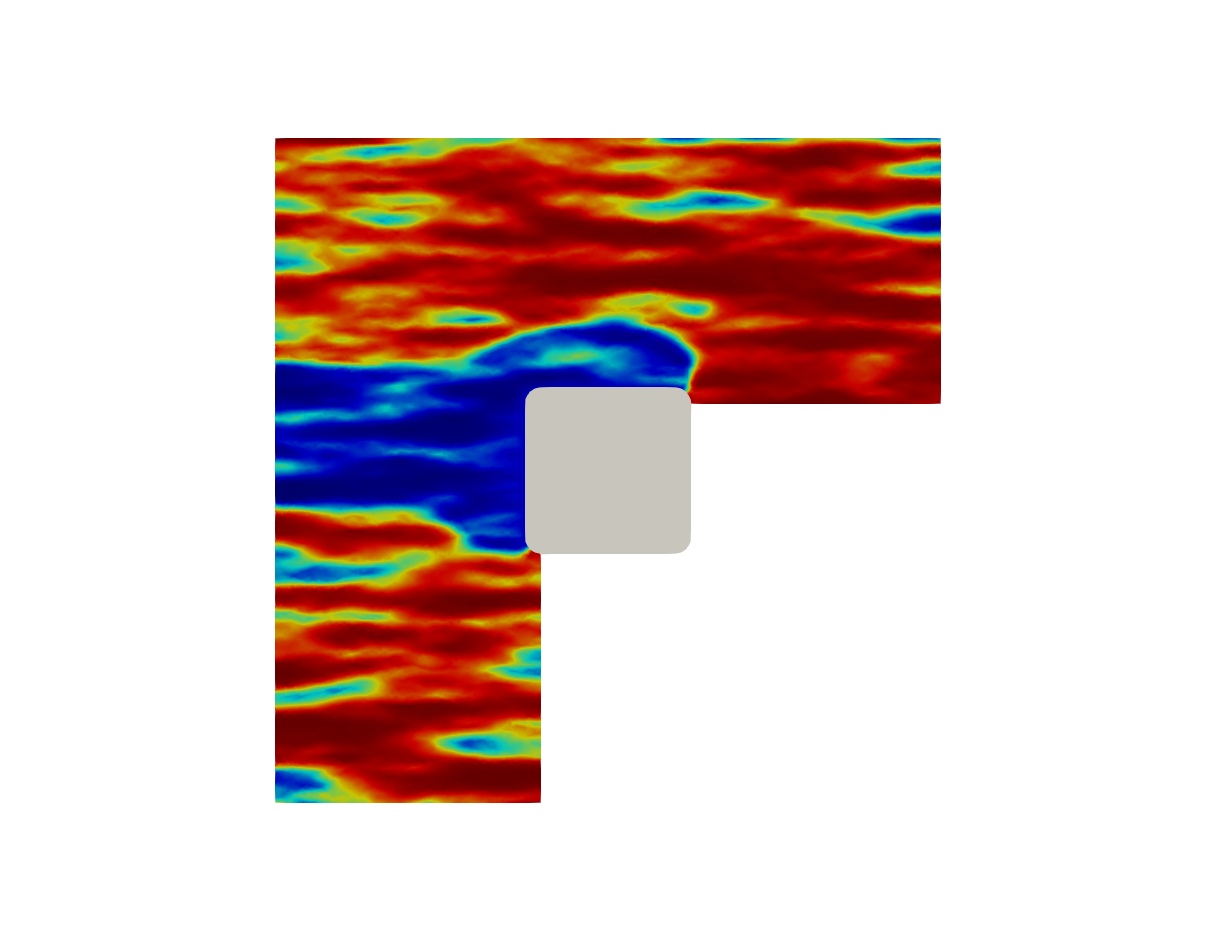}~
    \includegraphics[trim={2.5in 1.5in 2.5in 1.5in},clip,width=0.23\linewidth]{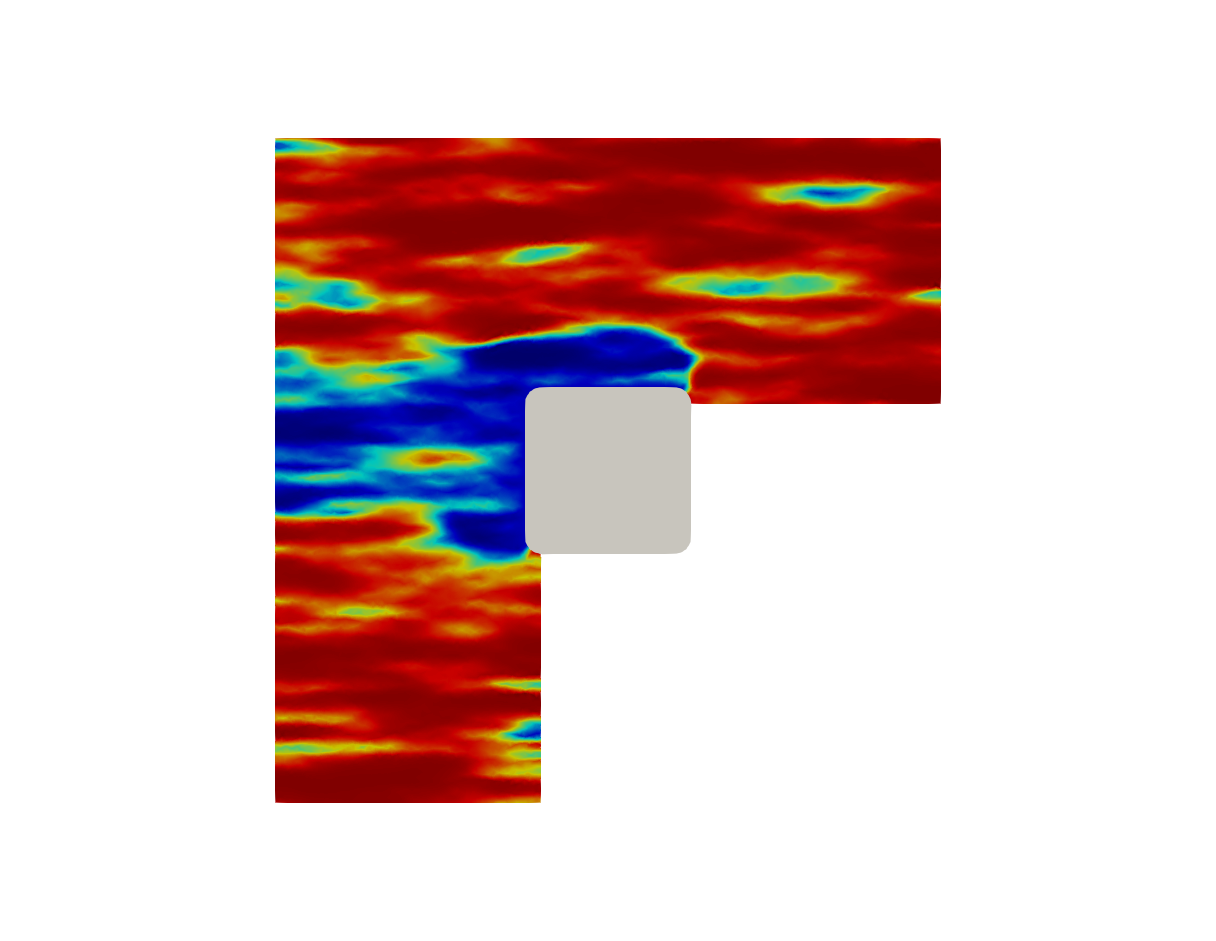}~
    \includegraphics[trim={2.5in 1.5in 2.5in 1.5in},clip,width=0.23\linewidth]{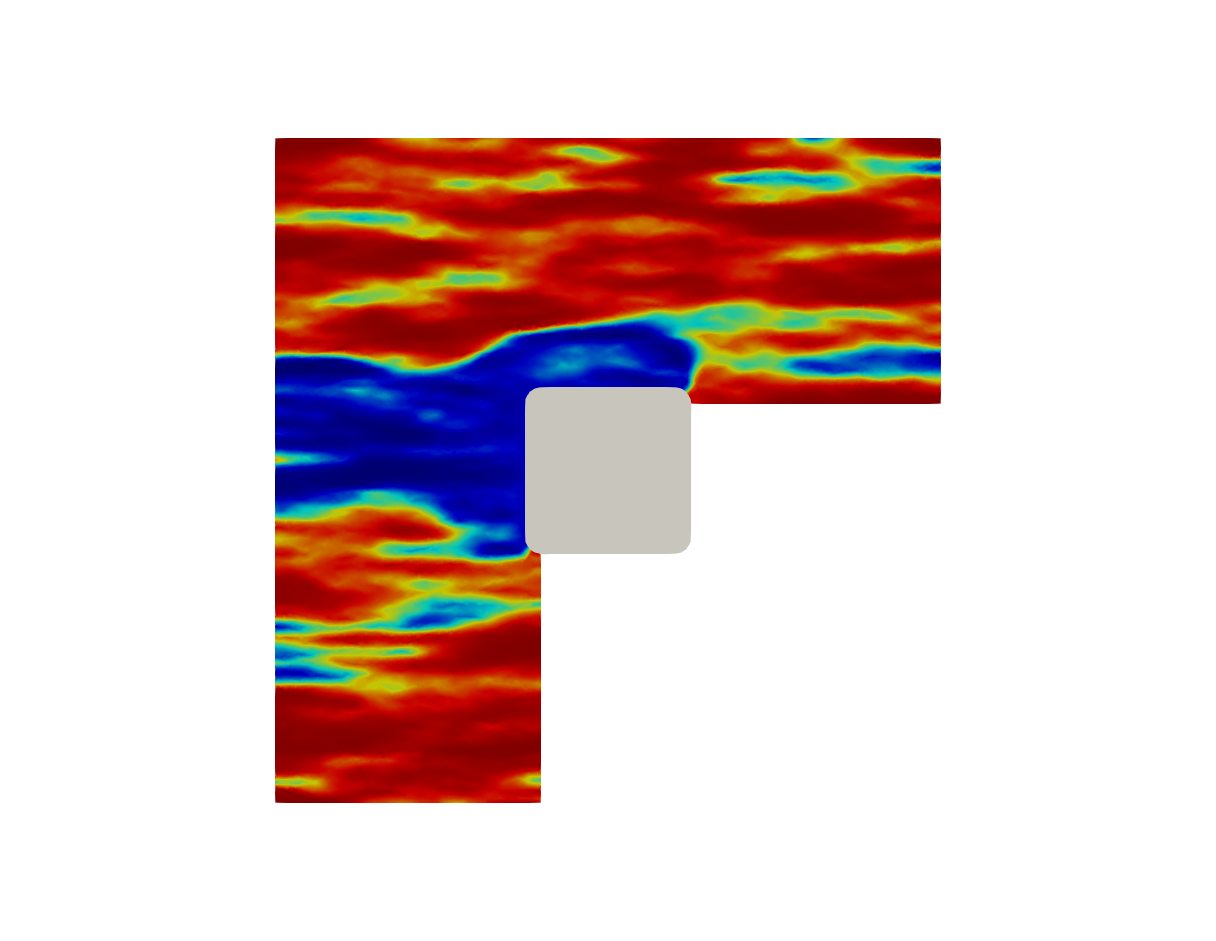}~
    \includegraphics[trim={2.5in 1.5in 2.5in 1.5in},clip,width=0.23\linewidth]{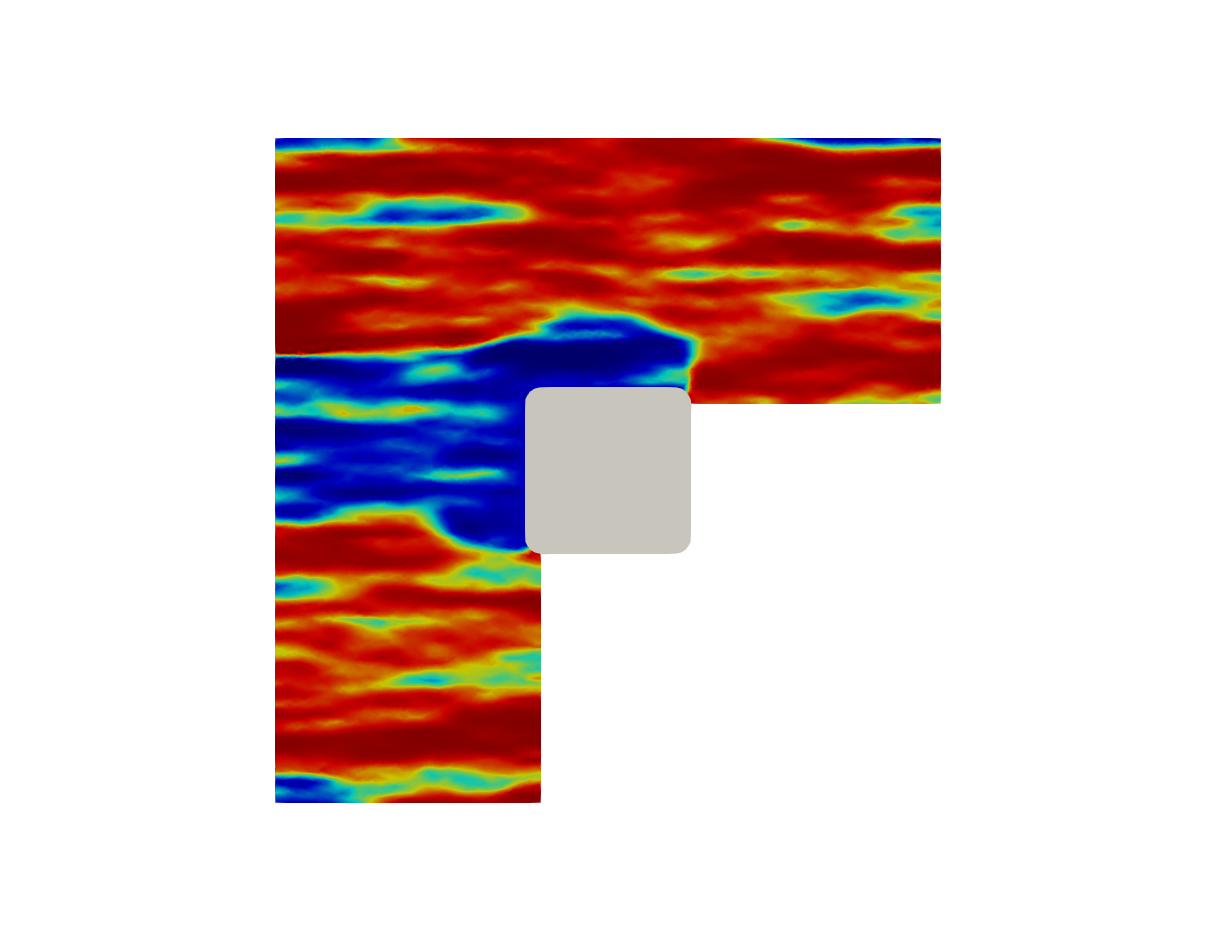}\\
    \caption{Samples of uncertain parameter $m$ and corresponding aerogel porosity field $\phi_f(\boldsymbol{x})$ for correlation length $L_{CR} = 0.25$ and variance of $\sigma^2 = {0.5}^2$. The mean of the uncertain parameter is $\Tilde{m}=0$ and anisotropy is $\upvartheta_x = 1, \upvartheta_y = 1 \times 10^{-4} $ for both cases. }
    \label{fig:porosity_samples}
\end{figure}
\subsection{Spatially-correlated uncertain parameter}
\noindent
%

\noindent
Figure \ref{fig:porosity_samples} illustrates the influence of the uncertain parameter $m$, modeled as an anisotropic Gaussian random field following \eqref{eq:matern_pde}, on the porosity field for the optimized design parameter field shown in Figure \ref{fig:HU}. In these simulations, the mean of the uncertain parameter is set to $\Bar{m}=0$, with a correlation length of $L_{CR} = 0.25$ and a variance of $\sigma^2 = {0.5}^2$.
The anisotropy tensor $\boldsymbol{\Theta}$ in \eqref{eq:matern_pde} is parameterized as $\upvartheta_x = 1, \upvartheta_y = 1 \times 10^{-4} $, indicating significantly greater variability along the $x$-direction, which corresponds to the horizontal layer-by-layer manufacturing process.
This effectively captures the heterogeneity in silica aerogel properties both between layers and within individual layers, arising from variations in pre-aerogel ink composition and differences in the drying stage during extrusion-based 3D printing of silica aerogel \cite{zhou}.

\subsection{Quadratic and Monte Carlo approximations}
\noindent
\begin{figure}[ht]
    \centering
    \includegraphics[trim={0.0in 1.5in 0.7in 2.0in},clip,width=0.48\linewidth]{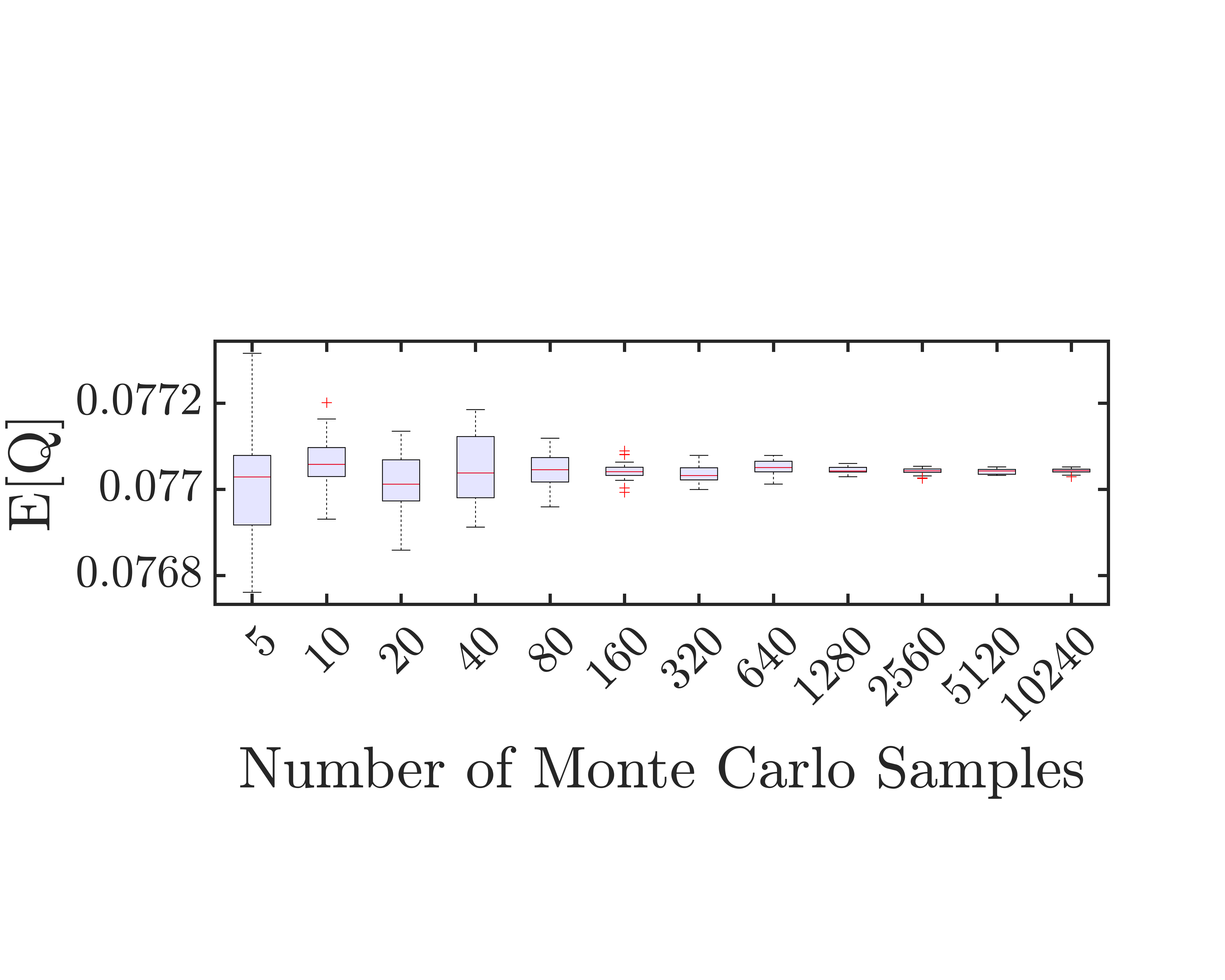}~
    \includegraphics[trim={0.0in 1.5in 0.7in 2.0in},clip,width=0.48\linewidth]{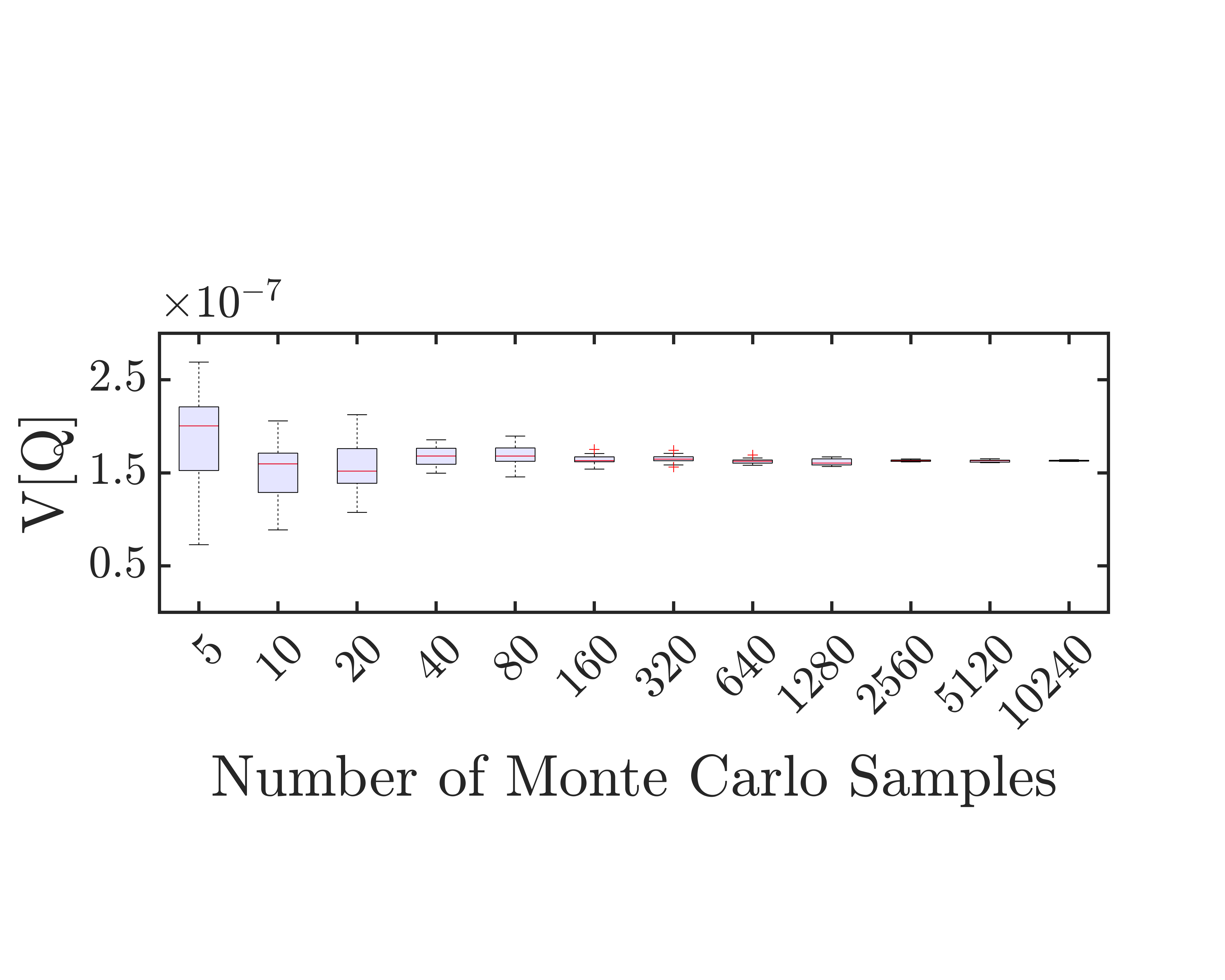}
    \\ (a) \\
    \includegraphics[trim={0.0in 1.5in 0.7in 2.0in},clip,width=0.48\linewidth]{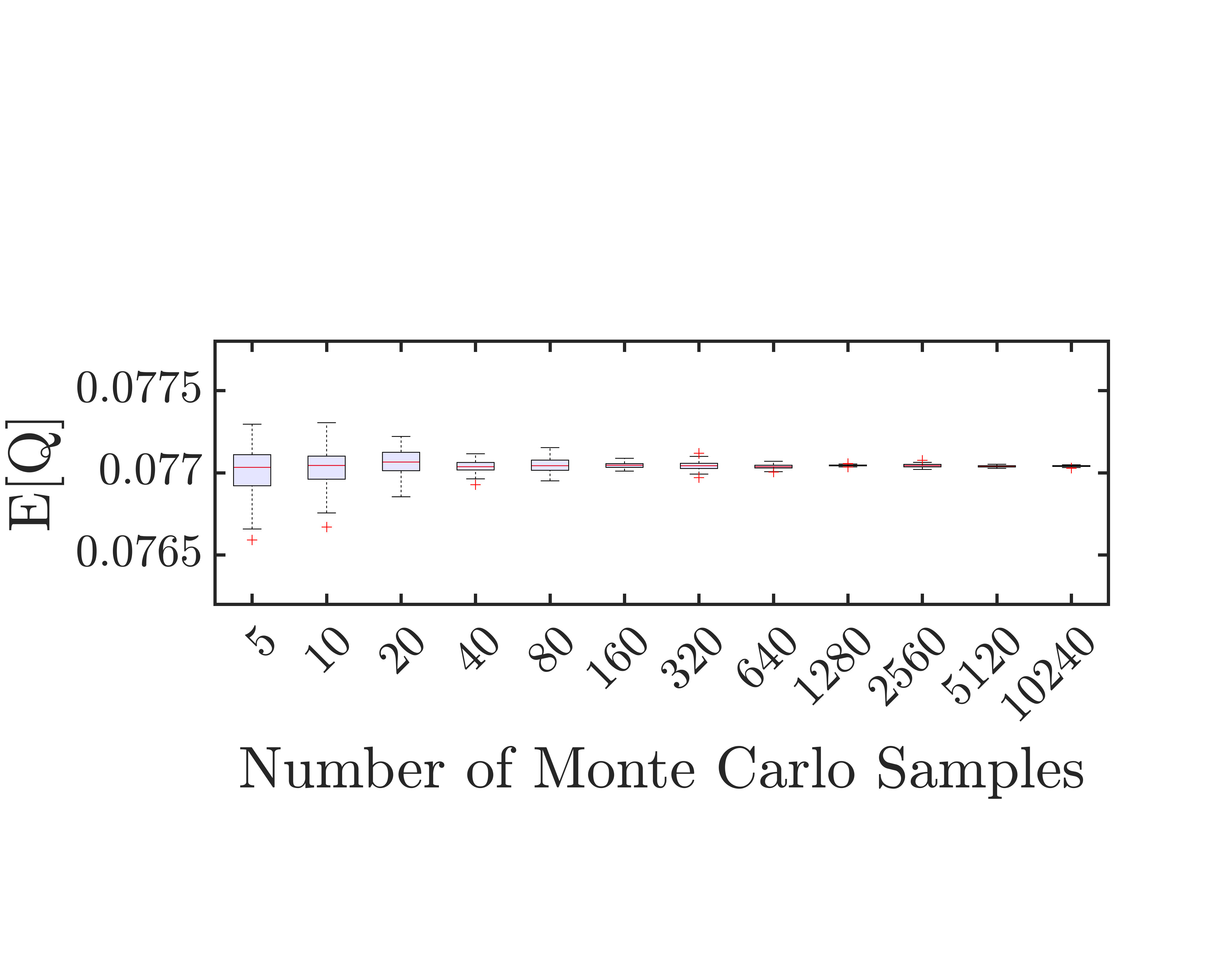}~
    \includegraphics[trim={0.0in 1.5in 0.7in 2.0in},clip,width=0.48\linewidth]{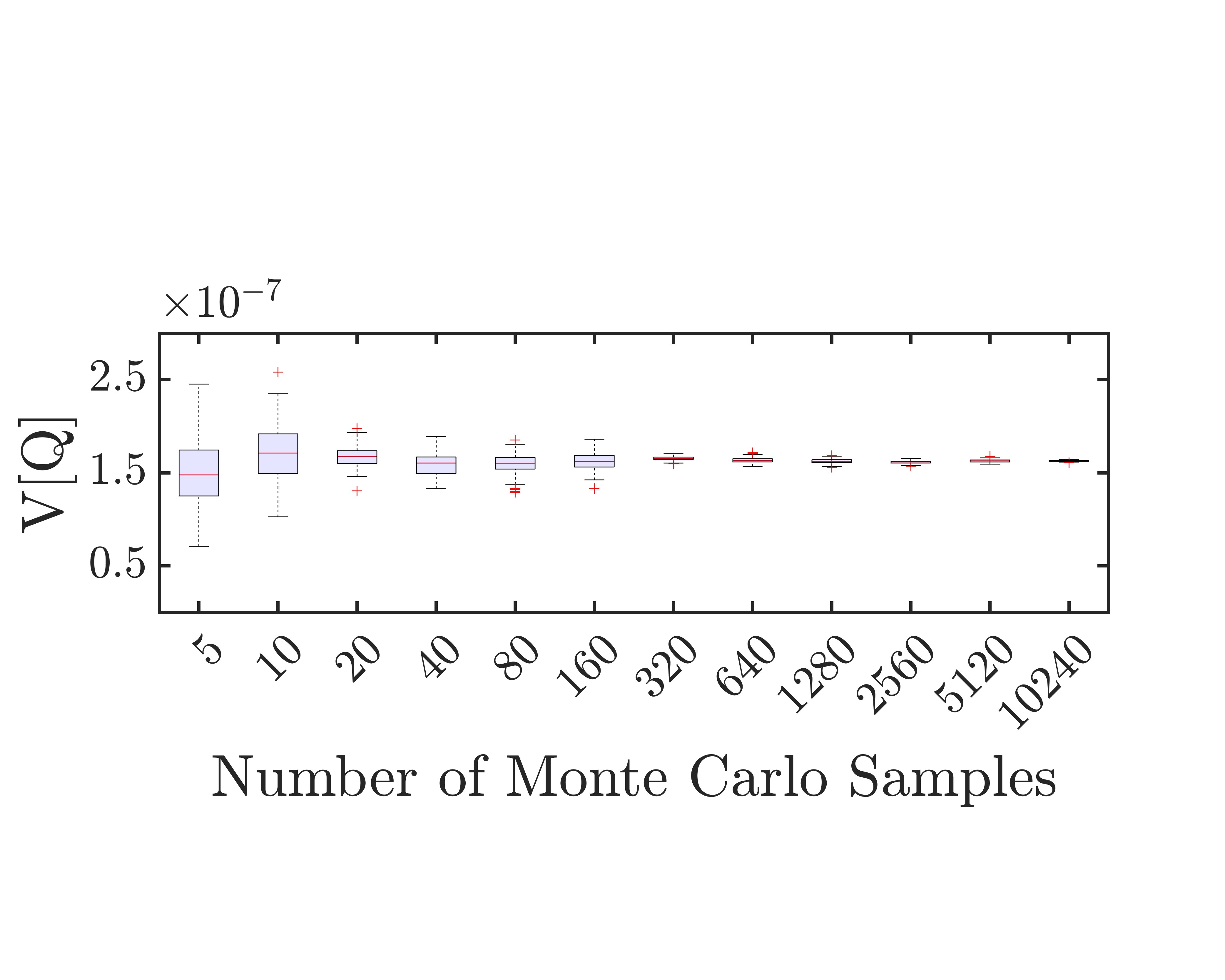}\\ (b) \\
    \includegraphics[trim={0.0in 2in 0.7in 2.0in},clip,width=0.48\linewidth]{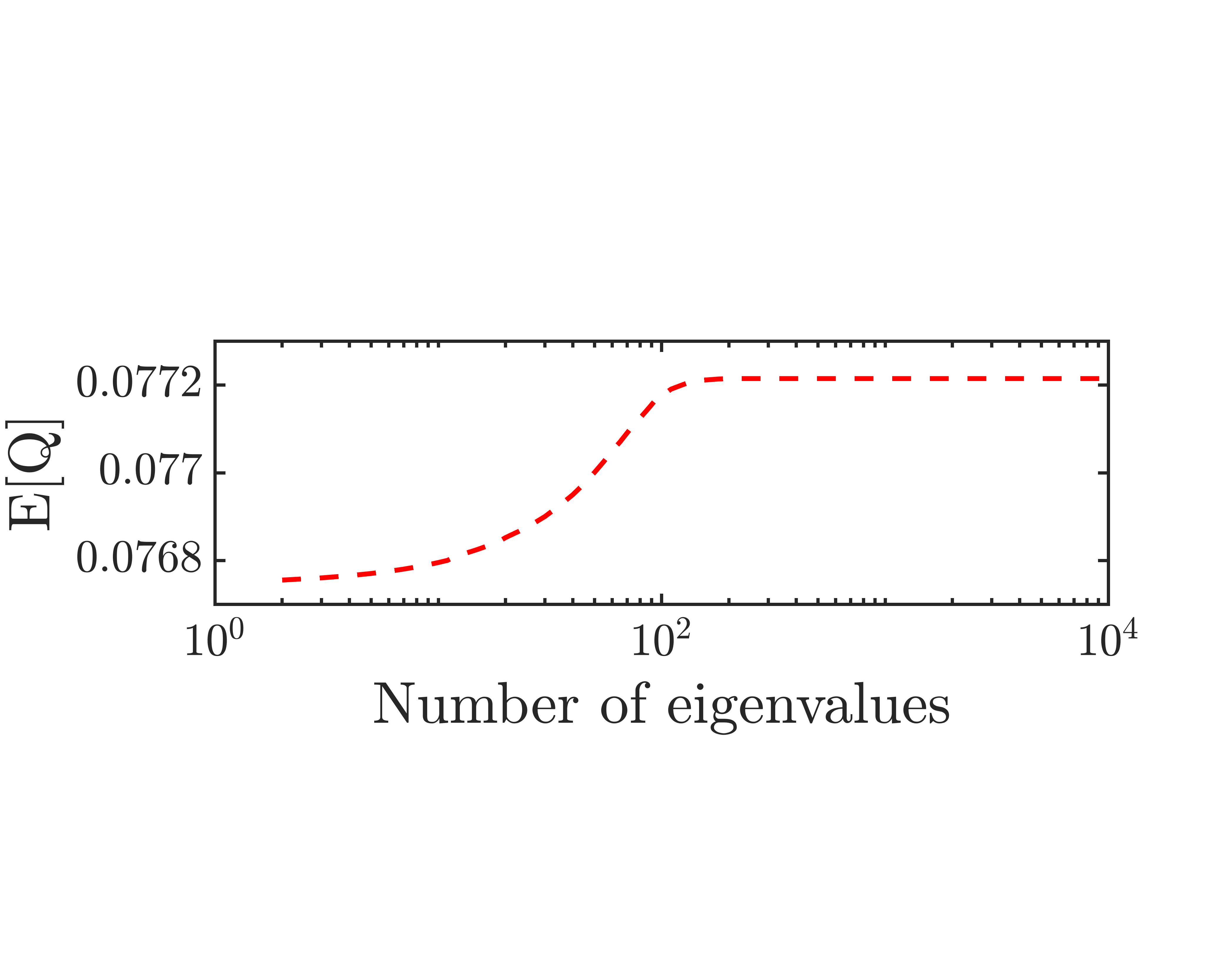}~
    \includegraphics[trim={0.0in 2in 0.7in 2.0in},clip,width=0.48\linewidth]{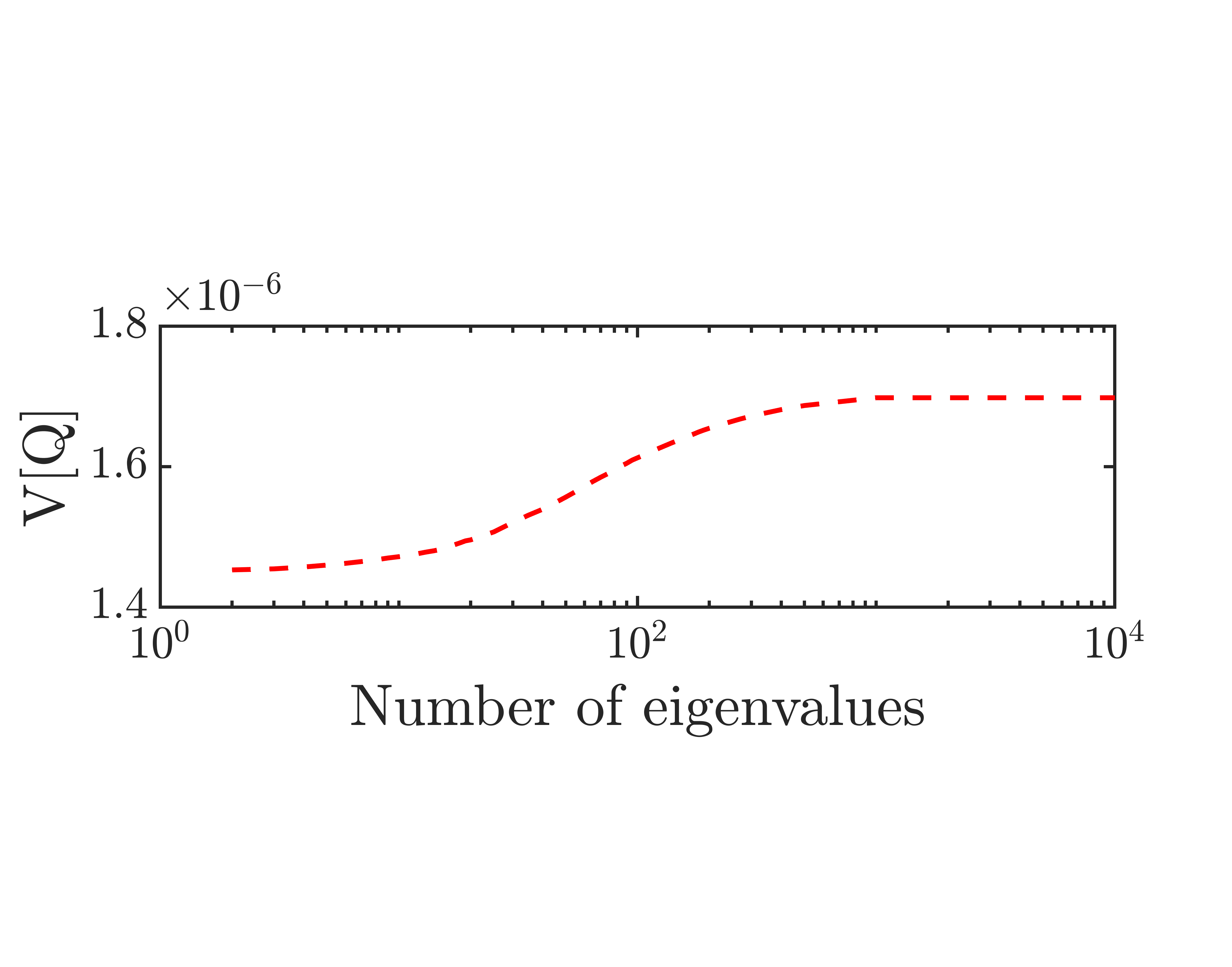}\\ (c) \\
    \caption{Convergence plots corresponding to scenario in Figure \ref{fig:LU} for the mean and variance of the design objective $Q$ : (a) standard Monte Carlo method (b) Monte Carlo with control variate and (c) Quadratic approximation. The uncertain parameter $m$ has the correlation length $L_{CR} =0.25$ and a variance of $\sigma^2 ={0.5}^2$. }
    \label{fig:convergence_plot}
\end{figure}
\begin{figure}[h]
    \centering
    \includegraphics[trim={3.5in 1.75in 3.5in 1.75in},clip,width=0.25\linewidth]{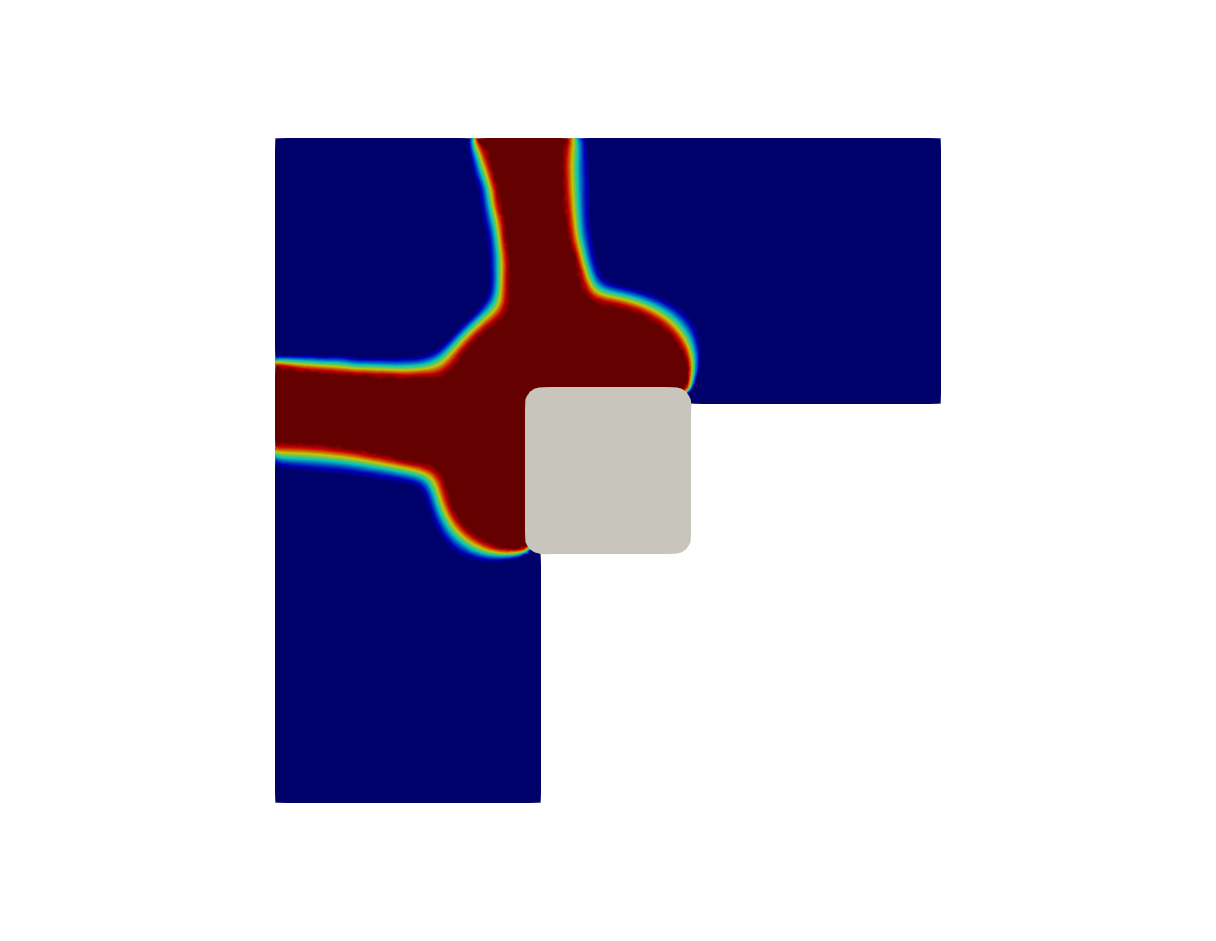}~
    \includegraphics[trim={3.5in 1.75in 3.5in 1.75in},clip,width=0.25\linewidth]{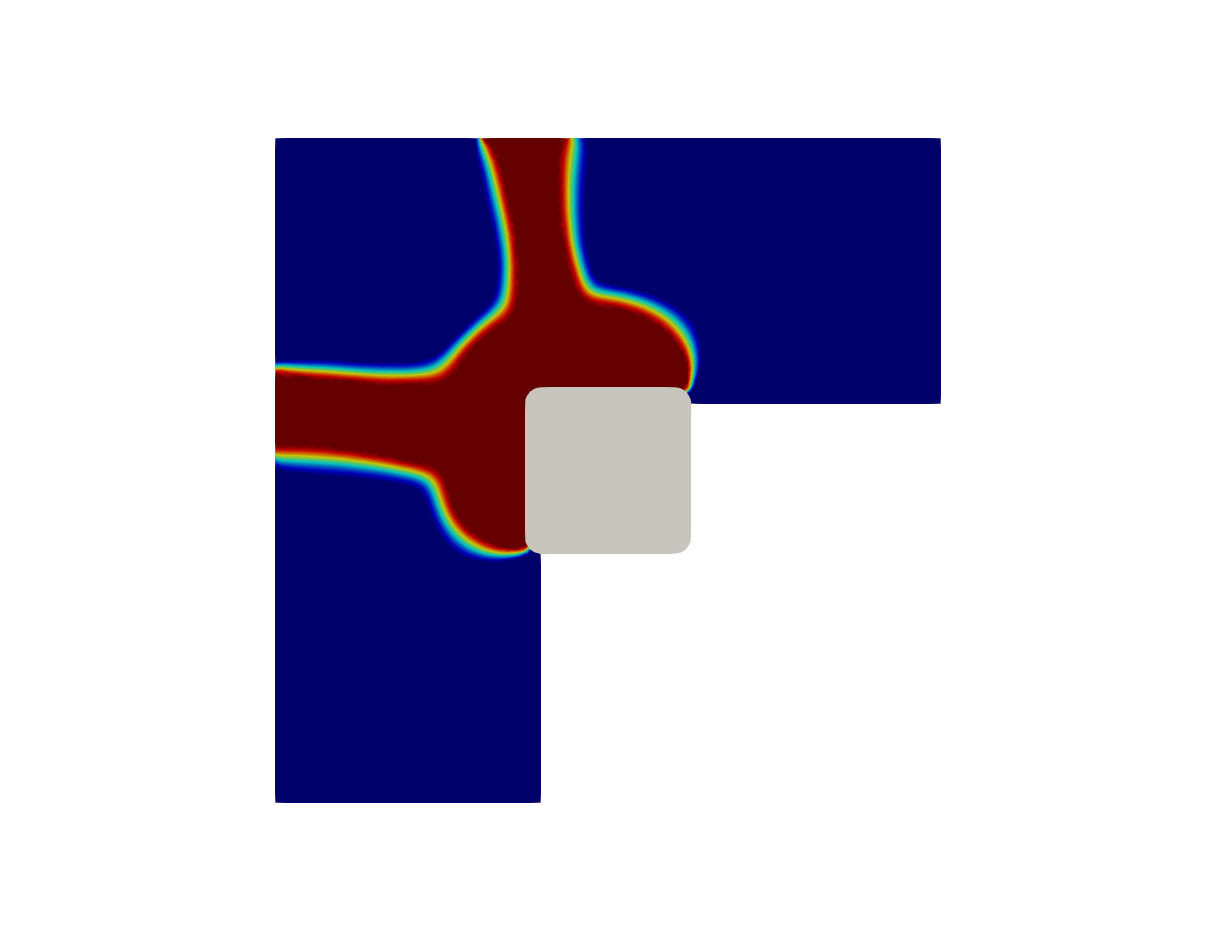}~
    \includegraphics[trim={3.5in 1.75in 3.5in 1.75in},clip,width=0.25\linewidth]{Figures/LU2.png}~
    \includegraphics[width=0.08\linewidth]{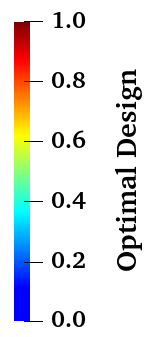}\\
    \hspace{-0.35in} (a) \hspace{1.2in} (b) \hspace{1.2in} (c) \\
    \includegraphics[trim={0in 0.4in 0in 0.8in},clip,width=0.45\linewidth]{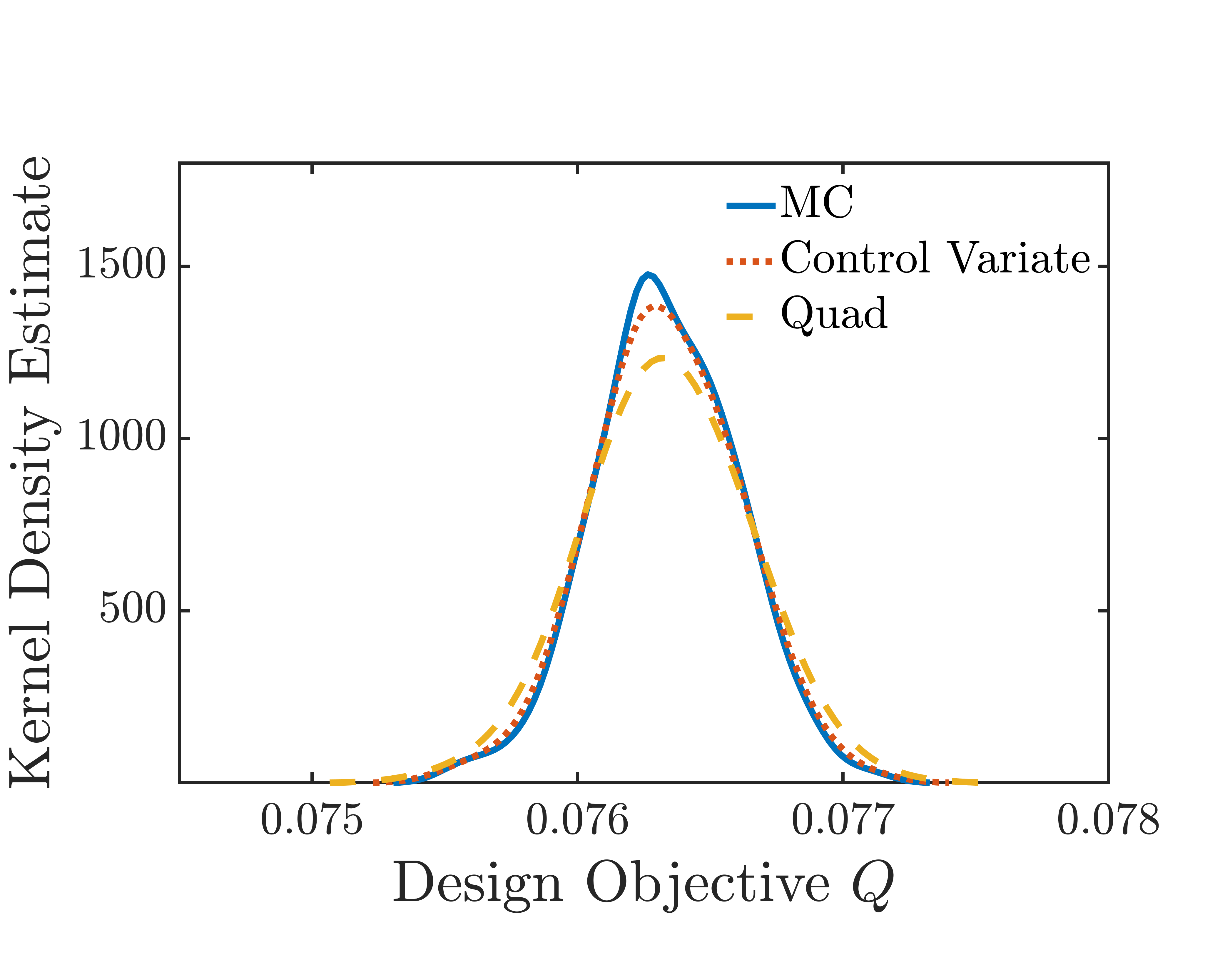}
    \vspace{-0.02in}
    \caption{The top row shows the optimal design obtained for $\sigma=0.25$ for the three different approaches: (a) Sampled Average approximation (500 samples) (b) Monte Carlo correction (100 samples) and (c) Quadratic approximation ($N_{eig}^{Q}=25$). The bottom row shows the QoI distribution of the design objective $Q$ for the three different cases above. The uncertain parameter $m$ has a correlation length $L_{CR} = 0.05$ for all three cases.}
    \label{fig:LU}
\end{figure}
\begin{figure}[h]
    \centering
    \includegraphics[trim={3.5in 1.75in 3.5in 1.75in},clip,width=0.25\linewidth]{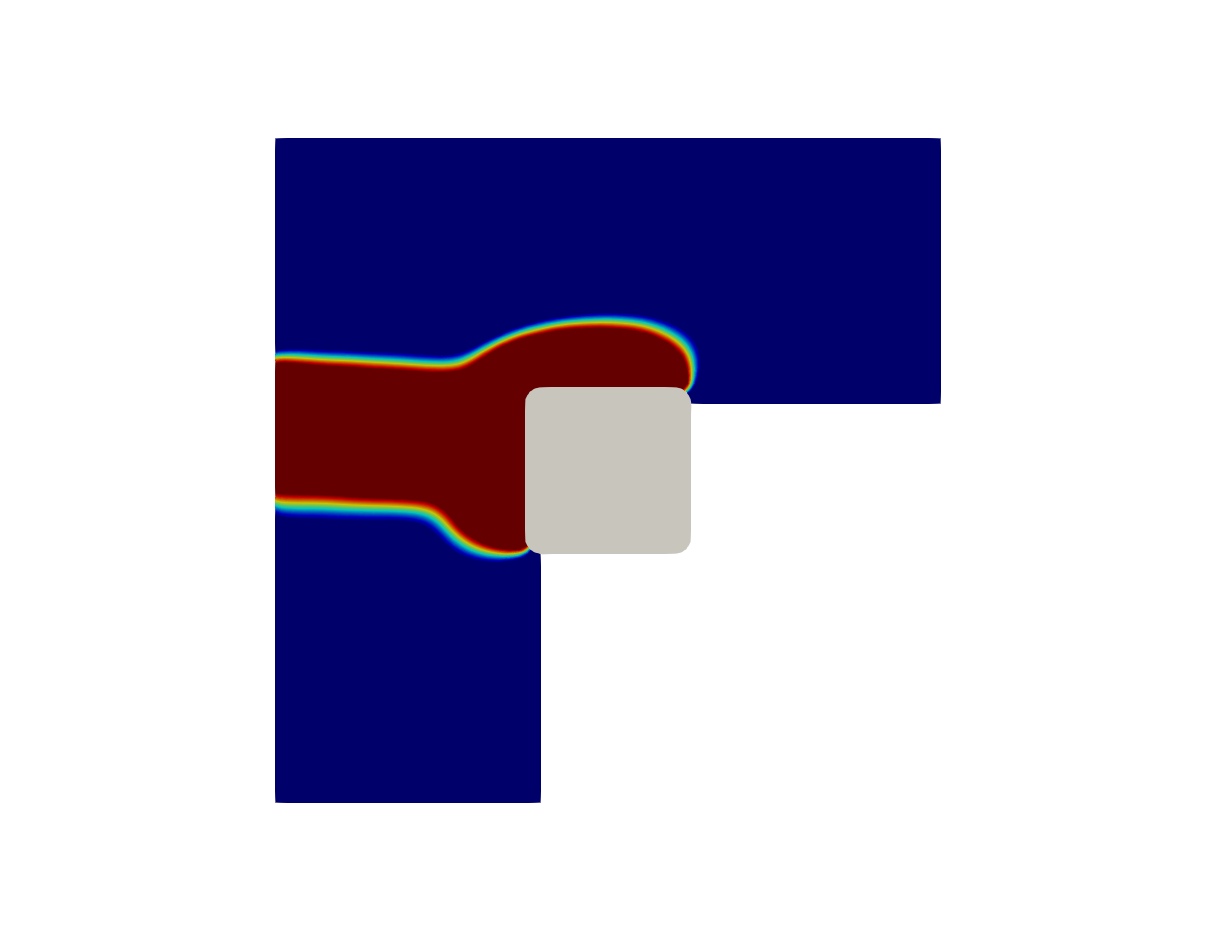}~
    \includegraphics[trim={3.5in 1.75in 3.5in 1.75in},clip,width=0.25\linewidth]{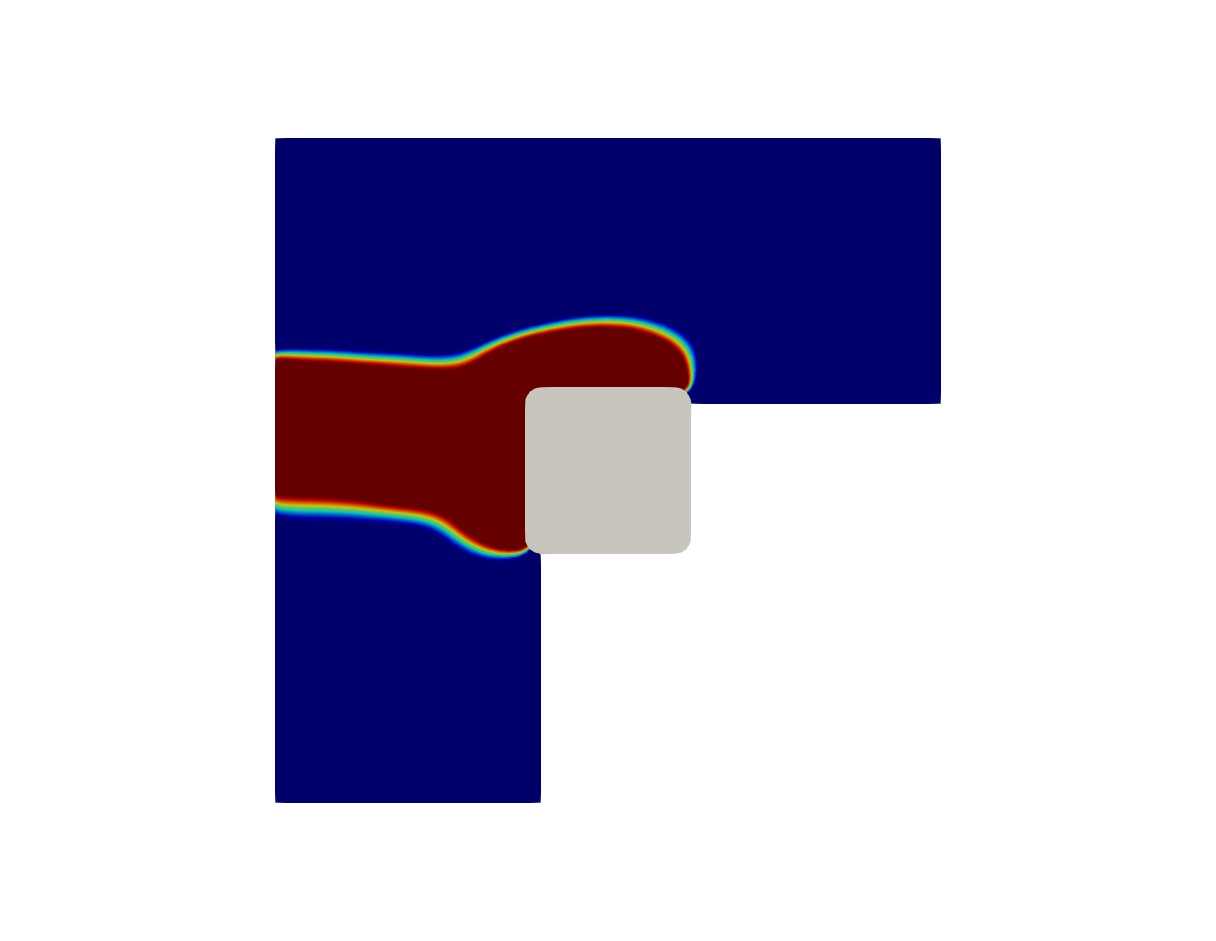}~
    \includegraphics[trim={3.5in 1.75in 3.5in 1.75in},clip,width=0.25\linewidth]{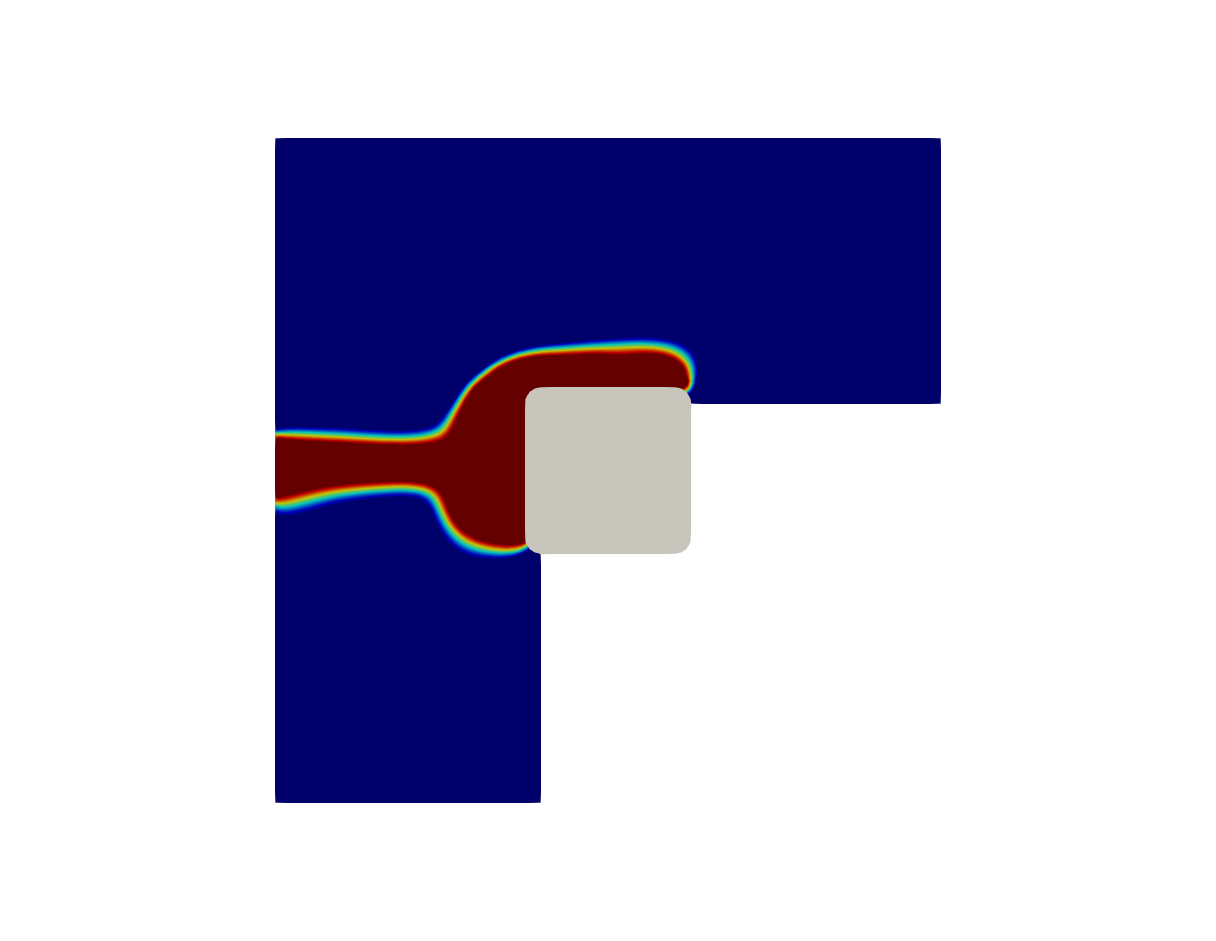}~
    \includegraphics[width=0.08\linewidth]{Figures/design_contour.png}\\
    \includegraphics[trim={0in 0.4in 0in 0.8in},clip,width=0.45\linewidth]{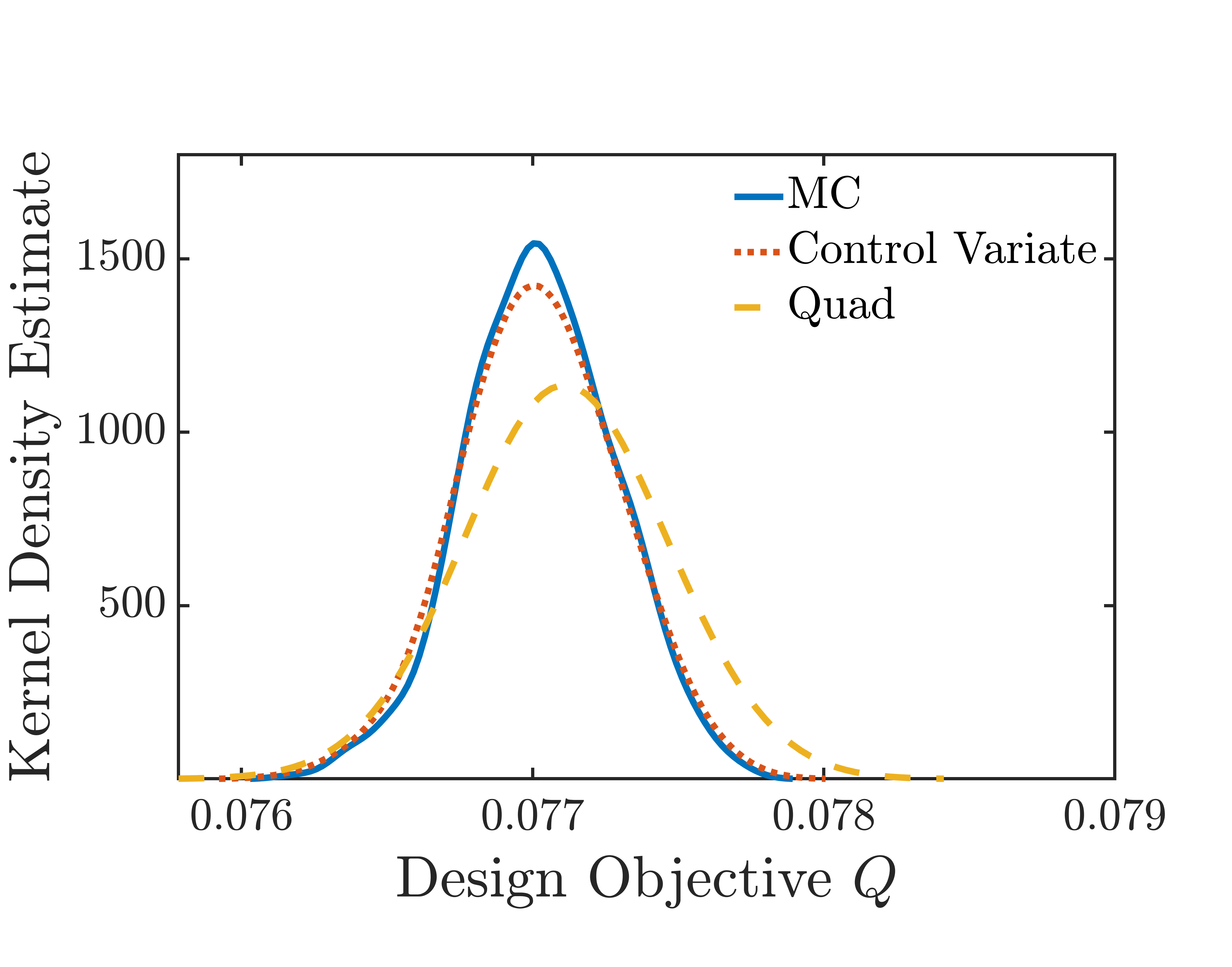}
    \vspace{-0.02in}
    \caption{The top row shows the optimal design obtained for $\sigma=0.5$ for the three different approaches: (a) standard Monte Carlo using 500 samples, (b) Monte Carlo with control variate using 100 samples, and (c) Quadratic approximation using $N_{eig}^{Q}=25$. The bottom row shows the QoI distribution of the design objective $Q$ for the three different cases above. The uncertain parameter $m$ has a correlation length of $L_{CR} = 0.25$ for all three cases.}
    \label{fig:HU}
\end{figure}
To evaluate the accuracy and efficiency of the quadratic approximation and Monte Carlo estimations with a control variate in optimal design, we perform comparative analyses. Figure \ref{fig:convergence_plot} shows that, as the sample size and number of dominant eigenvalues increase, both methods converge to similar mean and variance values, aligning with standard Monte Carlo estimations
For large sample sizes, the estimated mean and variance from standard Monte Carlo (Figure \ref{fig:convergence_plot}(a)) and Monte Carlo with a control variate (Figure \ref{fig:convergence_plot}(b)) are nearly identical. However, the quadratic approximation as a control variate achieves lower error bounds with significantly fewer samples, as shown in the box plots. Notably, with only $N_{eig}^{Q}=25$ trace estimators, the quadratic approximation closely matches Monte Carlo results (Figure \ref{fig:convergence_plot}(c)), substantially reducing computational cost while maintaining accuracy.

Figures \ref{fig:LU} and \ref{fig:HU} compare the effects of different variances and correlation lengths of the uncertain parameter $m$ on the optimal design $d_{\text{opt}}$ indicating the spatial distribution of aerogel porosity and the probability distributions of the design objective $Q$ representing the thermal insulation performance of the thermal break. The optimal design is evaluated using standard Monte Carlo, Monte Carlo with control variate, and a quadratic Taylor approximation of the mean and variance of $Q$.
In Figure \ref{fig:LU} (low variance and short correlation length), the morphologies of $d_{\text{opt}}$ and probability distributions of $Q$ are nearly identical across all three methods. However, in Figure \ref{fig:HU} (high variance and long correlation length), the quadratic approximation predicts different optimal design pattern with higher $Q$, while Monte Carlo with control variate yields results closely matching standard Monte Carlo but with significantly fewer samples (PDEs solved).
Interestingly, while the spatial pattern of the optimal design in Figure \ref{fig:LU} appears nearly symmetric, the higher uncertainty in Figure \ref{fig:HU} results in optimal designs that consistently position stronger material (lower aerogel porosity) along the horizontal axis, across all three approximation methods. This pattern reflects the anisotropy of the uncertain parameter, modeled with a higher correlation length along the layer deposition direction (see Figure \ref{fig:porosity_samples}). Also, the optimal designs at mean $\Bar{m}=0$ are much similar in Figure \ref{fig:LU} as compared to Figure \ref{fig:HU} where there is a deviation in the design patterns, which shows that there is a trade-off favoring the reduction of the uncertainty over mean performance for our design formulation \eqref{eq:cost}. These results demonstrate that the proposed optimization under uncertainty framework effectively captures spatial uncertainty, producing more reliable design solutions.

\subsection{Scalability with respect to number of design parameters}
\begin{figure}[ht]
    \centering
    \includegraphics[width=0.31\linewidth]{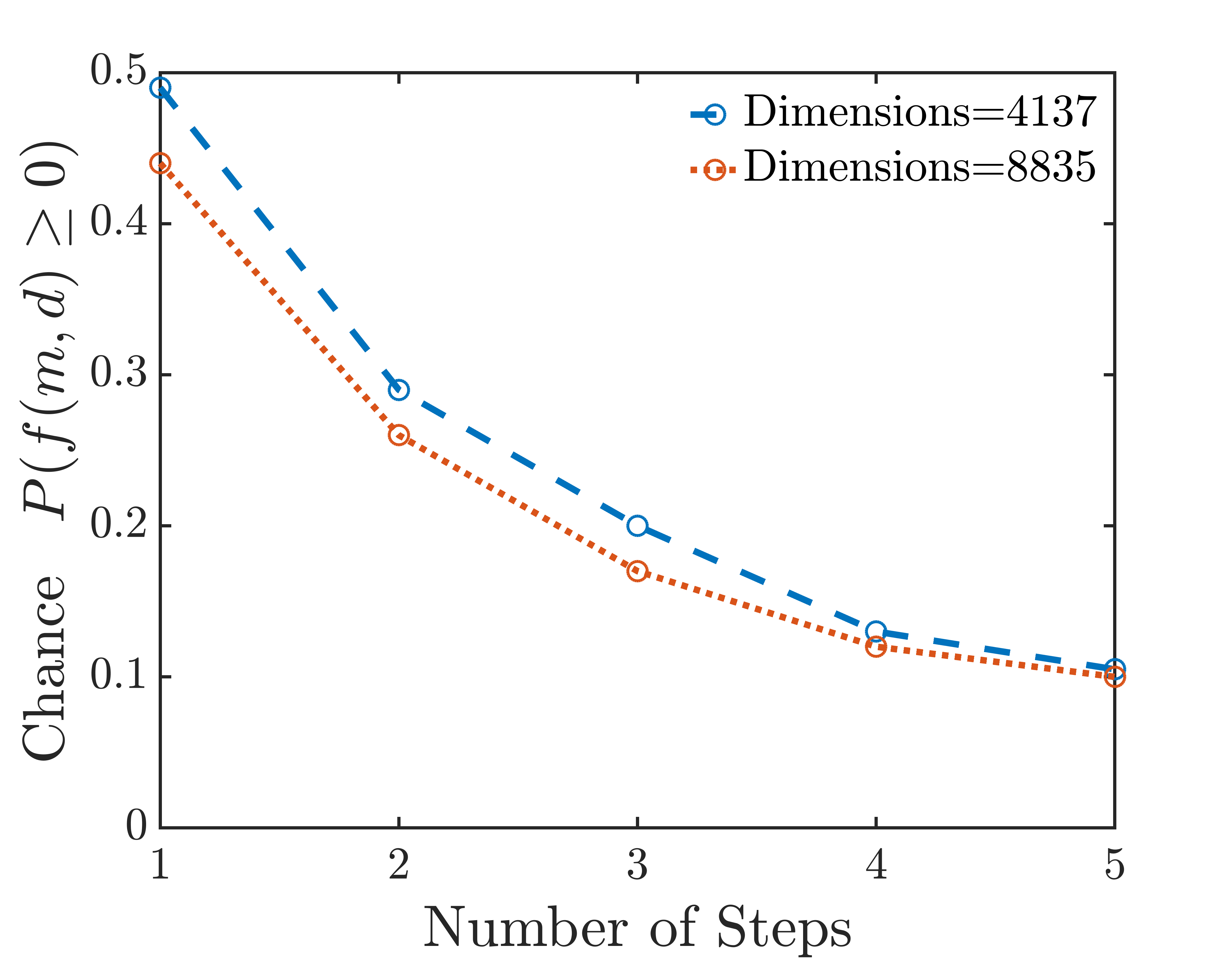}
    ~
    \includegraphics[width=0.31\linewidth]{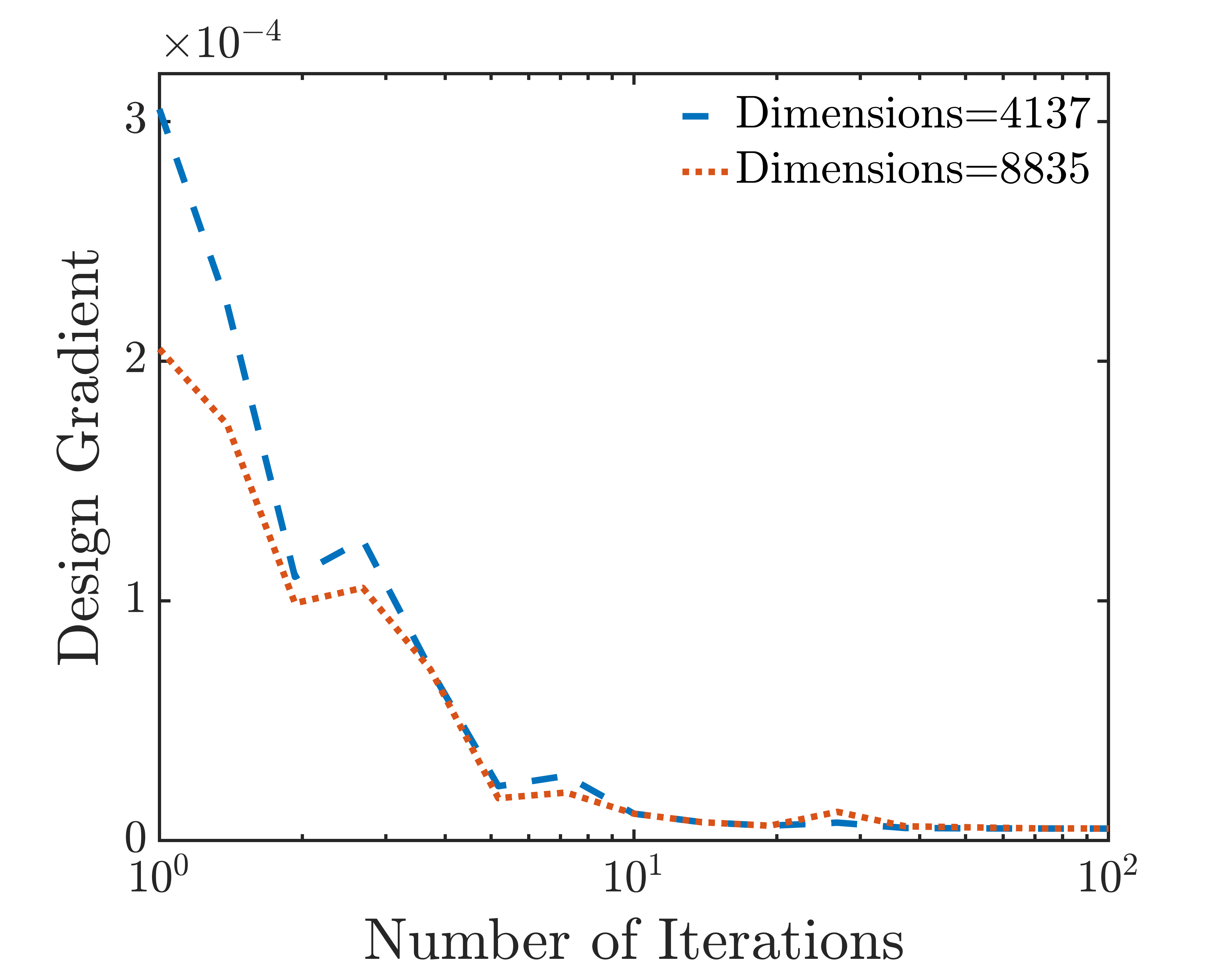}
    ~
    \includegraphics[width=0.31\linewidth]{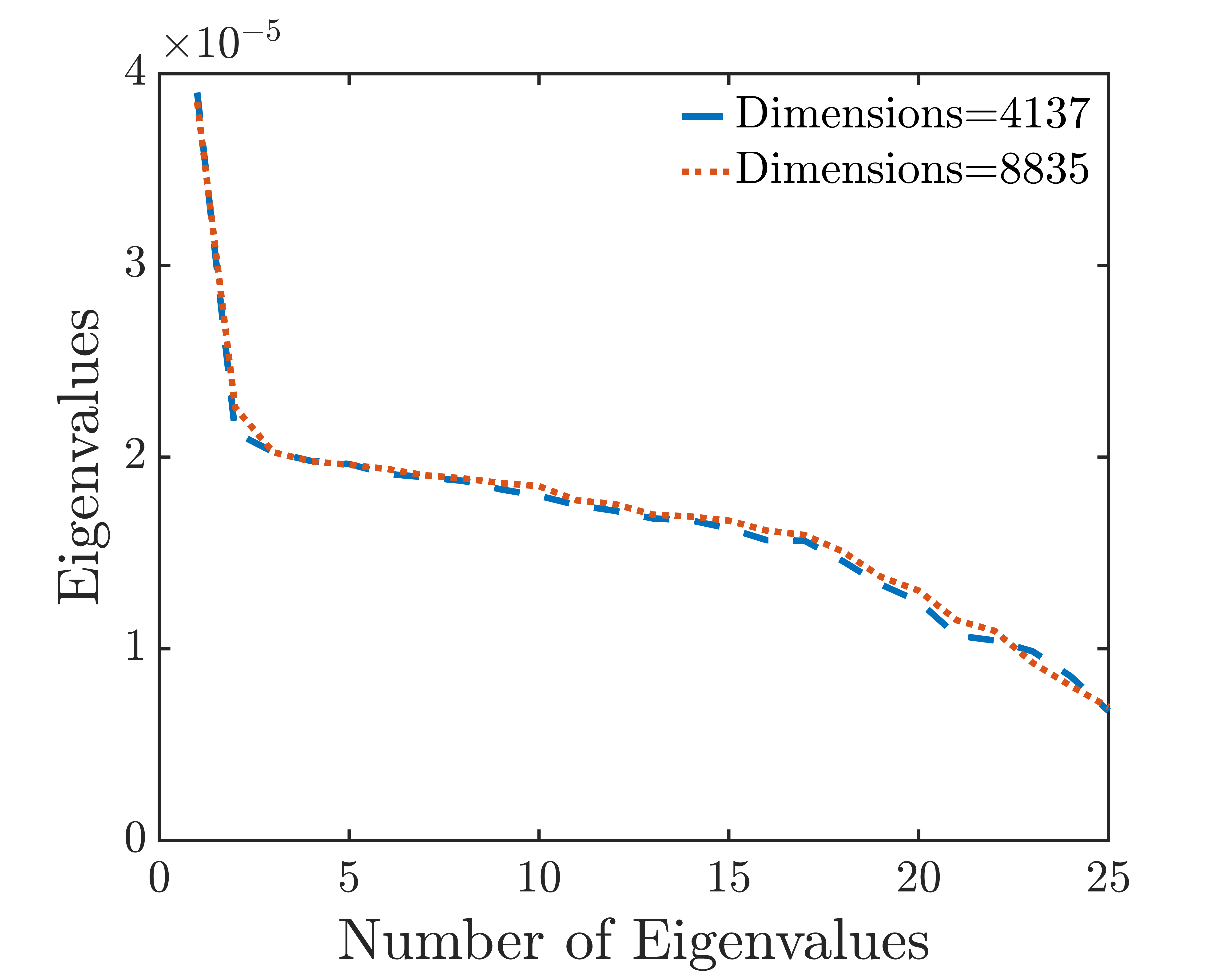}
    \\
    (a) \hspace{1.5 in} (b) \hspace{1.5in} (c)\\
    \caption{Optimization solution corresponding to Figure \ref{fig:LU} (c) for different dimensions of design parameter: (a) Convergence of the chance $P(f(m,d) \geq 0)$ with the number of steps, (b) convergence of the norm of the design gradient with number of iterations of INCG optimizer, (c) eigenvalue decay of the covariance-preconditioned Hessian}
    \label{fig:scalability}
\end{figure}
%

We assess the scalability of the algorithm for optimization under uncertainty with respect to the design dimension, determined by the number of nodes in the spatial discretization of $d$, and governed by two primary factors: 
Firstly, the eigenvalue decay rate of the covariance-preconditioned Hessian, estimated via the randomized trace estimator, and secondly the convergence behavior of the INCG optimizer. Figure \ref{fig:scalability} illustrates the behavior of the chance constraint function in \eqref{eq:chance_function} with the number of steps in the continuation scheme, the gradient norm \ref{eq:design_gradient} and the absolute values of the generalized eigenvalues for the last step of chance convergence discussed in Algorithm \ref{algo:adaptive_optimization}. These metrics are presented for two design parameter dimensions, corresponding to the optimal design problem depicted in Figure \ref{fig:LU} (c).

These results demonstrate that the quadratic approximation enables rapid decay of the critical chance, with the gradient norm converging below the tolerance threshold within 100 iterations for both mesh resolutions. The minimal dependence of convergence on mesh resolution highlights that the gradient-based INCG optimizer is effectively independent of the number of uncertain parameters. Additionally, Figure \ref{fig:scalability}(c) shows a rapid eigenvalue decay, unaffected by the dimension of uncertain parameters, supporting the scalability of the randomized eigensolver used for trace approximation.
Together, these dimension-independent factors suggest that leveraging quadratic Taylor approximations and stochastic optimization techniques enables a scalable approach for optimal design under high-dimensional uncertainty with the enforcement of chance constraints.

\subsection{Optimal design: effect of chance constraint}
\noindent
\begin{figure}[ht]
    \centering
    \footnotesize
    \includegraphics[trim={3.5in 1.75in 3.5in 1.75in},clip,width=0.2\linewidth]{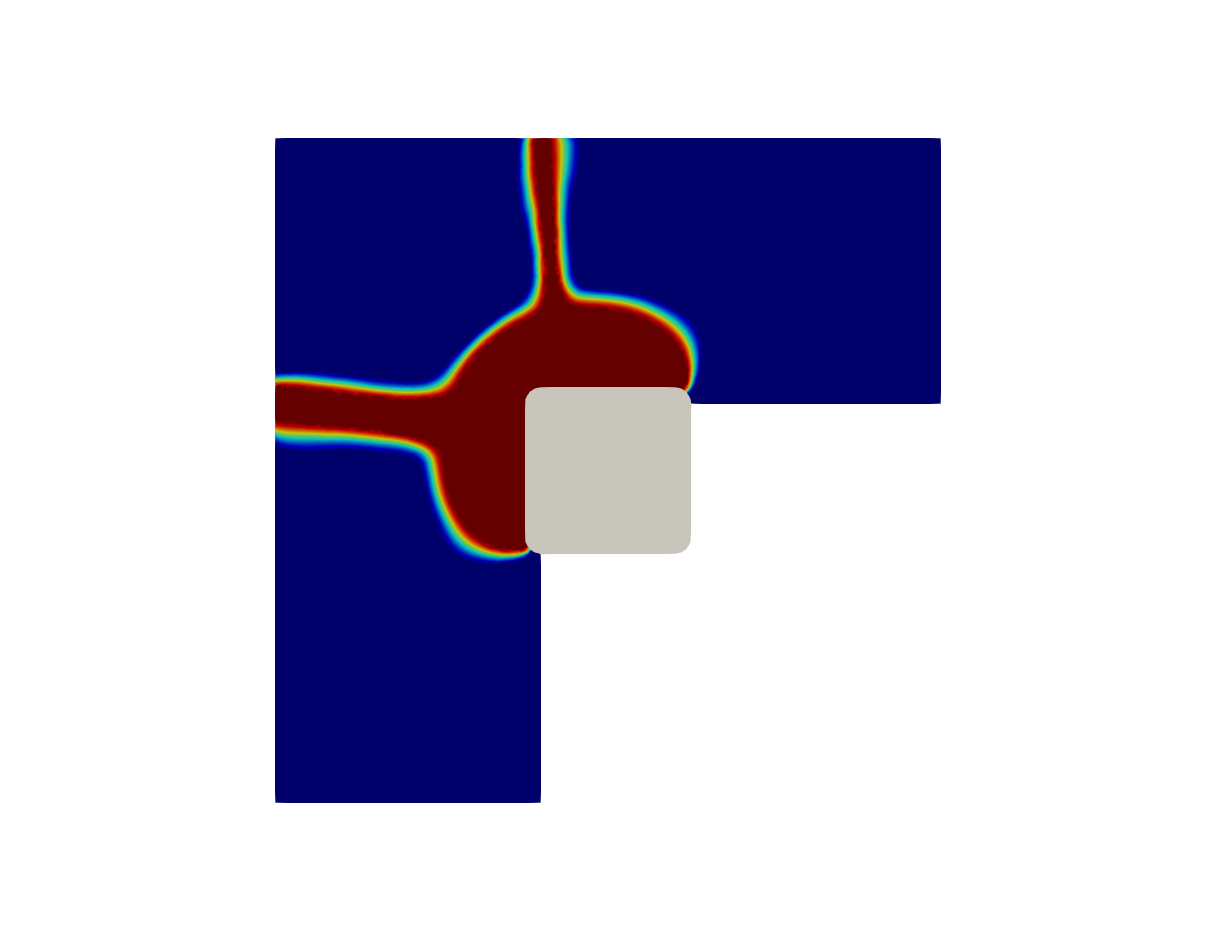}~
    \includegraphics[trim={3.5in 1.75in 3.5in 1.75in},clip,width=0.2\linewidth]{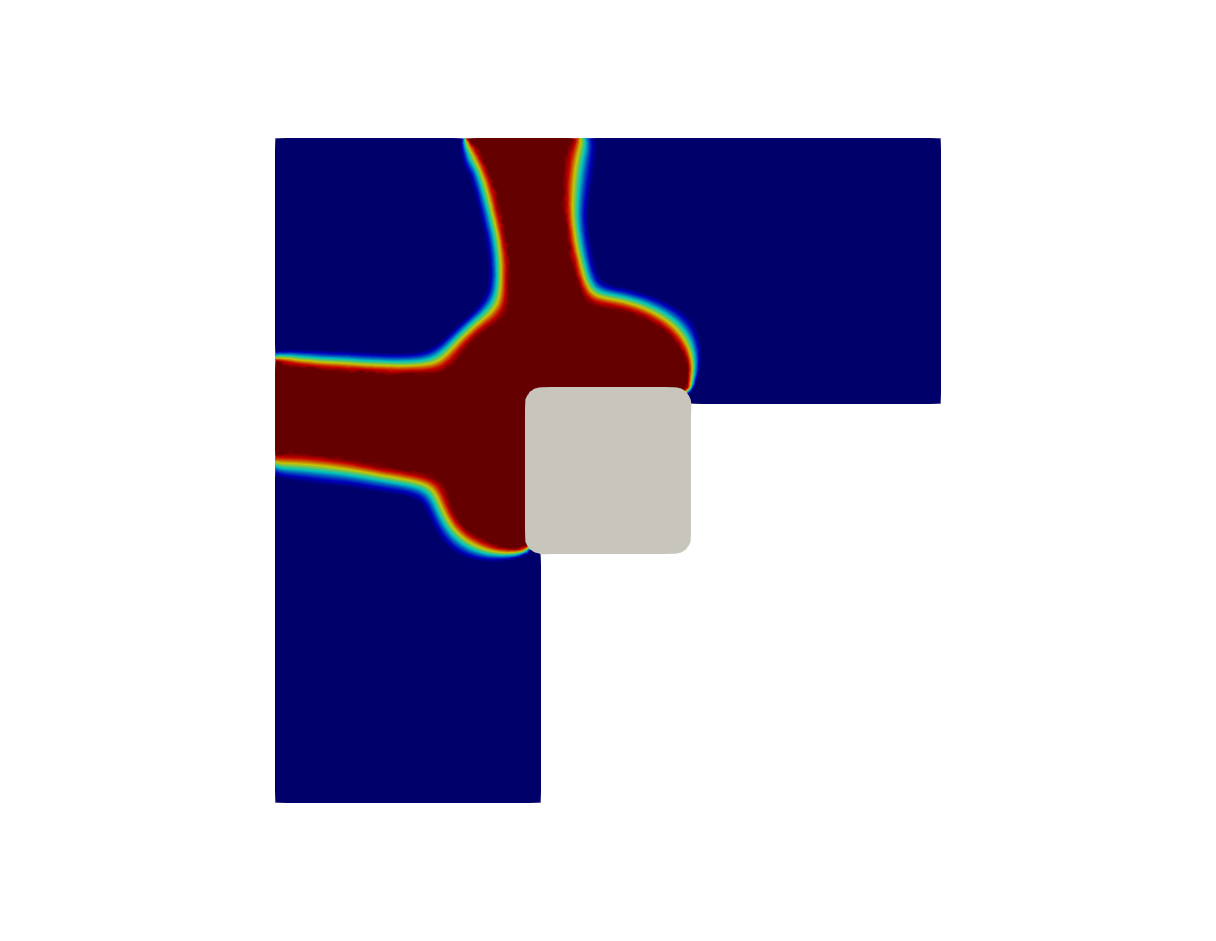}~
    \includegraphics[trim={3.5in 1.75in 3.5in 1.75in},clip,width=0.2\linewidth]{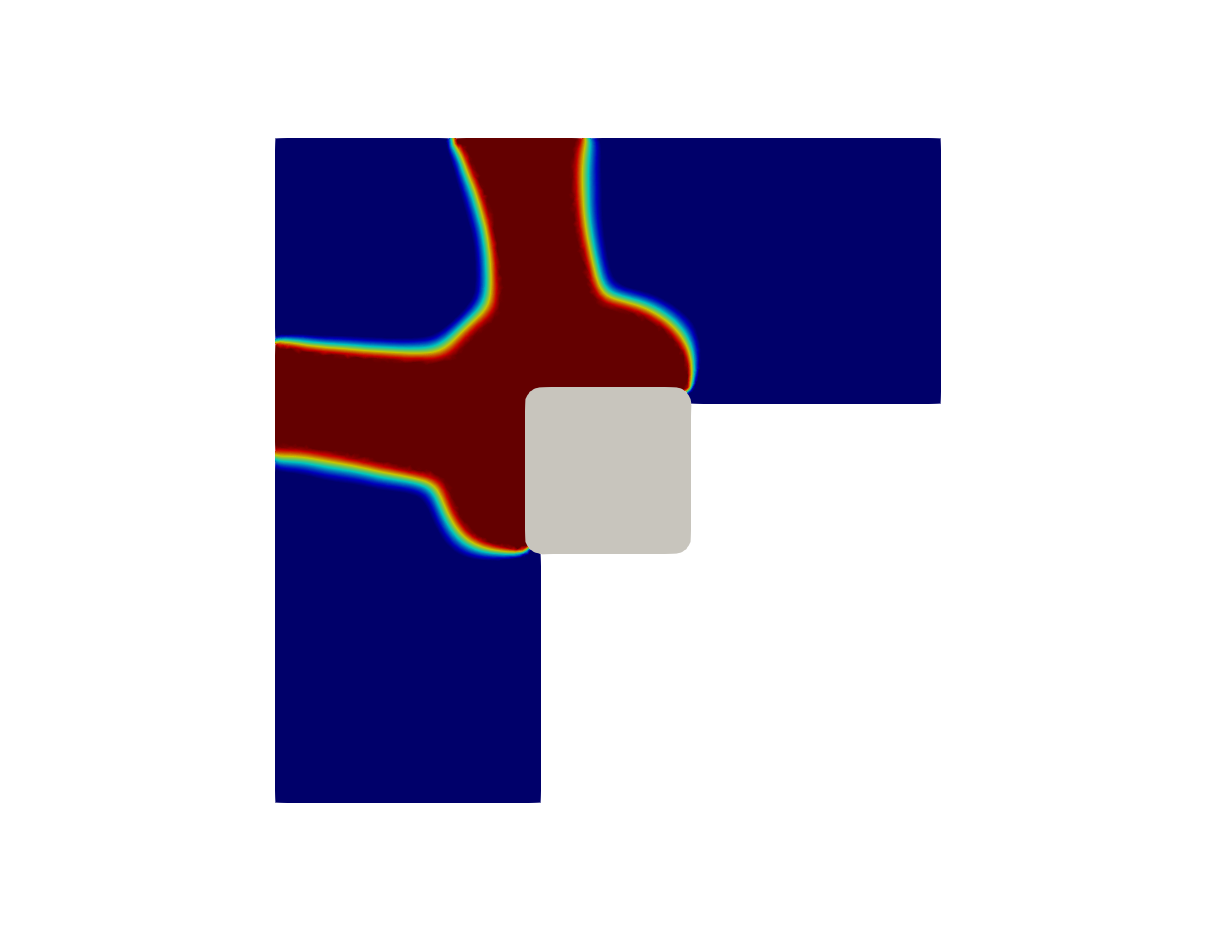}~
    \includegraphics[width=0.08\linewidth]{Figures/design_contour.png}\\
    \includegraphics[trim={3.5in 1.75in 3.5in 1.75in},clip,width=0.2\linewidth]{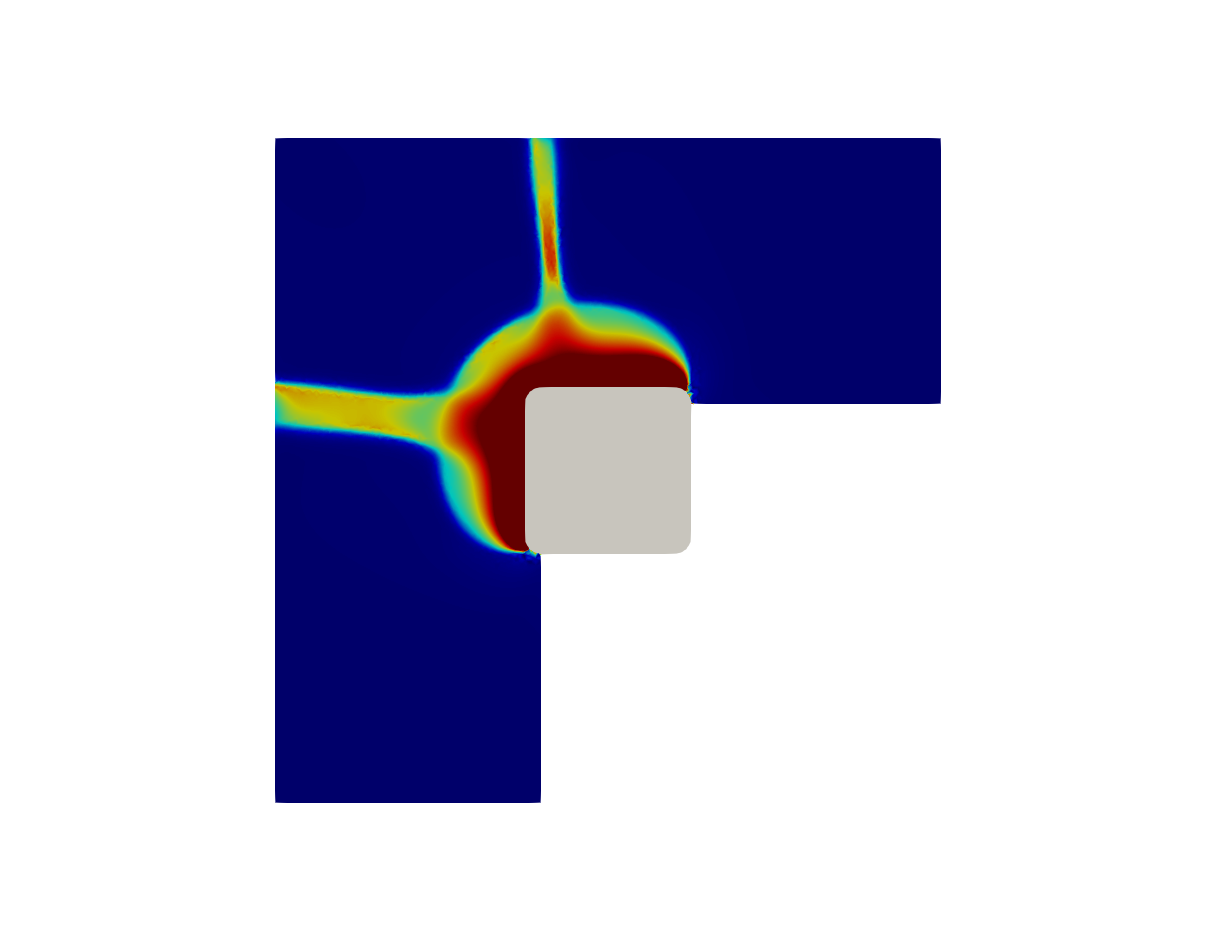}~
    \includegraphics[trim={3.5in 1.75in 3.5in 1.75in},clip,width=0.2\linewidth]{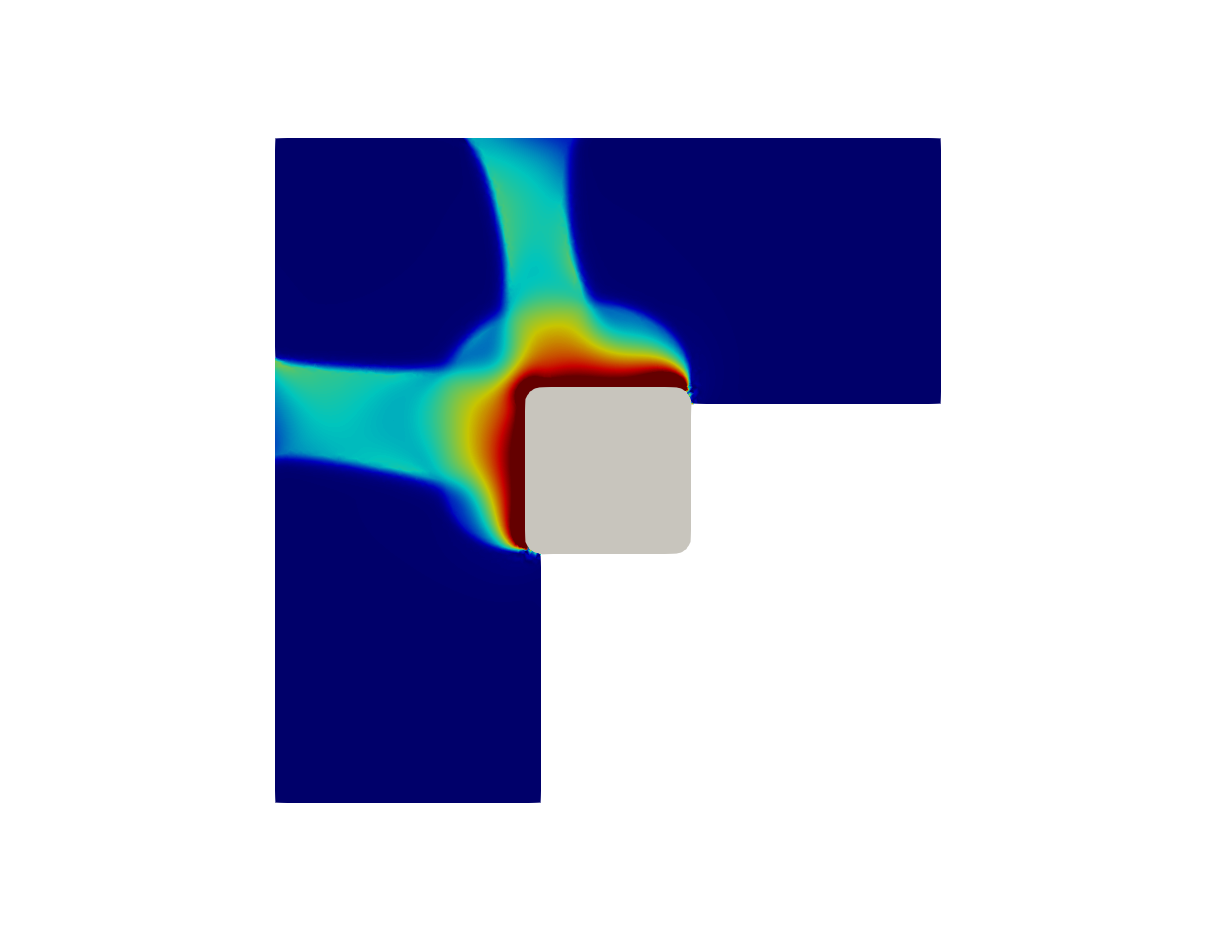}~
    \includegraphics[trim={3.5in 1.75in 3.5in 1.75in},clip,width=0.2\linewidth]{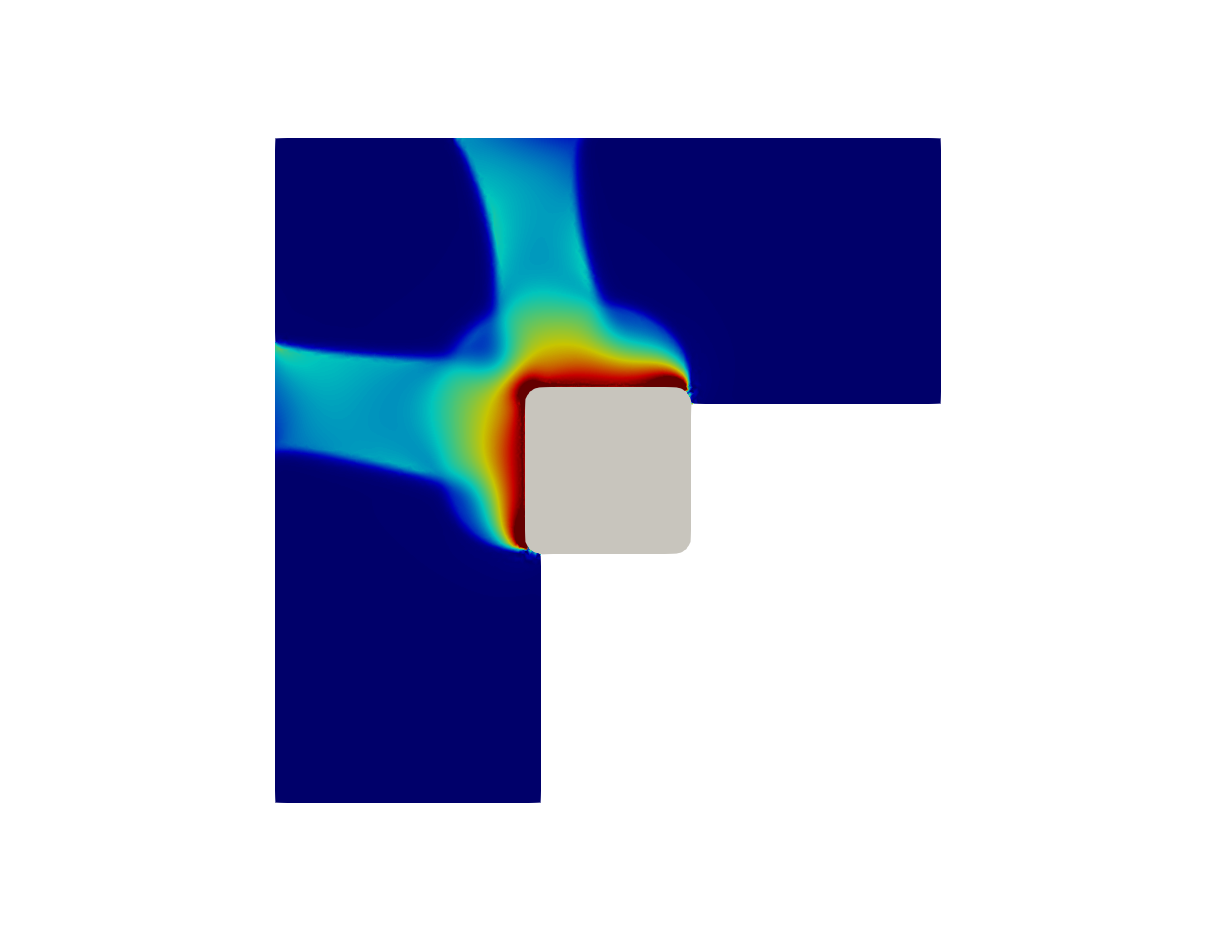}~
    \includegraphics[width=0.08\linewidth]{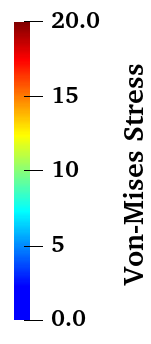}\\
    (a) \hspace{1 in} (b) \hspace{1in} (c)\\
    \caption{Effect of the critical chance $\alpha_c$ on optimal design. The top row shows the optimal design results obtained at the mean of the uncertain sample $\Tilde{m}=0$ for different values of critical chance: (a) $\alpha_c = 0.05$ (b) $\alpha_c = 0.08$ and (c) $\alpha_c = 0.1$. The bottom row represents the von Mises stress for the corresponding optimal designs in the top row. In all results $T_{cr} = 22.5 \: \text{MPa}, L_{CR}= 0.25, \sigma = 0.25, \beta_V = 0.1$ and $\beta_{RG} = 1 \times 10^{-5}$.}
    \label{fig:critical_chance}
    \vspace{-0.1in}
\end{figure}
\begin{figure}[h!]
    \centering
    \footnotesize
    \includegraphics[trim={3.5in 1.75in 3.5in 1.75in},clip,width=0.2\linewidth]{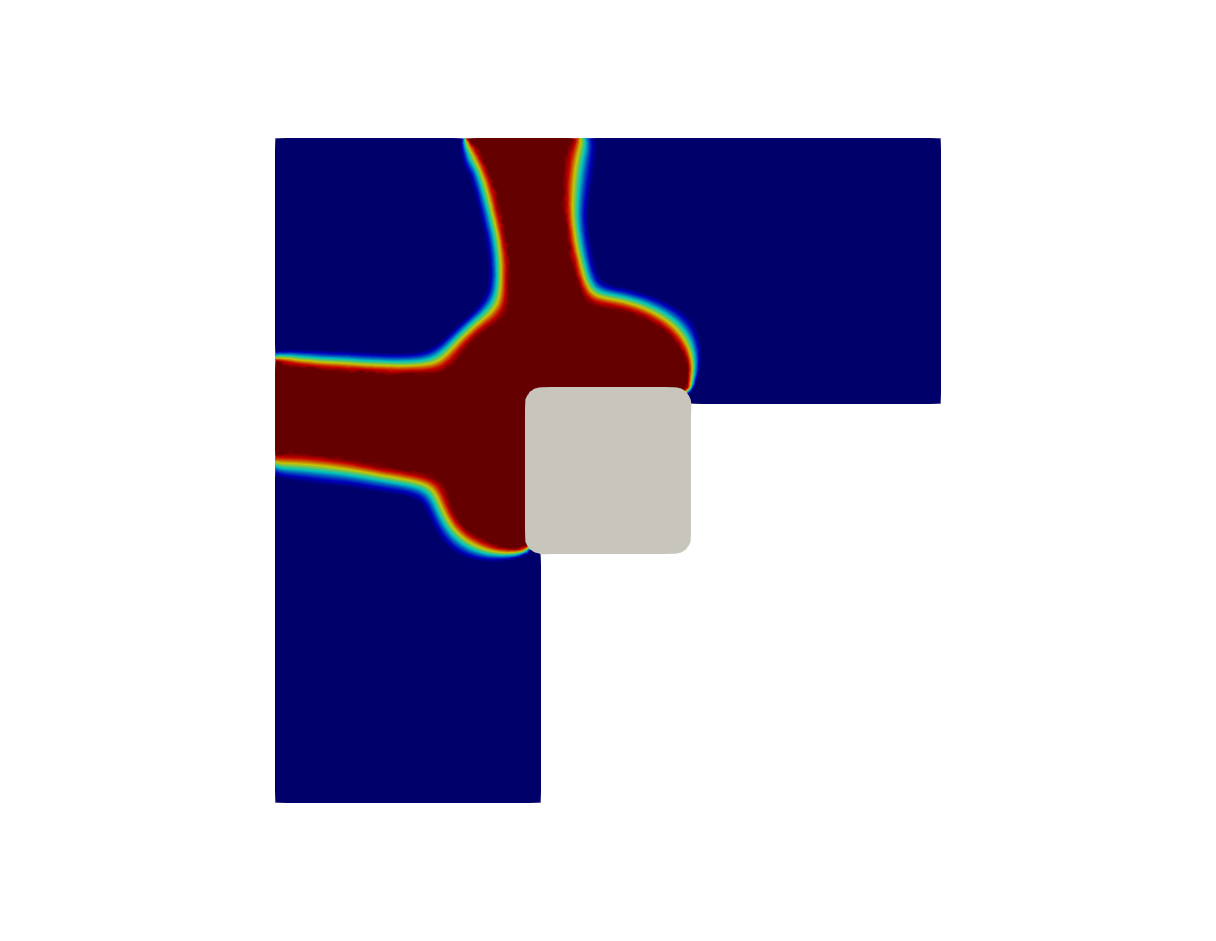}~
    \includegraphics[trim={3.5in 1.75in 3.5in 1.75in},clip,width=0.2\linewidth]{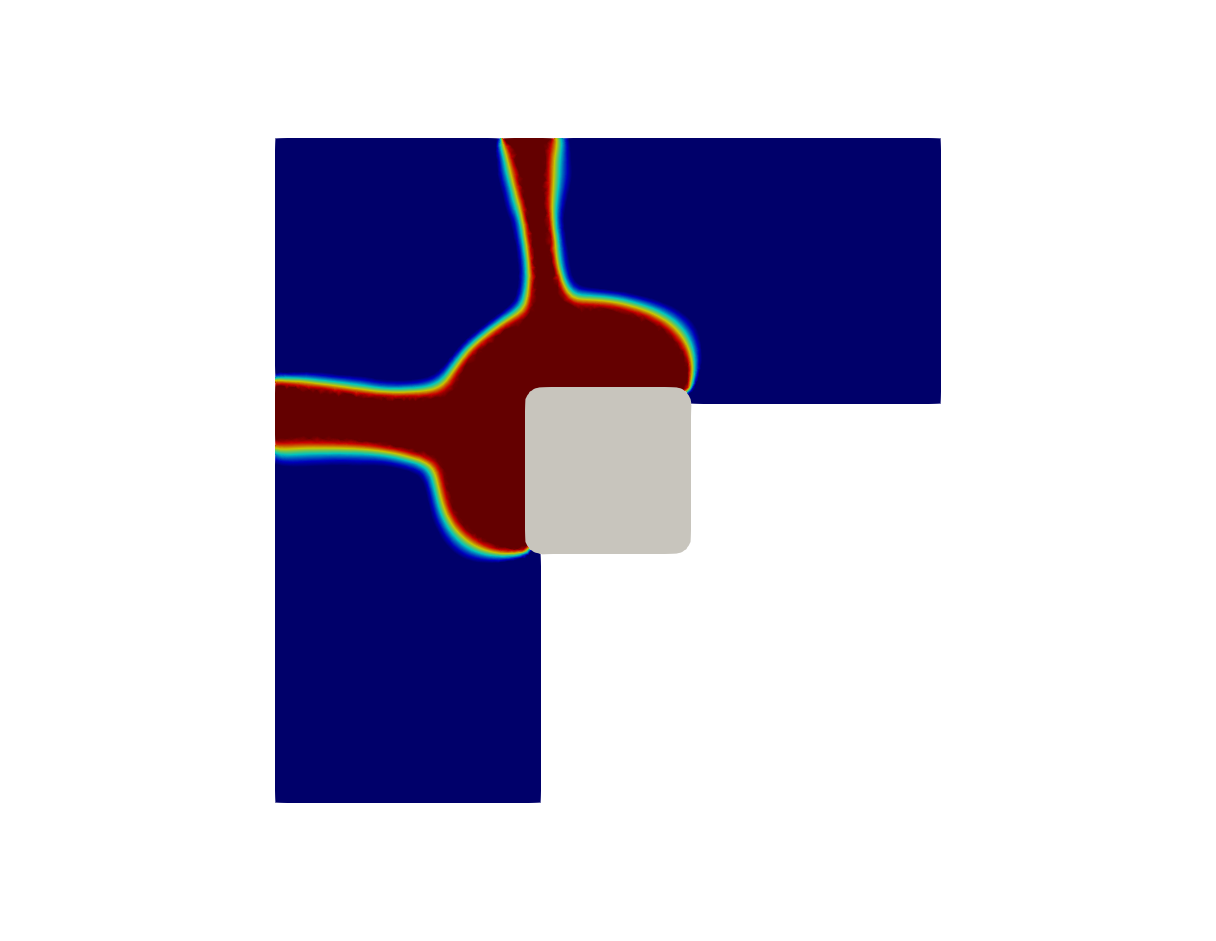}~
    \includegraphics[trim={3.5in 1.75in 3.5in 1.75in},clip,width=0.2\linewidth]{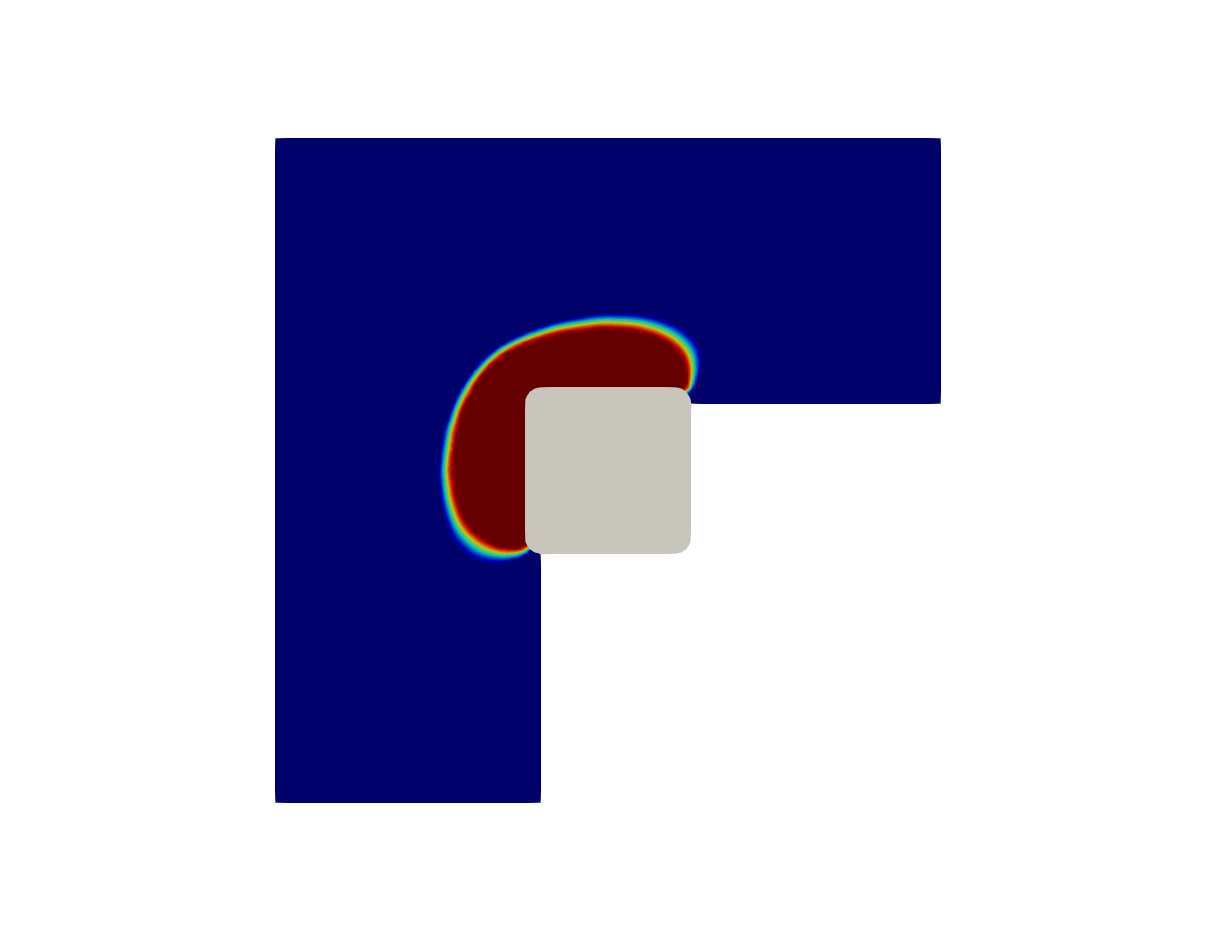}~
    \includegraphics[width=0.08\linewidth]{Figures/design_contour.png}\\
    \includegraphics[trim={3.5in 1.75in 3.5in 1.75in},clip,width=0.2\linewidth]{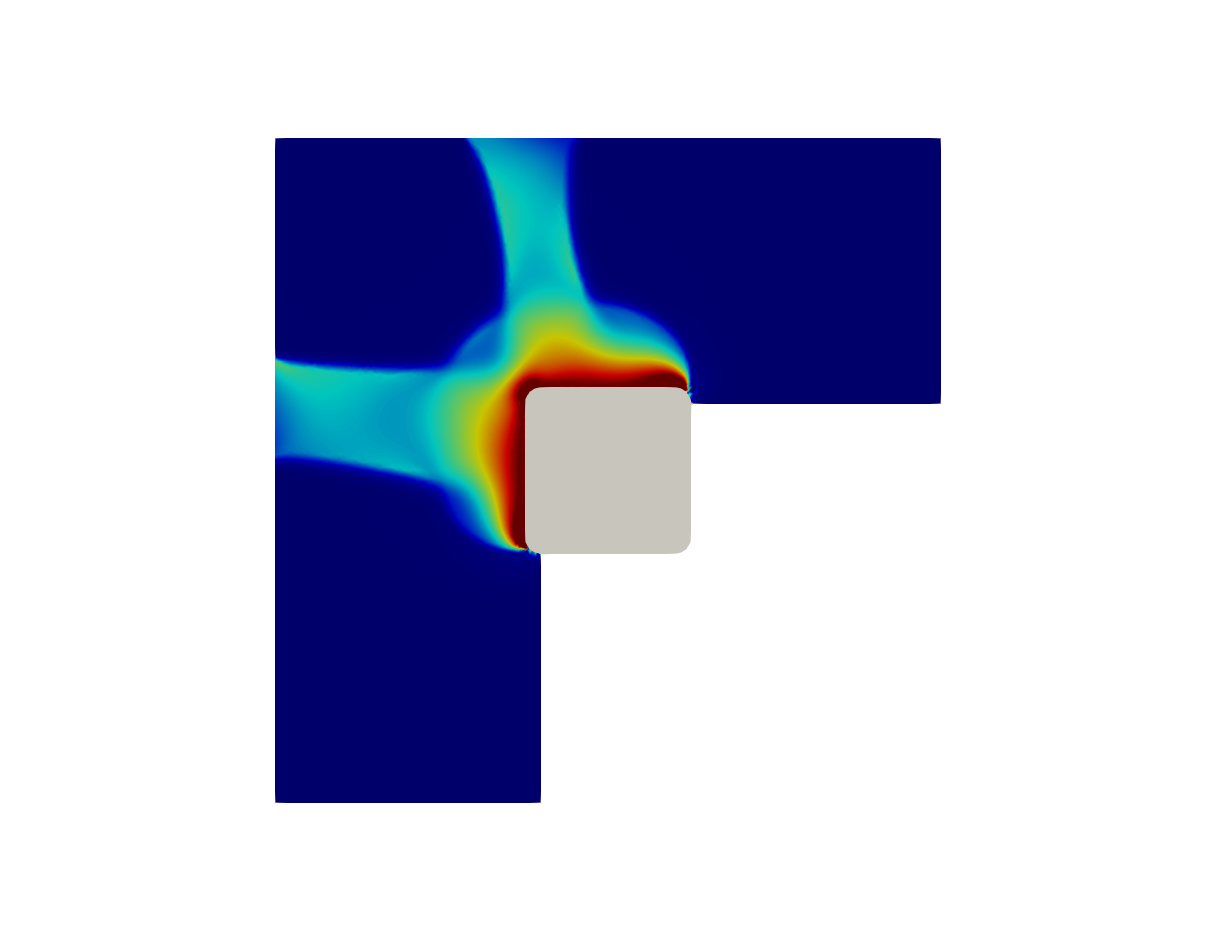}~
    \includegraphics[trim={3.5in 1.75in 3.5in 1.75in},clip,width=0.2\linewidth]{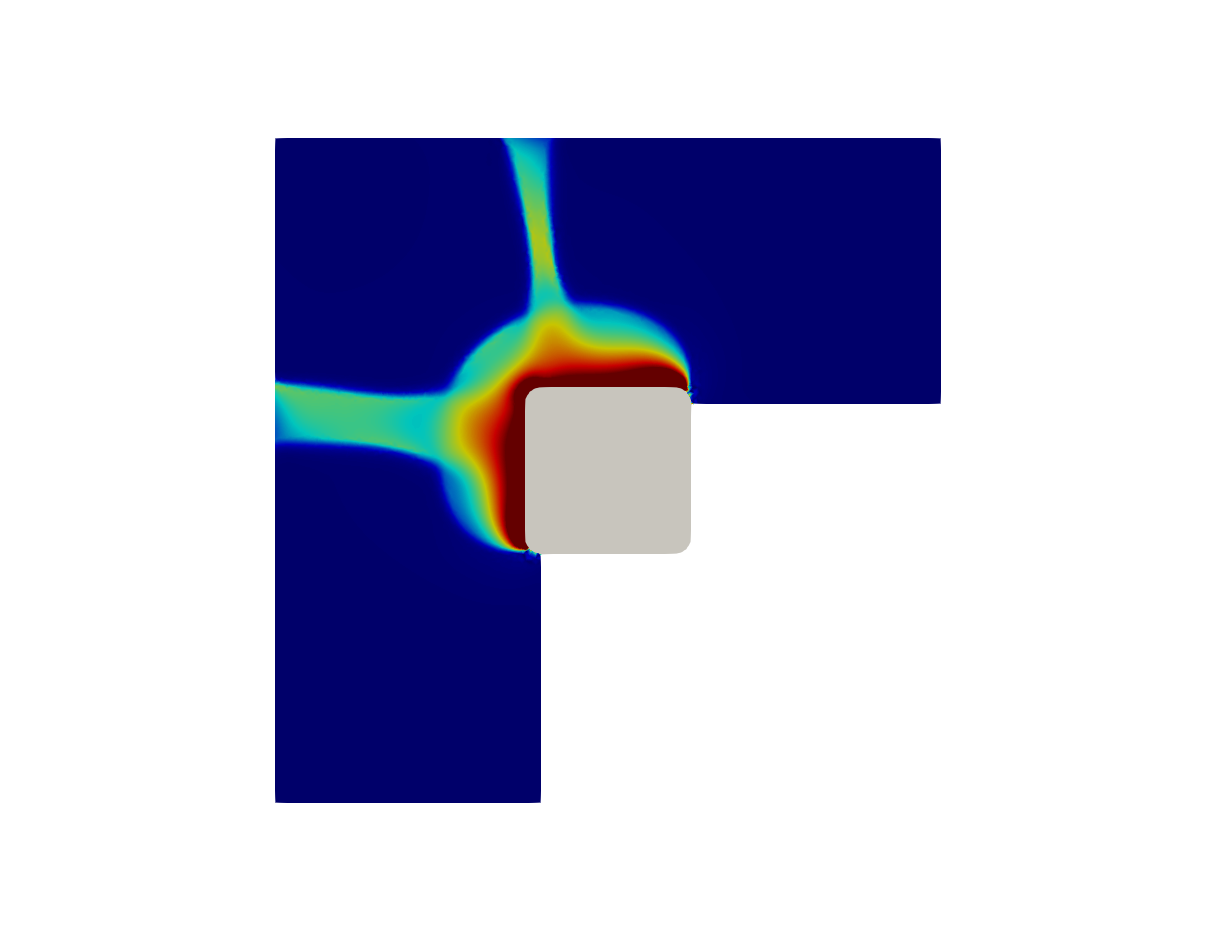}~
    \includegraphics[trim={3.5in 1.75in 3.5in 1.75in},clip,width=0.2\linewidth]{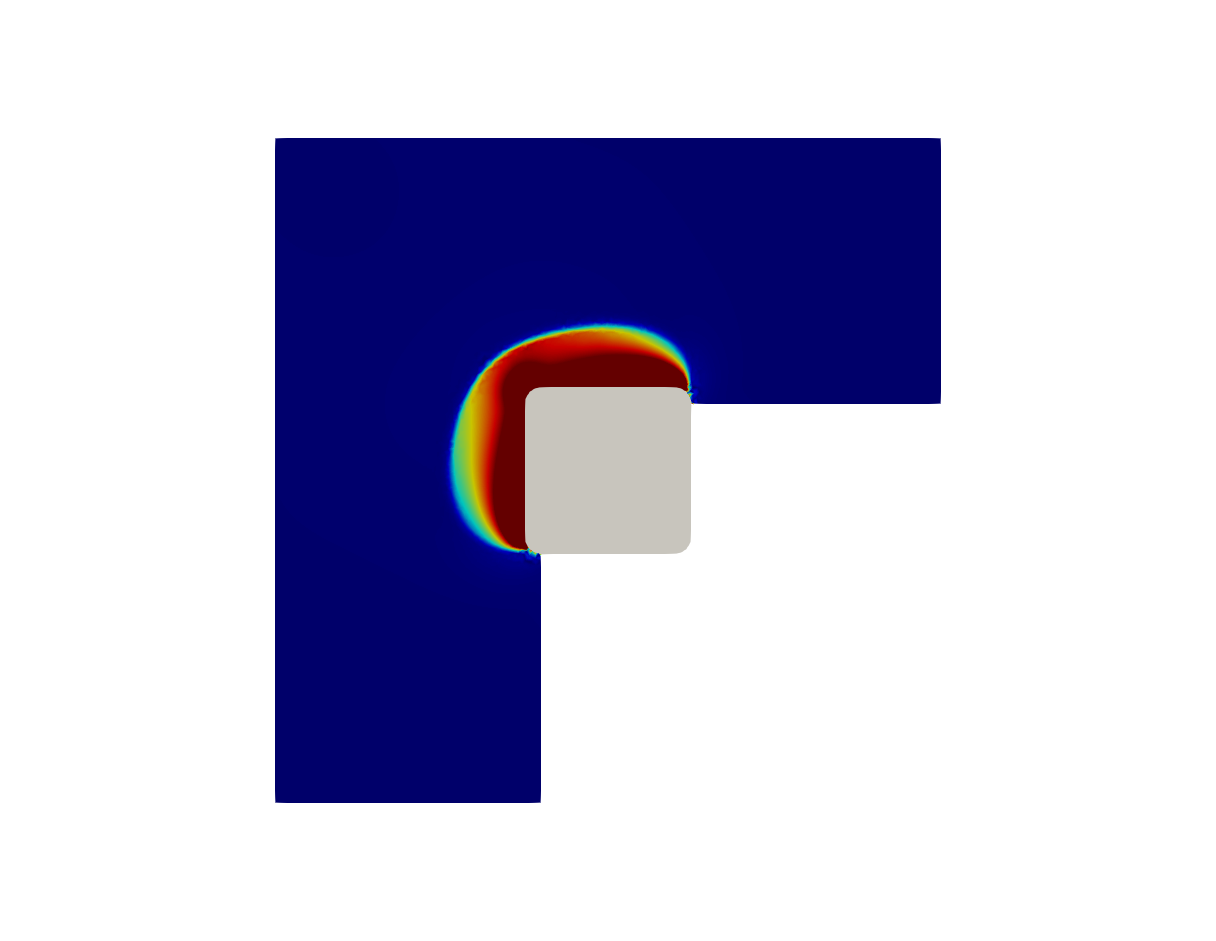}~
    \includegraphics[width=0.08\linewidth]{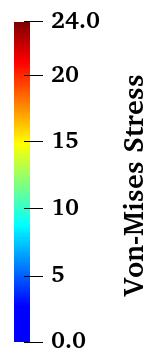}\\
    (a) \hspace{1 in} (b) \hspace{1 in} (c)\\
    \caption{Effect of the limiting critical stress $T_{cr}$ on optimal design. The top row shows the optimal design results obtained at the mean of the uncertain sample $\Tilde{m}=0$ for different values of limiting critical stress: (a) $T_{cr} = 20 \: \text{MPa}$ (b) $T_{cr} = 22.5 \: \text{MPa}$ and (c) $T_{cr} = 25 \: \text{MPa}$. The bottom row represents the von Mises stress for the corresponding optimal designs in the top row. In all results $\alpha_c =0.05, L_{CR}= 0.25,\sigma = 0.25, \beta_V = 0.1$ and $\beta_{RG} = 1 \times 10^{-5}$.}
    \label{fig:critical_stress}
    \vspace{-0.1in}
\end{figure}

\noindent
The chance constraint function defined in \eqref{eq:chance_function} incorporates parameters such as $\alpha_c$ and $T_{cr}$, which govern the enforcement of constrain to mitigate stress concentrations within the domain. In particular, the probability of chance $P(f(m,d) \geq 0)$ is limited by a threshold $\alpha_c$, where the function $f$ is the difference between the limiting critical stress $T_{cr}$ and the p-norm of the von Mises stress $T_{\text{pn}}$. 
To examine the influence of $\alpha_c$ on optimal design, a parametric study was conducted with constant parameters, as shown in Figure \ref{fig:critical_chance}. The critical stress limit was set at $T_{cr}=22.5 \: \text{MPa}$, while $\alpha_c$ was varied among 0.05, 0.08, and 0.1.
This figure presents the optimal design, and von Mises stress distribution evaluated at the mean of the uncertain parameter $\Bar{m} = 0$. The results indicate that increasing critical chance $\alpha_c$ expands the regions of mechanically stronger material (highlighted in red), effectively reducing localized stresses across the component. However, this improvement in mechanical stability comes at the expense of insulation performance of the thermal break. 

\begin{figure}[ht]
    \centering
    \begin{tabular}{c:c:c:c}
    \multicolumn{4}{c}{\includegraphics[width=0.35\linewidth]{Figures/Porosity_contour.png}}\\
    \includegraphics[trim={3.5in 1.5in 3.5in 1.5in},clip,width=0.16\linewidth]{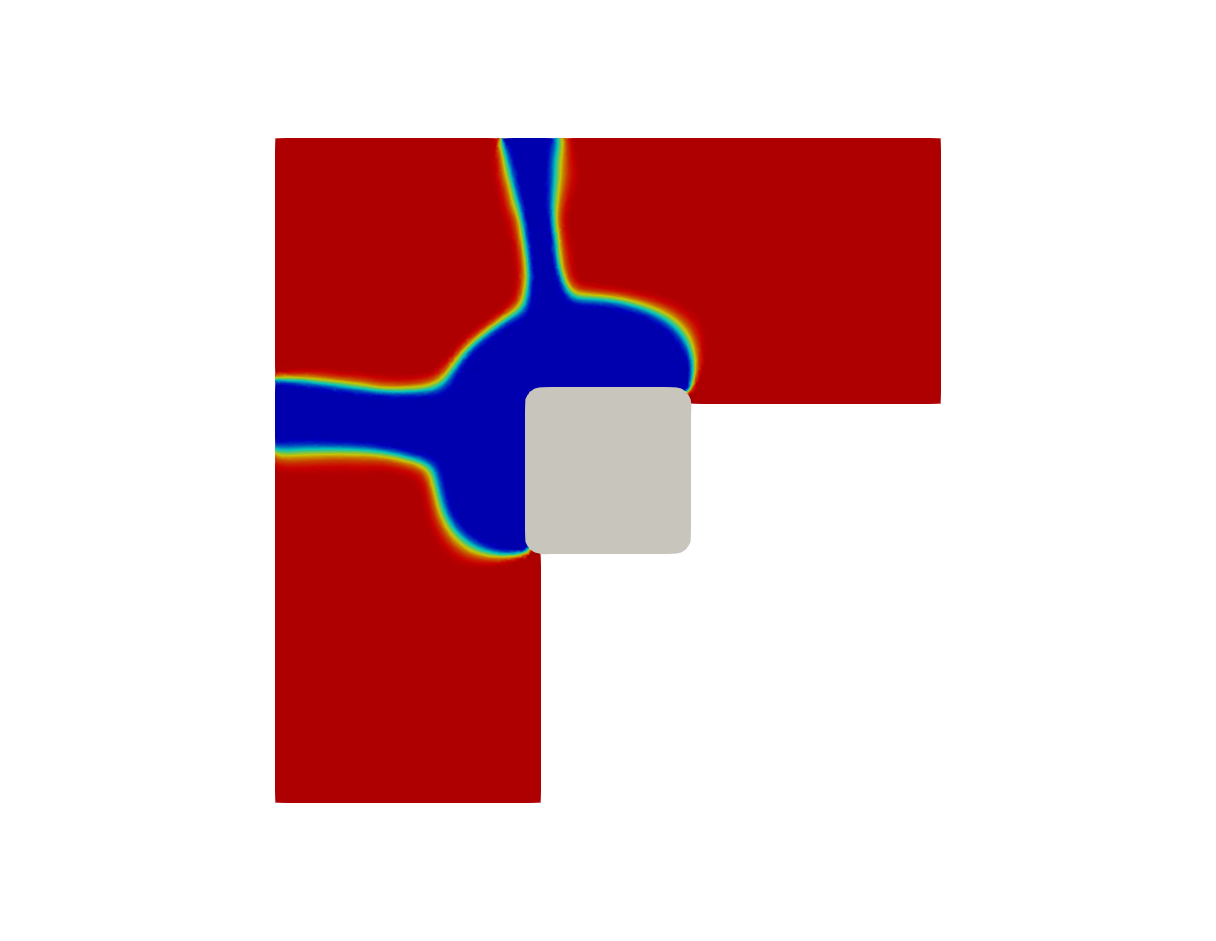}&
    \includegraphics[trim={3.5in 1.5in 3.5in 1.5in},clip,width=0.16\linewidth]{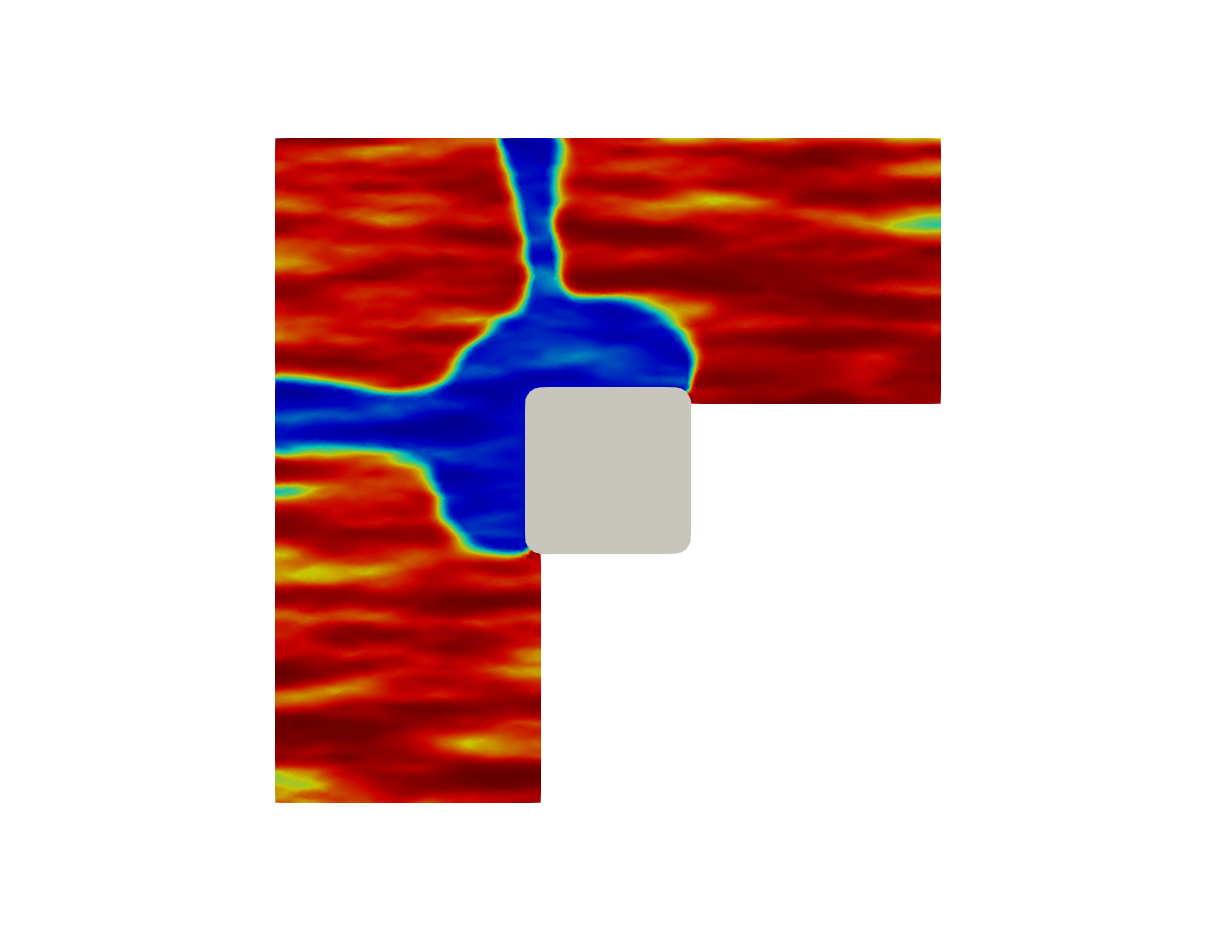}&
    \includegraphics[trim={3.5in 1.5in 3.5in 1.5in},clip,width=0.16\linewidth]{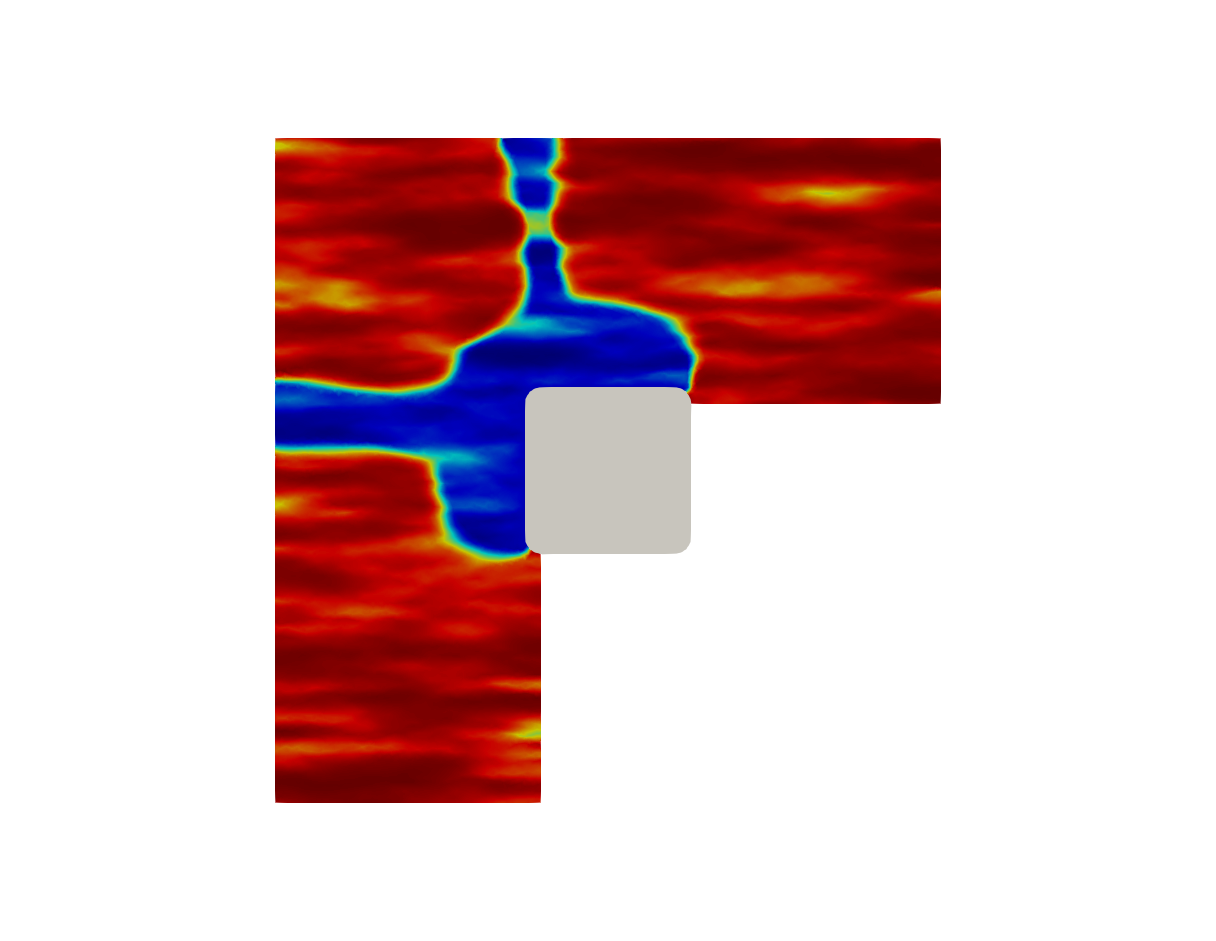}&
    \includegraphics[trim={3.5in 1.5in 3.5in 1.5in},clip,width=0.16\linewidth]{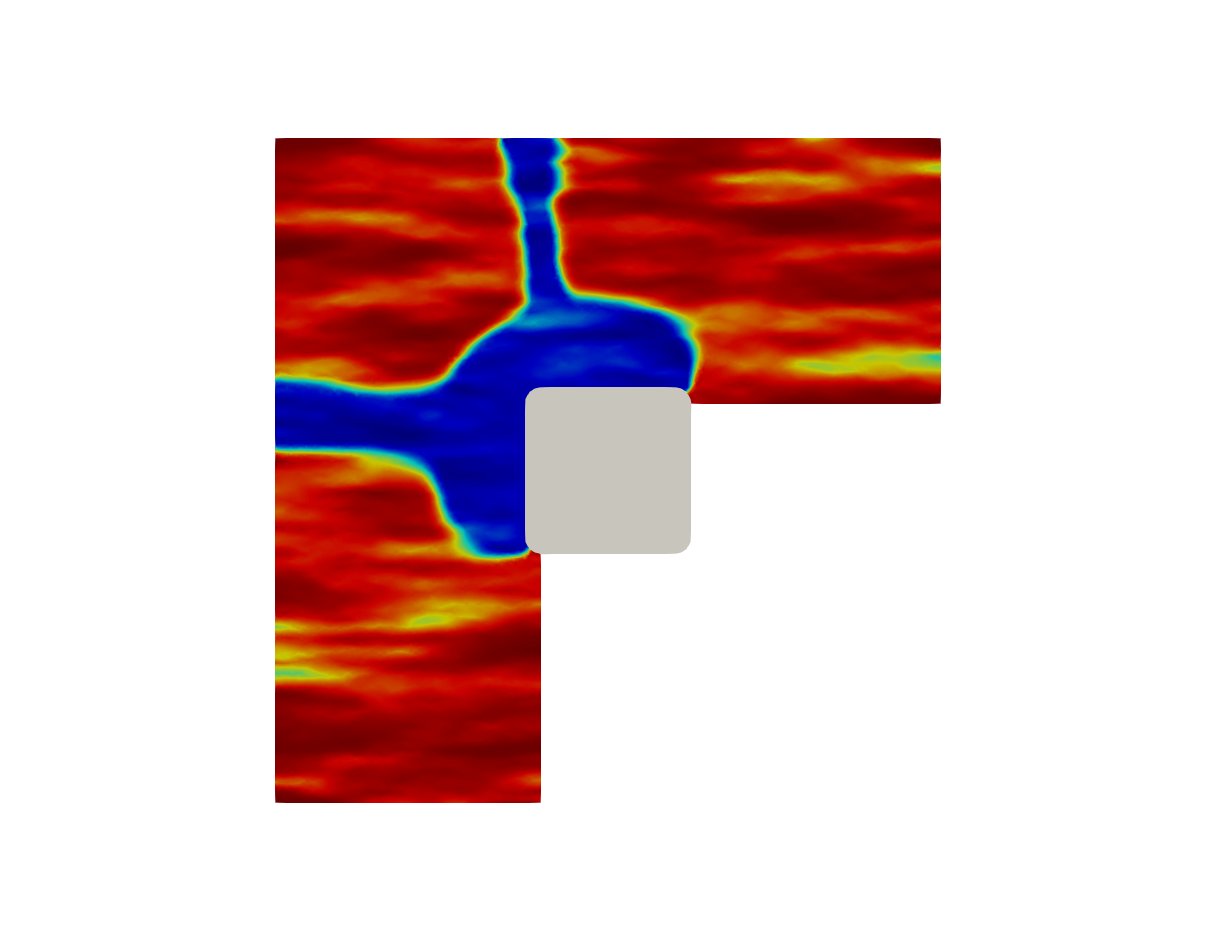}\\
    \multicolumn{4}{c}{\includegraphics[width=0.35\linewidth]{Figures/Disp_contour.png}}\\
    \includegraphics[trim={3.5in 1.5in 3.5in 1.5in},clip,width=0.16\linewidth]{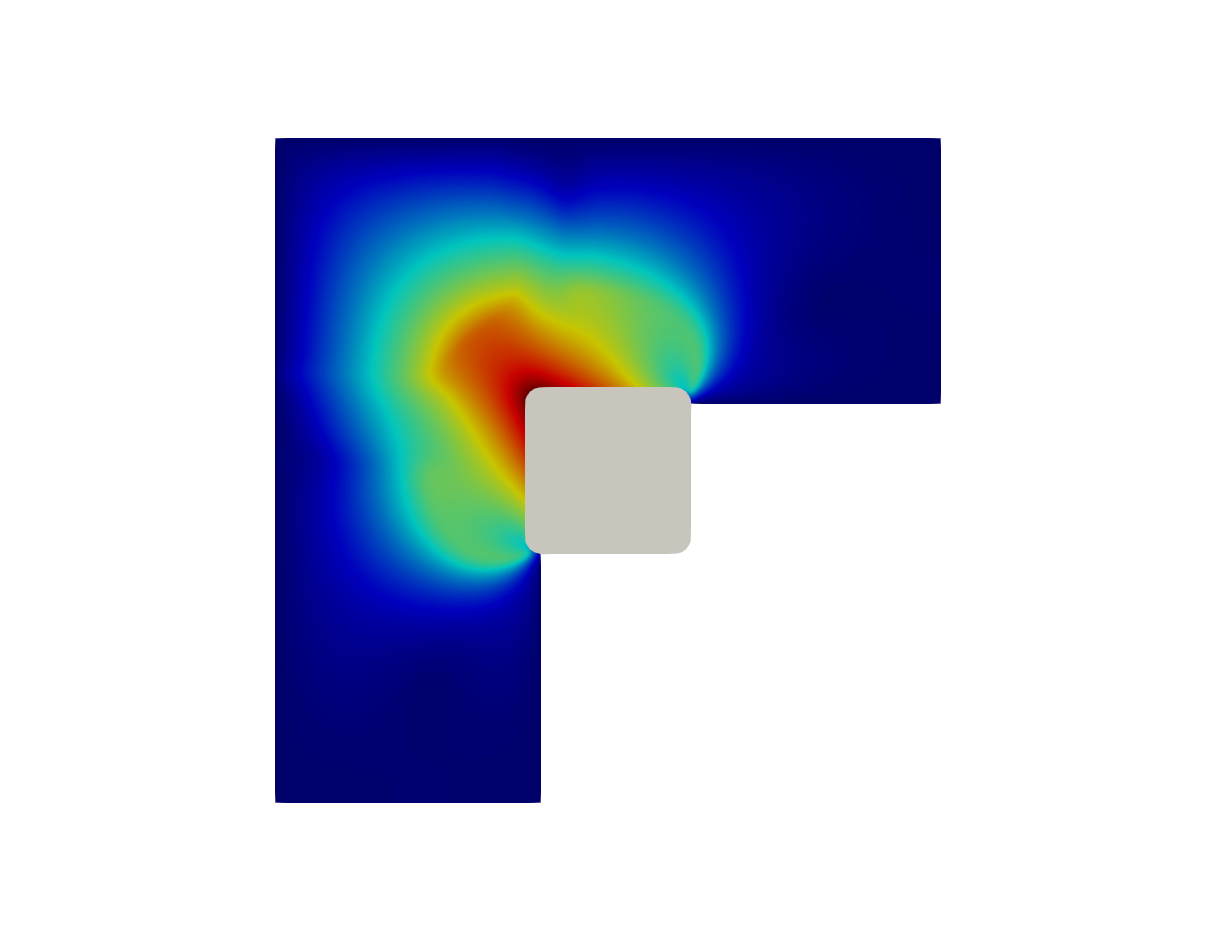}&
    \includegraphics[trim={3.5in 1.5in 3.5in 1.5in},clip,width=0.16\linewidth]{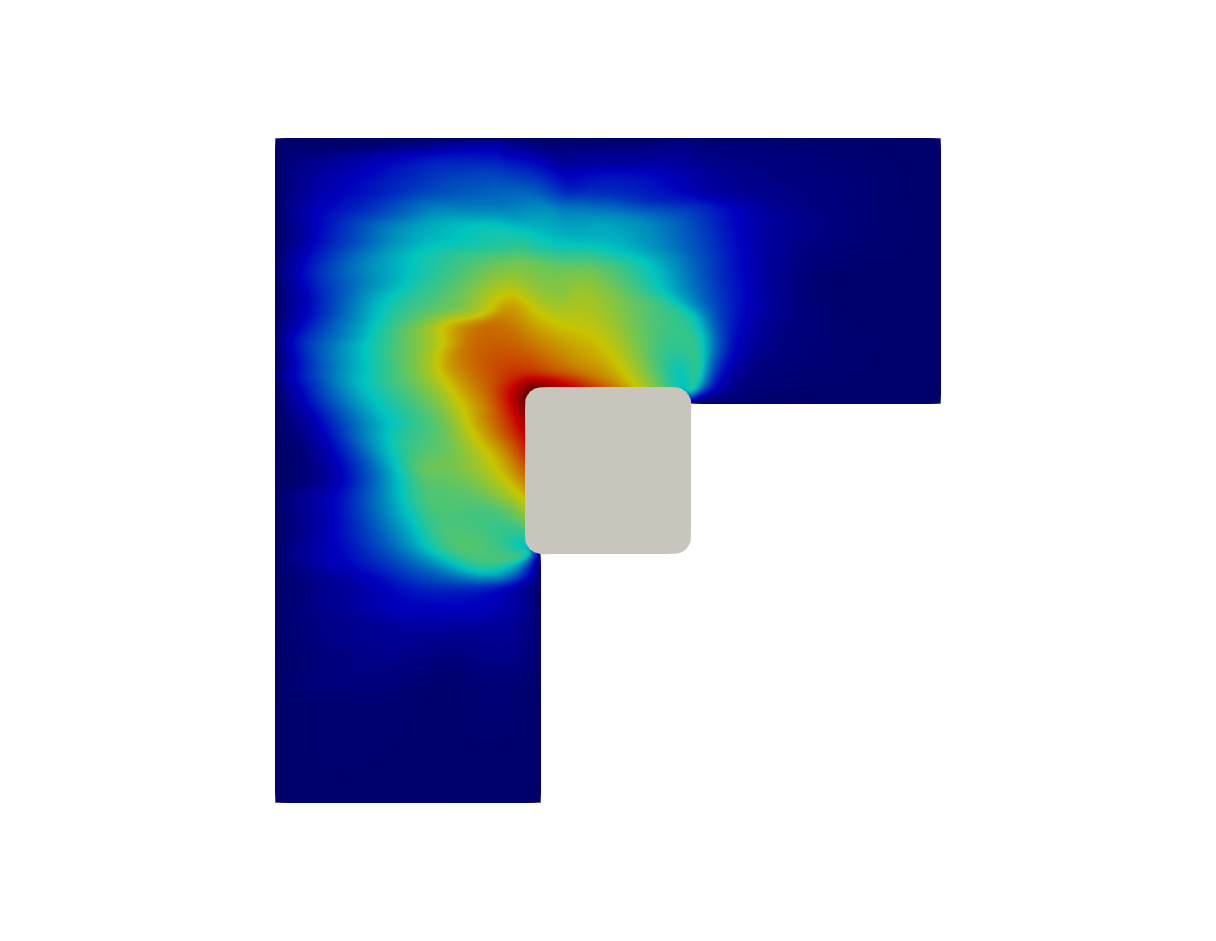}&
    \includegraphics[trim={3.5in 1.5in 3.5in 1.5in},clip,width=0.16\linewidth]{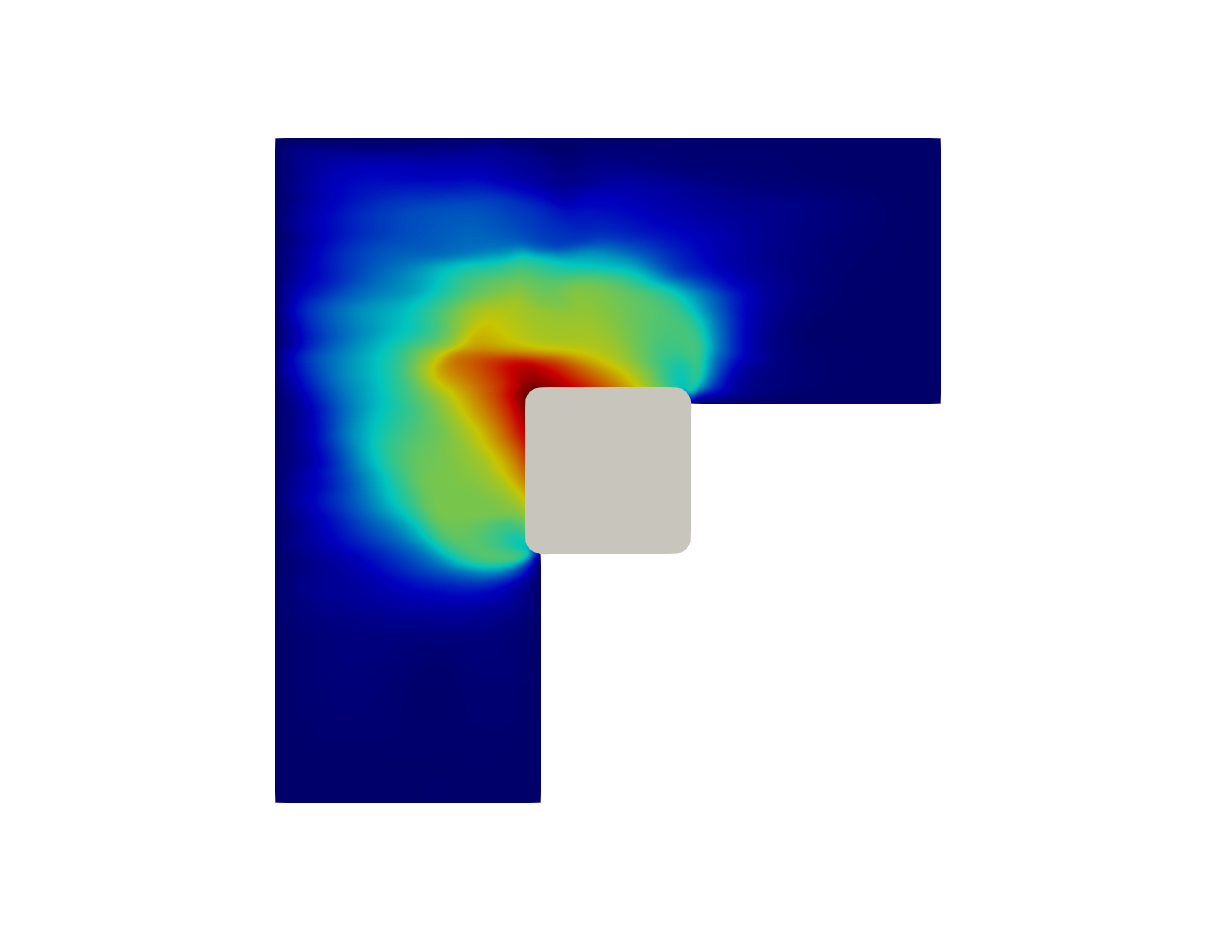}&
    \includegraphics[trim={3.5in 1.5in 3.5in 1.5in},clip,width=0.16\linewidth]{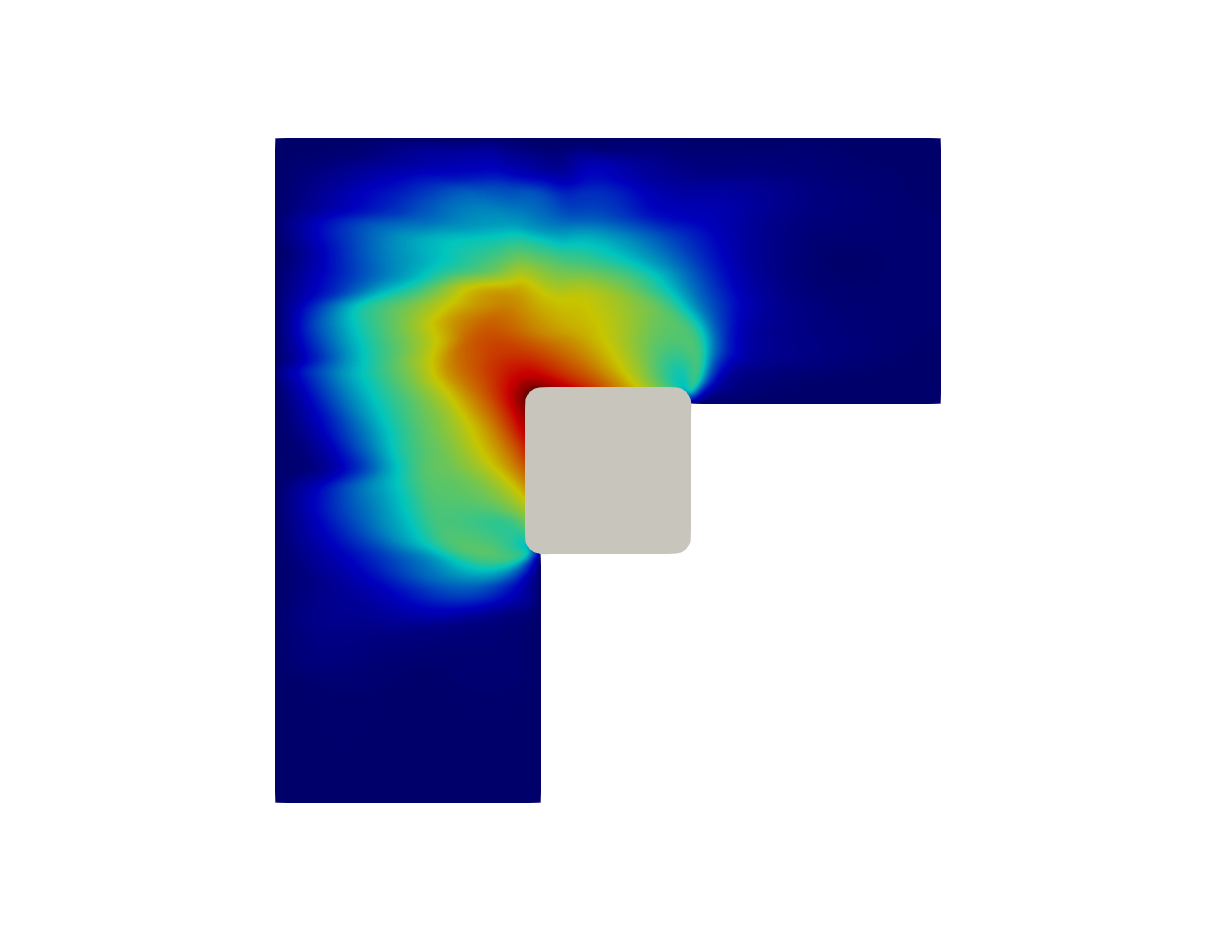}\\
    \multicolumn{4}{c}{\includegraphics[width=0.35\linewidth]{Figures/Fluid_temp_contour.png}}\\
    \includegraphics[trim={3.5in 1.5in 3.5in 1.5in},clip,width=0.16\linewidth]{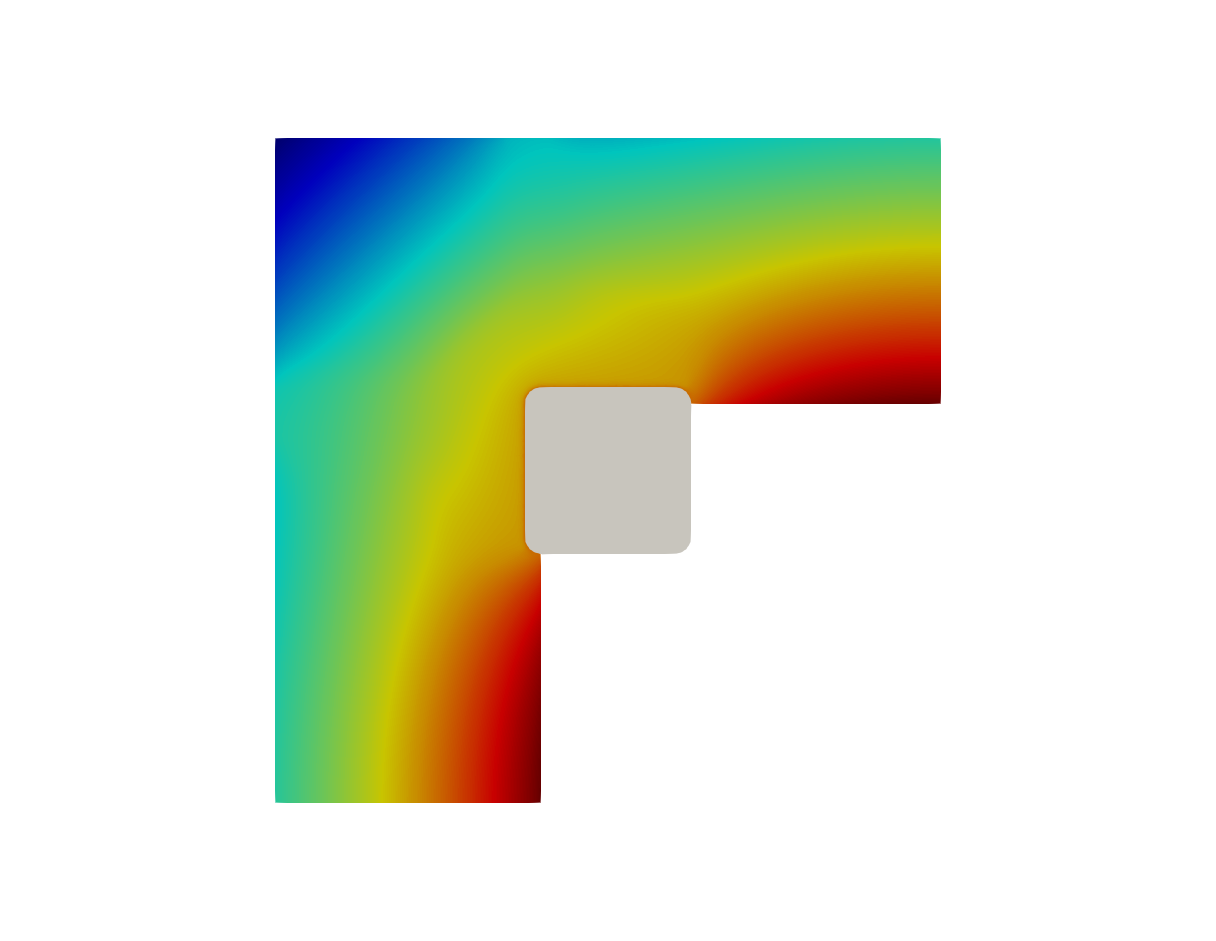}&
    \includegraphics[trim={3.5in 1.5in 3.5in 1.5in},clip,width=0.16\linewidth]{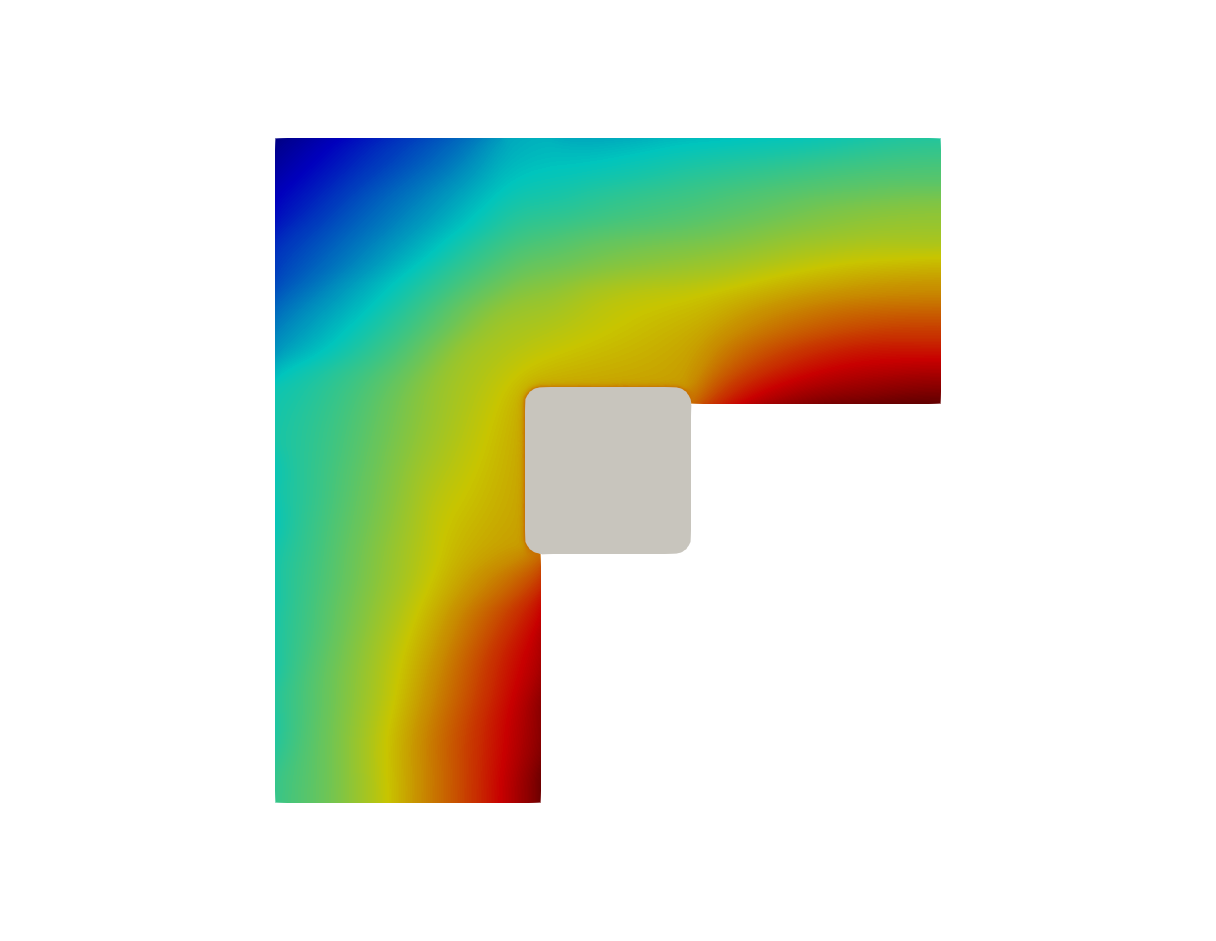}&
    \includegraphics[trim={3.5in 1.5in 3.5in 1.5in},clip,width=0.16\linewidth]{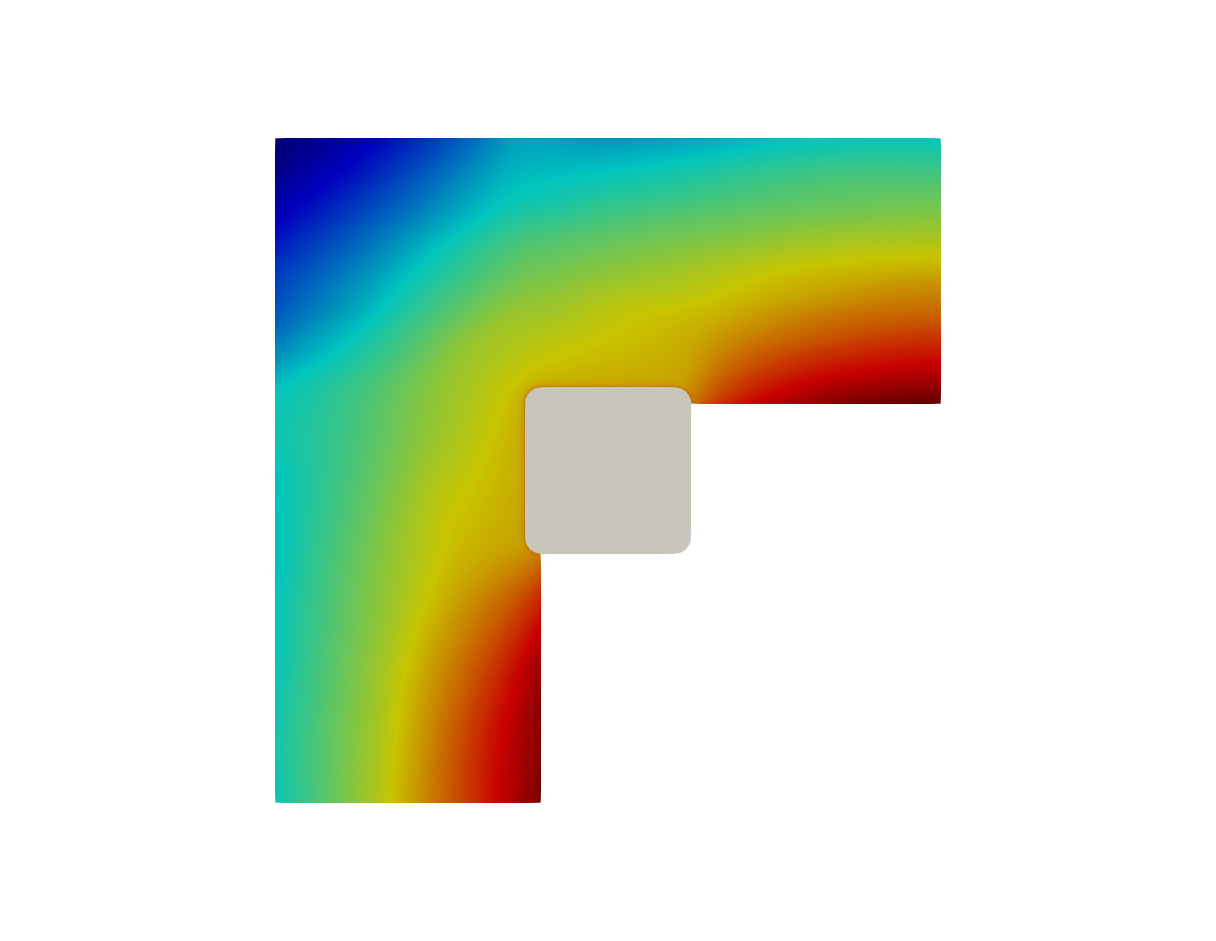}&
    \includegraphics[trim={3.5in 1.5in 3.5in 1.5in},clip,width=0.16\linewidth]{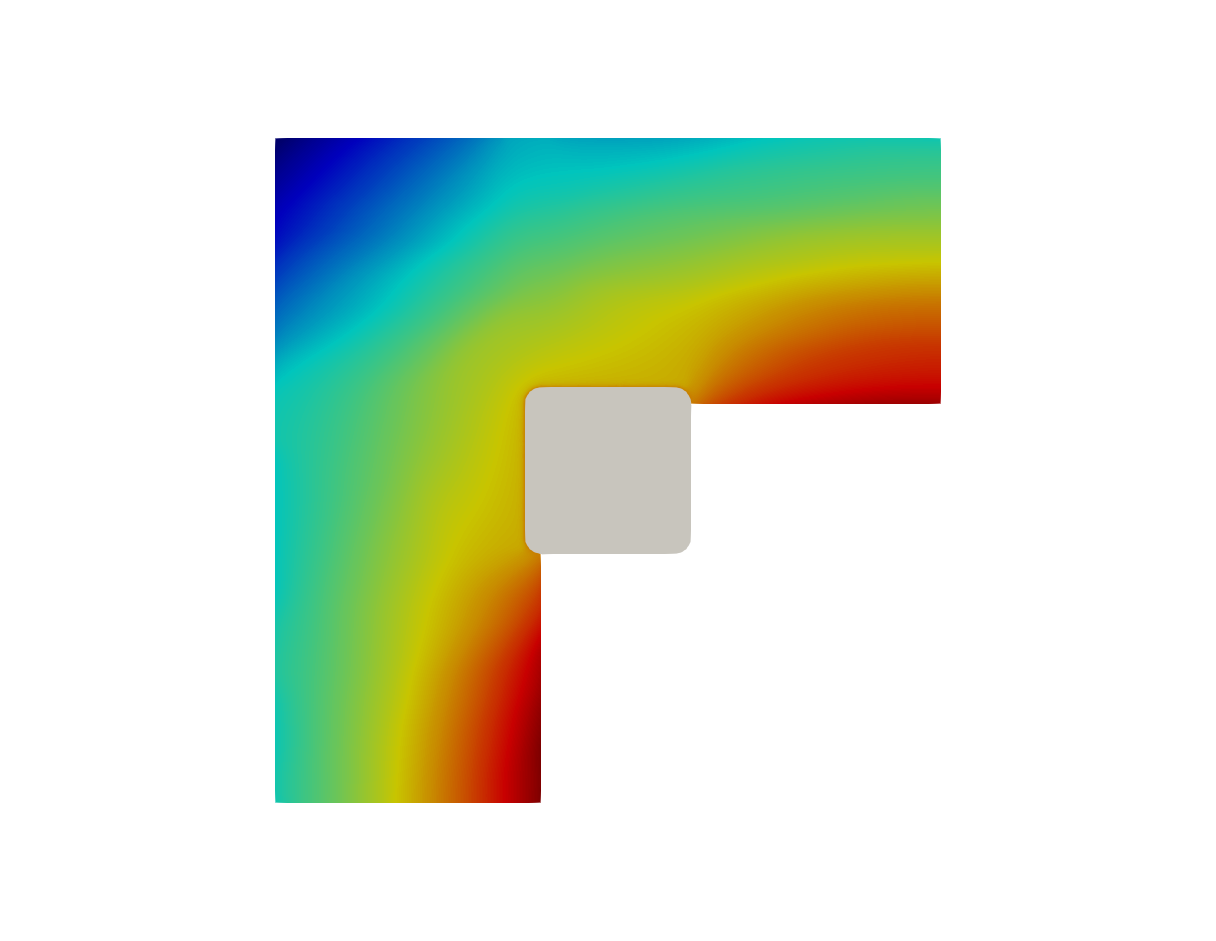}\\
    \multicolumn{4}{c}{\includegraphics[width=0.35\linewidth]{Figures/Solid_temp_contour.png}}\\
    \includegraphics[trim={3.5in 1.5in 3.5in 1.5in},clip,width=0.16\linewidth]{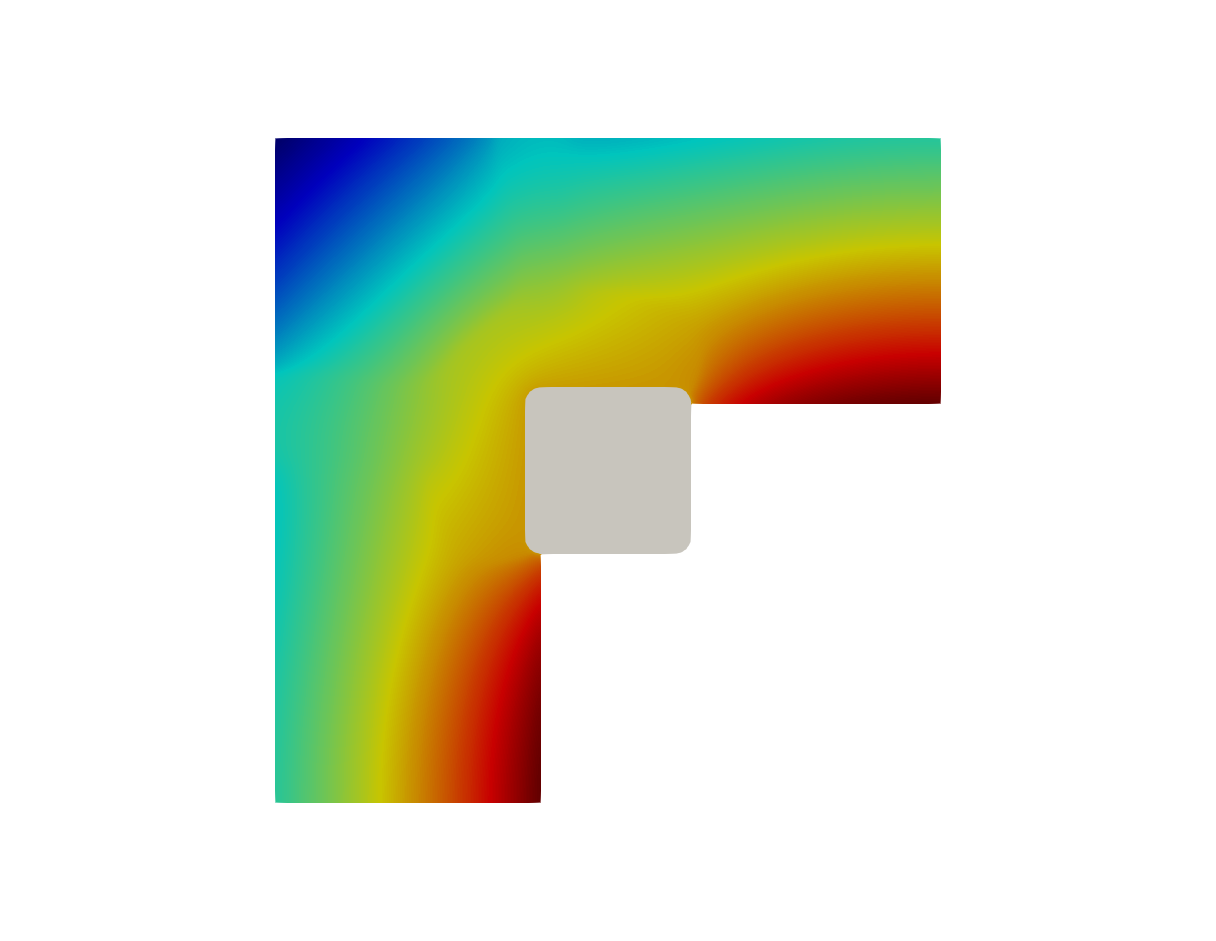}&
    \includegraphics[trim={3.5in 1.5in 3.5in 1.5in},clip,width=0.16\linewidth]{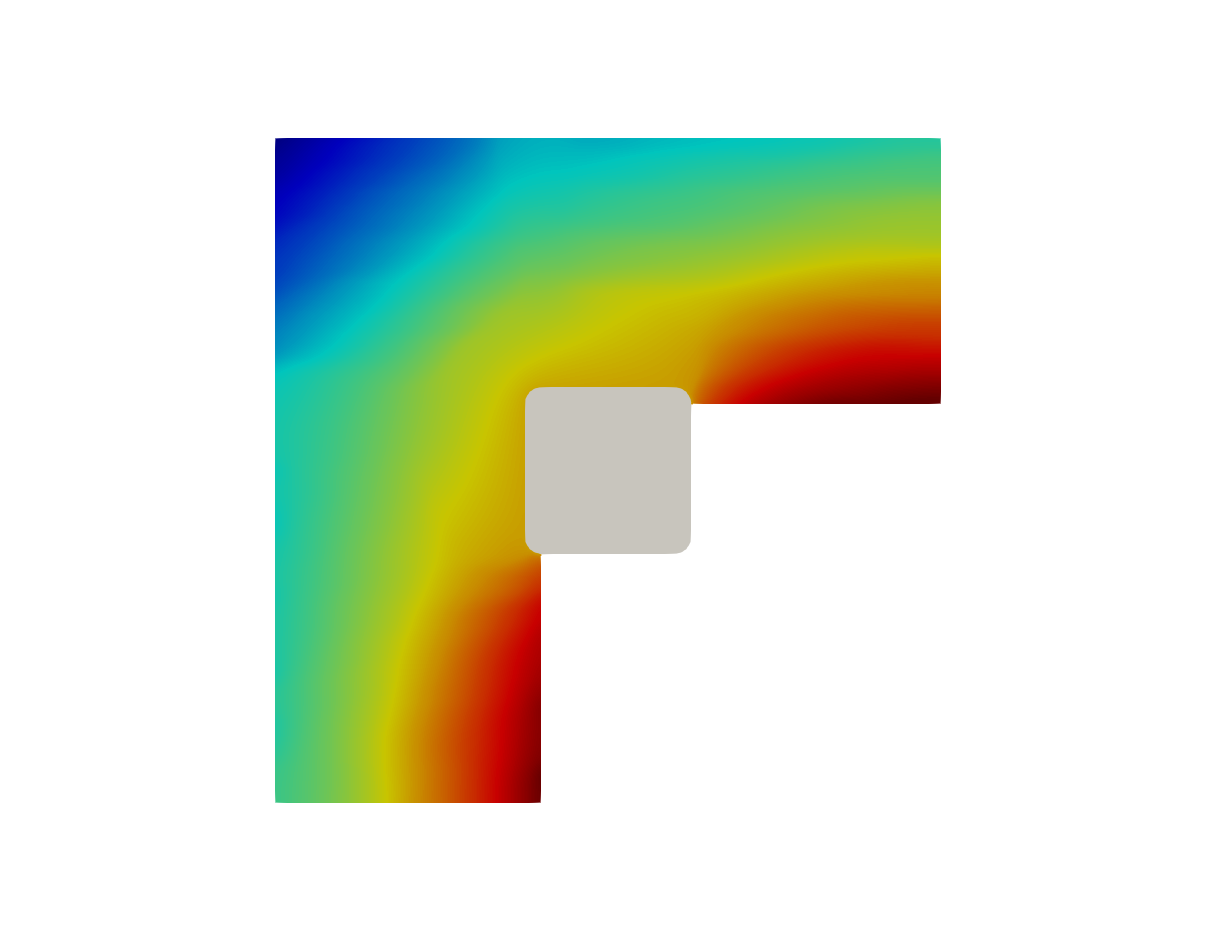}&
    \includegraphics[trim={3.5in 1.5in 3.5in 1.5in},clip,width=0.16\linewidth]{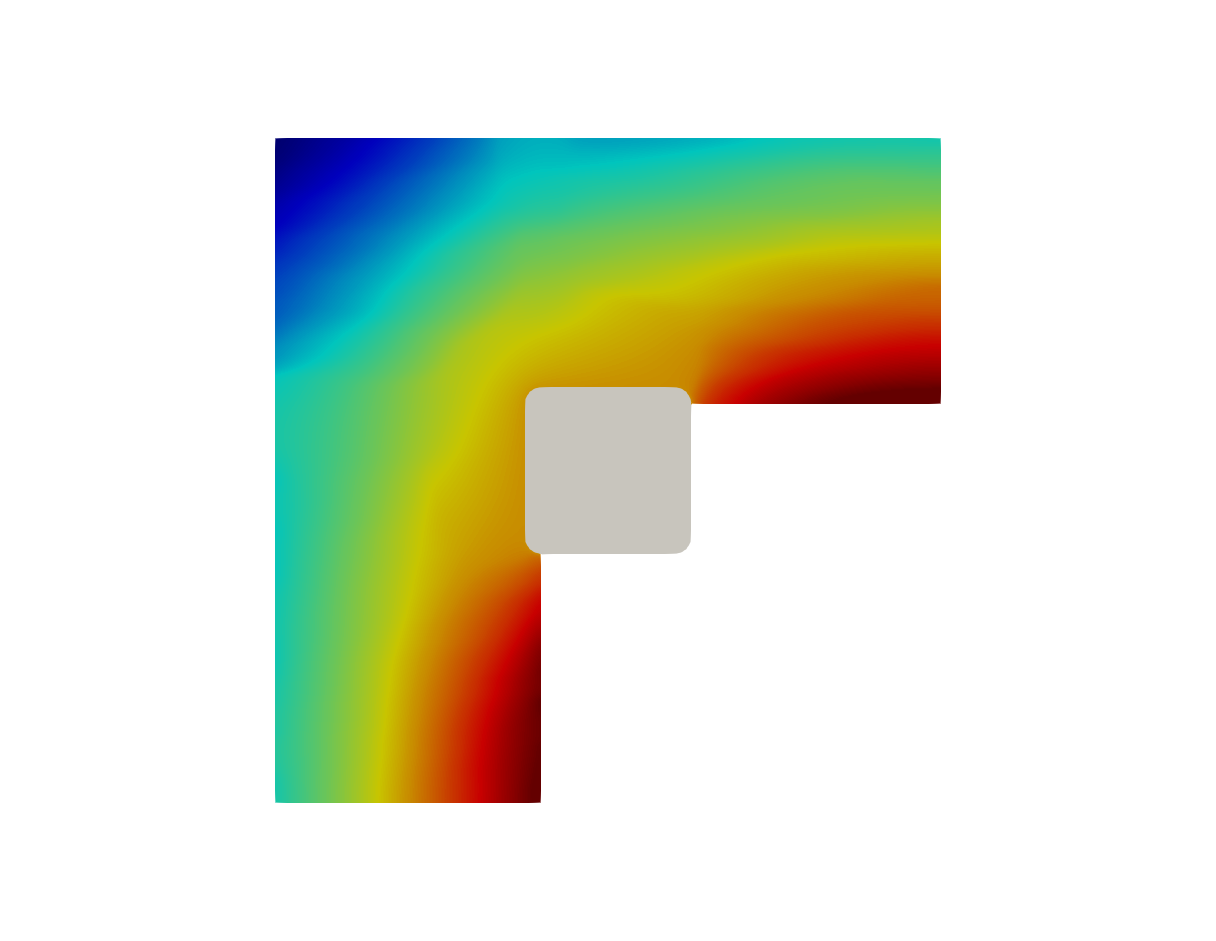}&
    \includegraphics[trim={3.5in 1.5in 3.5in 1.5in},clip,width=0.16\linewidth]{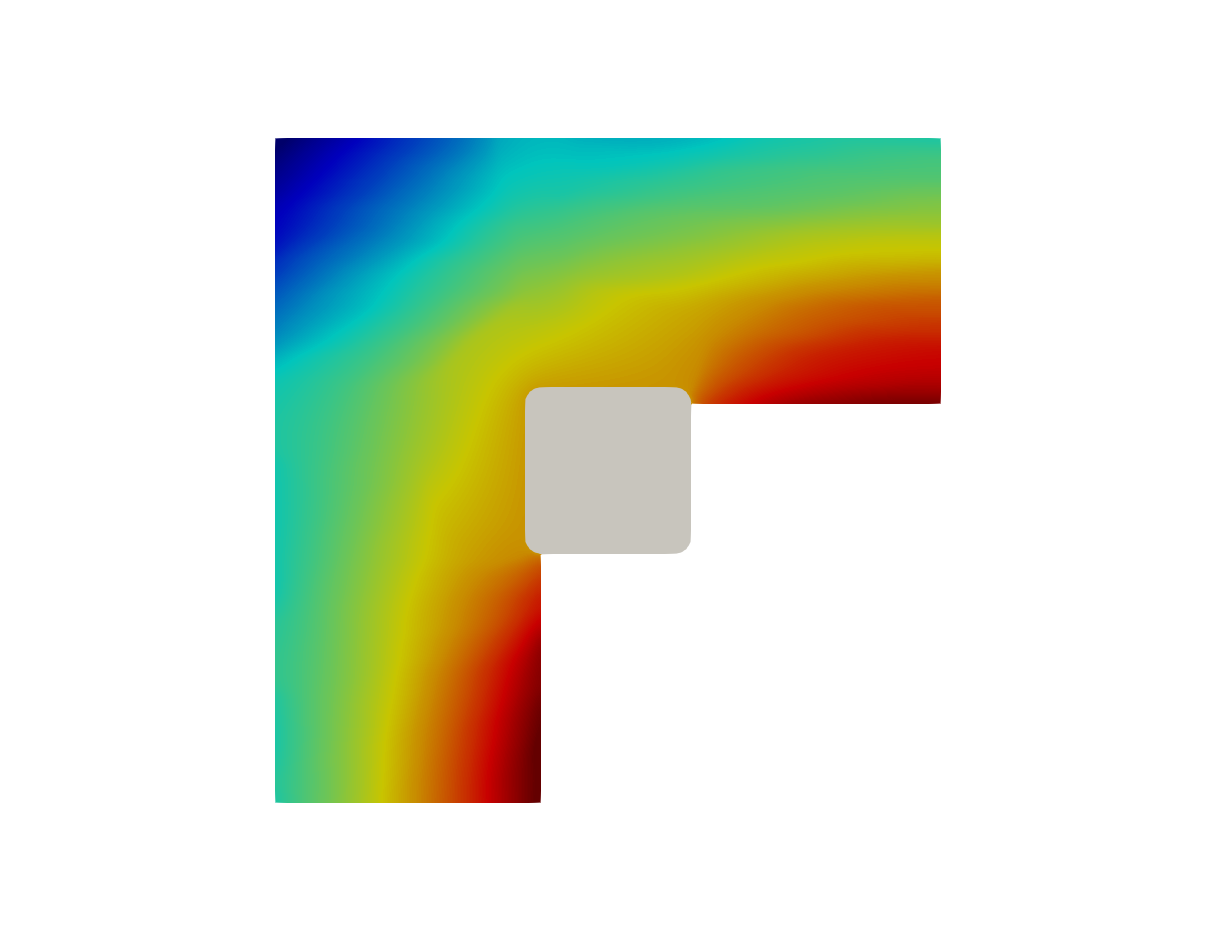}\\
    \multicolumn{4}{c}{\includegraphics[width=0.35\linewidth]{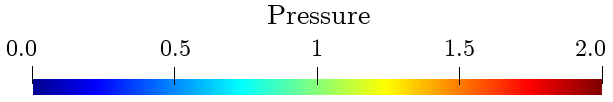}}\\
    \includegraphics[trim={3.5in 1.5in 3.5in 1.5in},clip,width=0.16\linewidth]{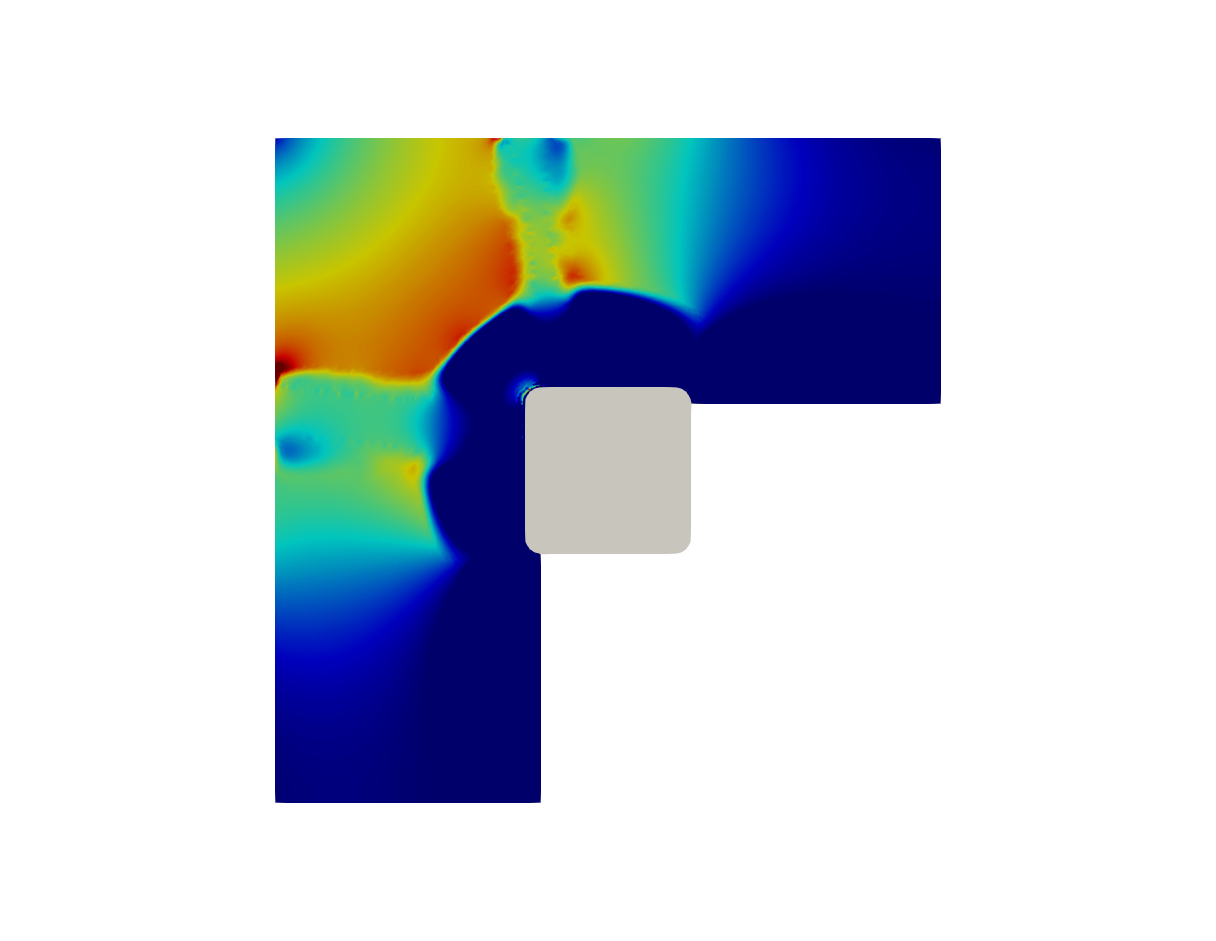}&
    \includegraphics[trim={3.5in 1.5in 3.5in 1.5in},clip,width=0.16\linewidth]{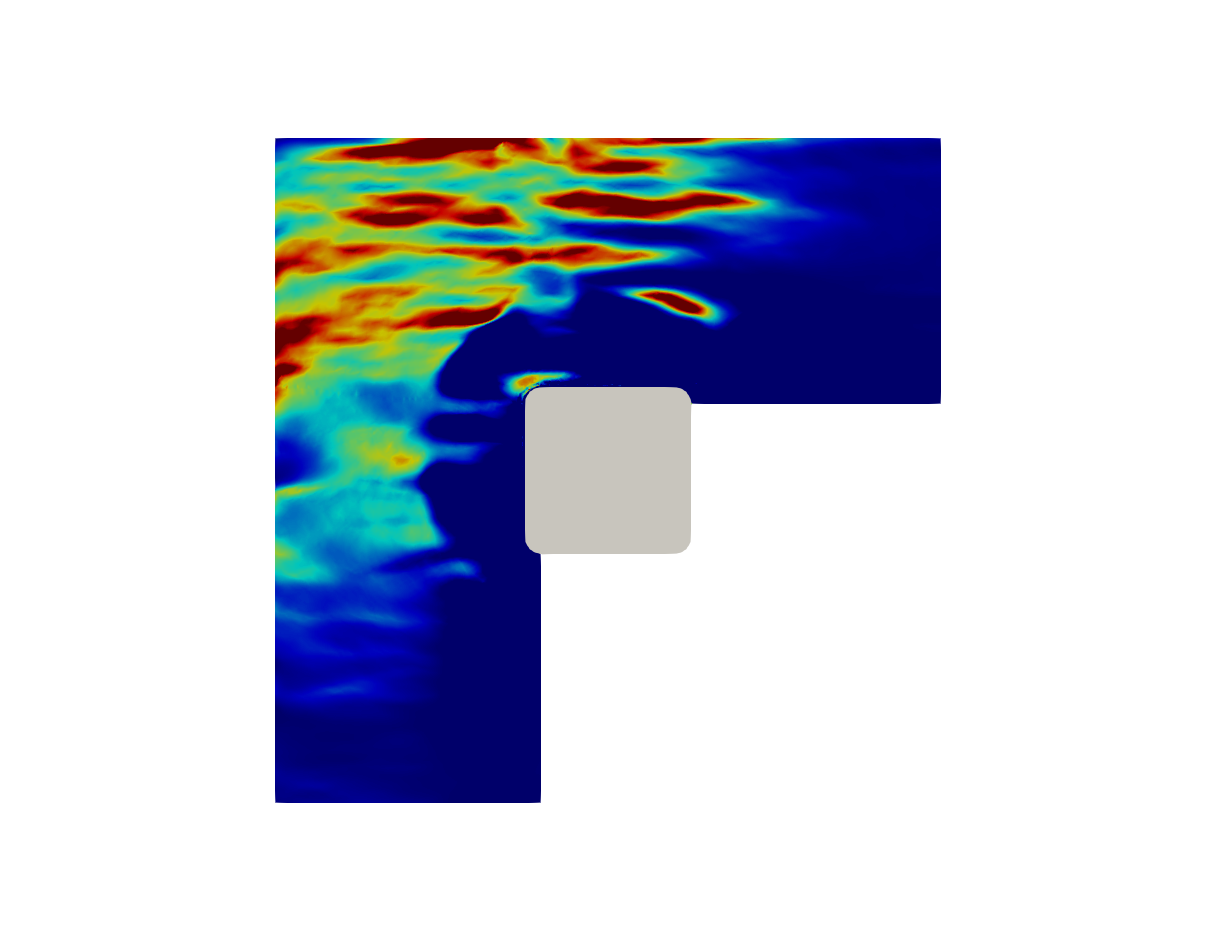}&
    \includegraphics[trim={3.5in 1.5in 3.5in 1.5in},clip,width=0.16\linewidth]{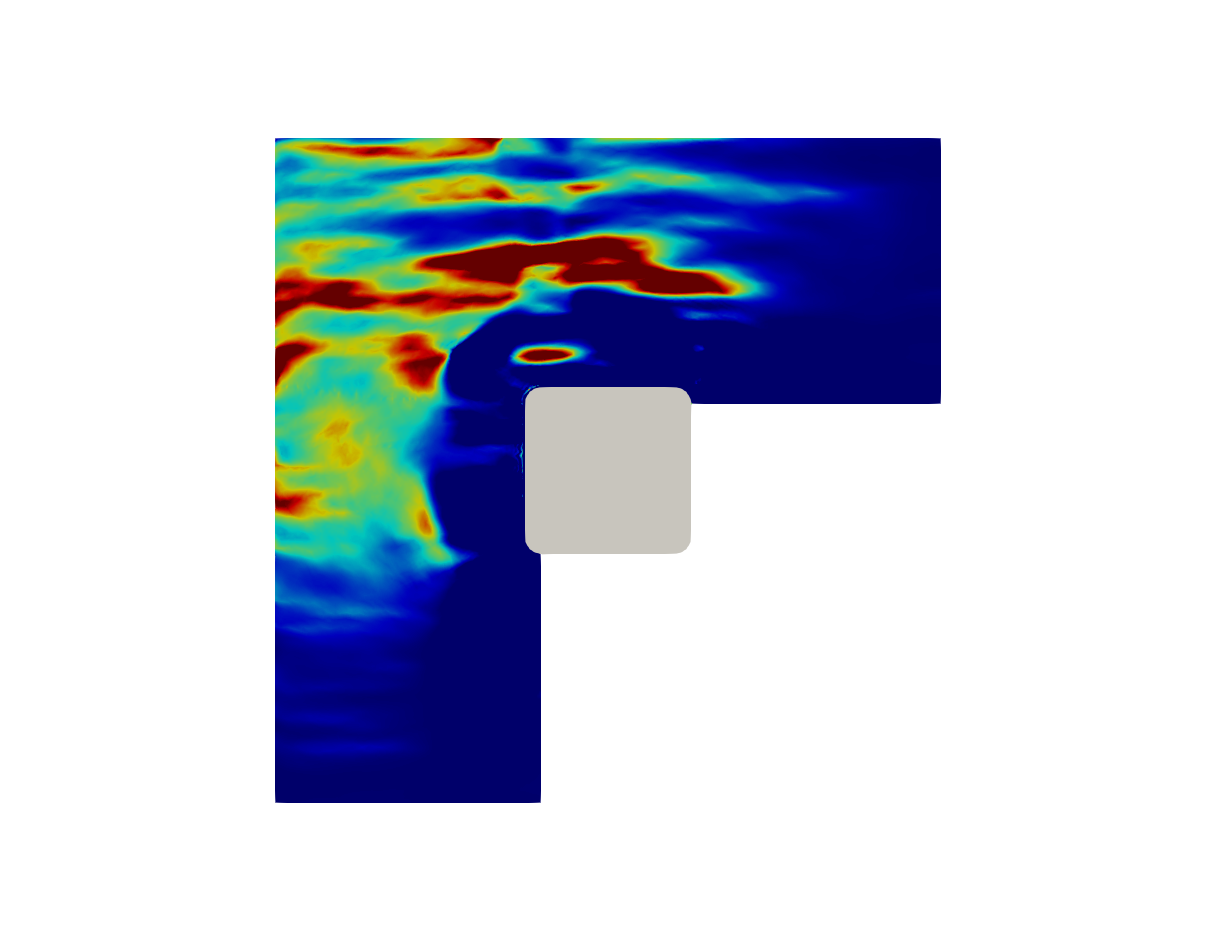}&
    \includegraphics[trim={3.5in 1.5in 3.5in 1.5in},clip,width=0.16\linewidth]{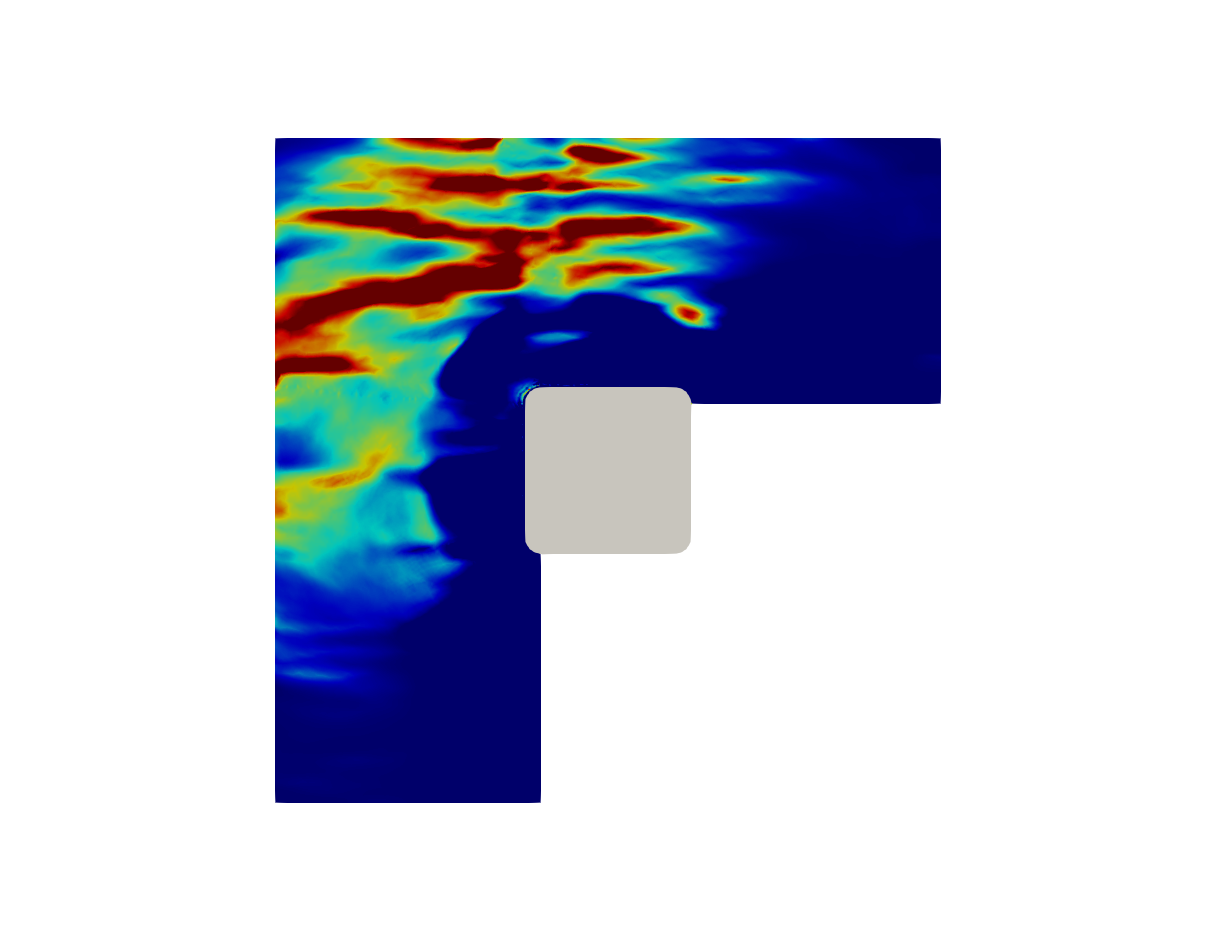}\\
    \end{tabular}
    \vspace{-0.1in}
    \caption{Porosity and states of the scenario at the optimal design shown in Figure \ref{fig:critical_chance} (b) obtained at the mean(first column) of the uncertain parameter $\Tilde{m}=0$, three i.i.d. samples (second, third and fourth column) of $m$. The rows show the porosity $\phi_f$, solid displacement magnitude $||\mathbf{u}_s||$, magnitude of effective solid stress $||\mathbf{T}_s||$, , solid temperature $\theta_s$, fluid temperature $\theta_f$ and fluid pressure $p$. }
    \label{fig:state_plots}
    \vspace{-0.2in}
\end{figure}
%

The limiting critical stress, $T_{cr}$, defines the maximum permissible von Mises stress for the beam structure. Figure \ref{fig:critical_stress} shows the influence of $T_{cr}$ on the optimal designs, evaluated at the mean of the uncertain parameter $\Bar{m} = 0$, along with the corresponding von Mises stress distributions. The findings indicate that increasing the critical stress $T_{cr}$ results in the allocation of stronger insulating material (depicted in blue) across a larger region of the domain. This enhances the insulation performance but compromises the mechanical strength of the component. Thus, a higher value of $T_{cr}$ improves insulation efficiency at the expense of reduced mechanical stability.
Figure \ref{fig:state_plots} shows the states and porosity field evaluated at $\Bar{m}=0$ and three samples for the scenario in Figure \ref{fig:critical_stress} (b). The states include the solid displacement magnitude $||\mathbf{u}_s||$, effective solid stress $||\mathbf{T}_s||$, solid temperature $\theta_s$, fluid temperature $\theta_f$ and fluid pressure $p$.

\clearpage
\subsection{Three-dimensional design of beam-insulator system}
\noindent
\begin{figure}[h!]
    \centering
    \begin{tabular}{c c}
       \includegraphics[trim={4.5in 2.0in 4.5in 1.0in},clip,width=0.4\linewidth]{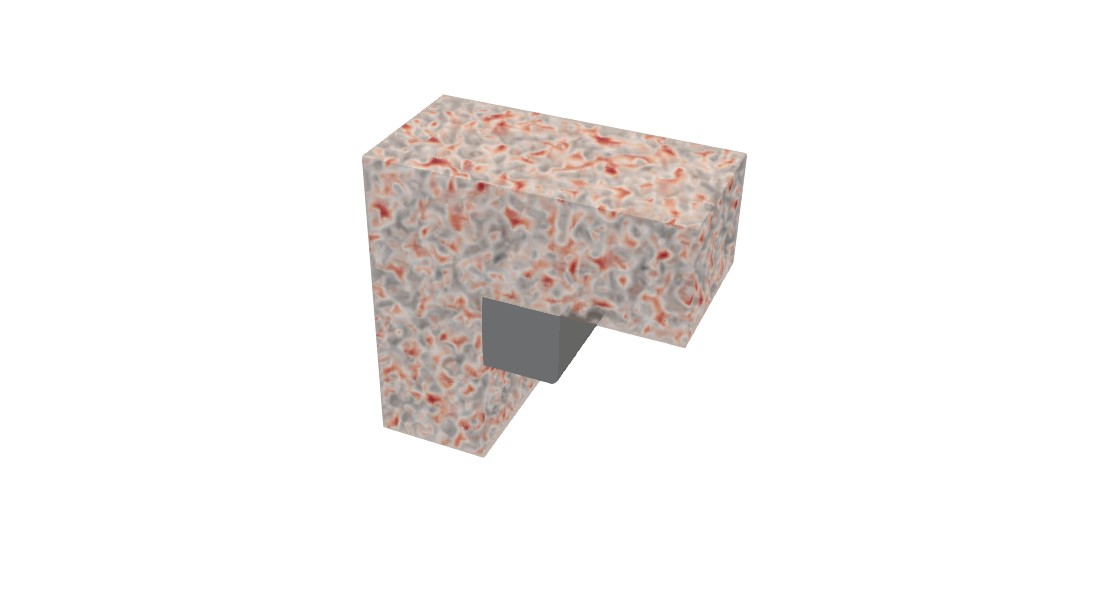}&\includegraphics[trim={4.5in 2.0in 4.5in 1.0in},clip,width=0.4\linewidth]{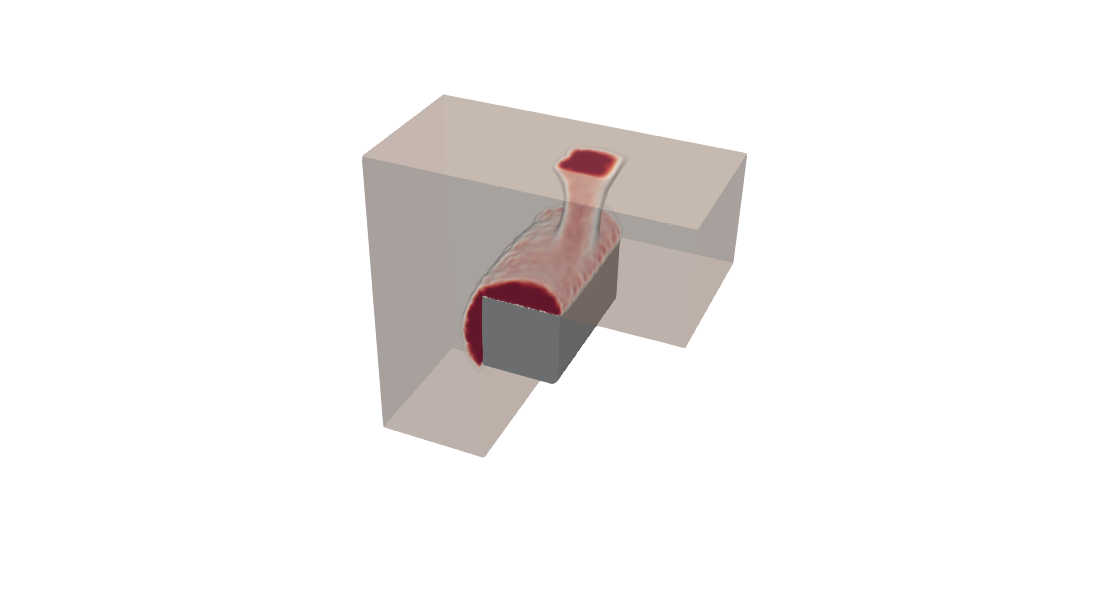}  \\
        (a) & (b)\\
         \includegraphics[trim={4.5in 2.0in 4.5in 1.0in},clip,width=0.4\linewidth]{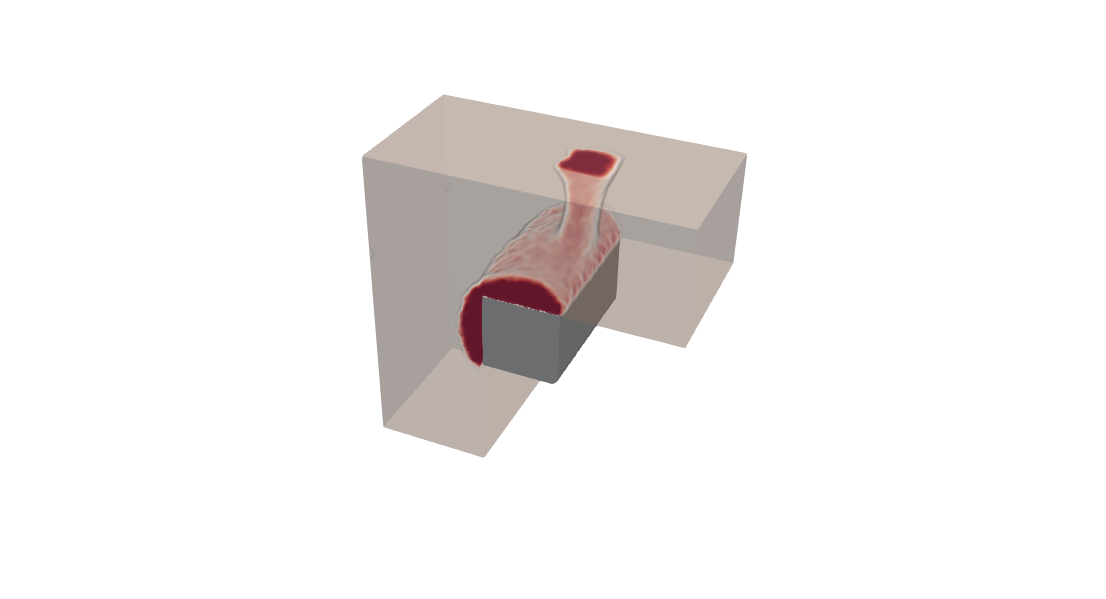}&\includegraphics[trim={4.5in 2.0in 4.5in 1.0in},clip,width=0.4\linewidth]{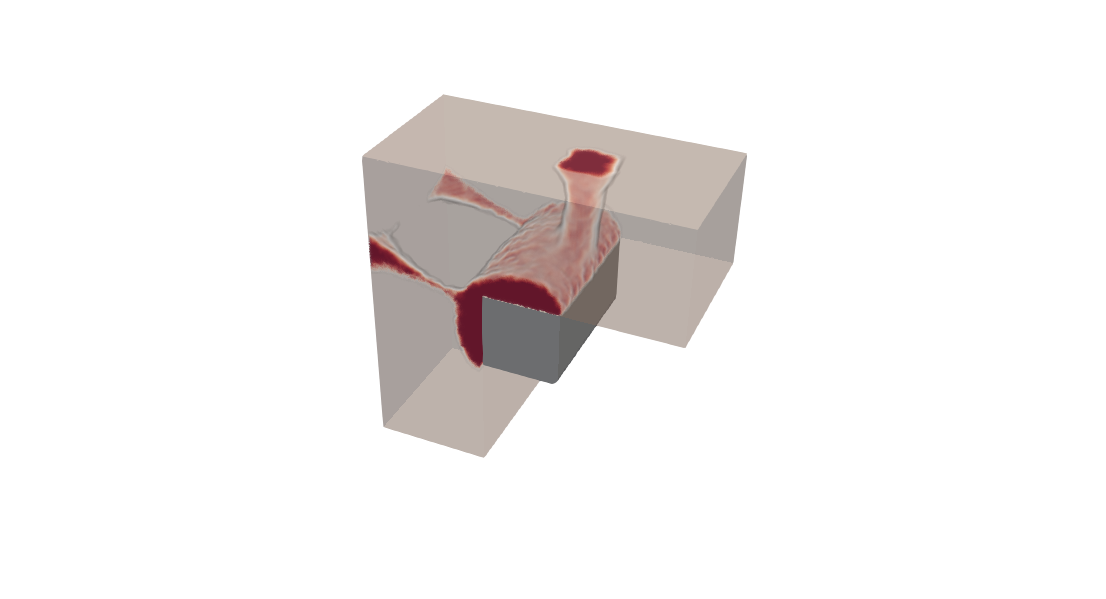} \\
         (c) & (d)\\
         \includegraphics[trim={4.5in 2.0in 4.5in 1.0in},clip,width=0.4\linewidth]{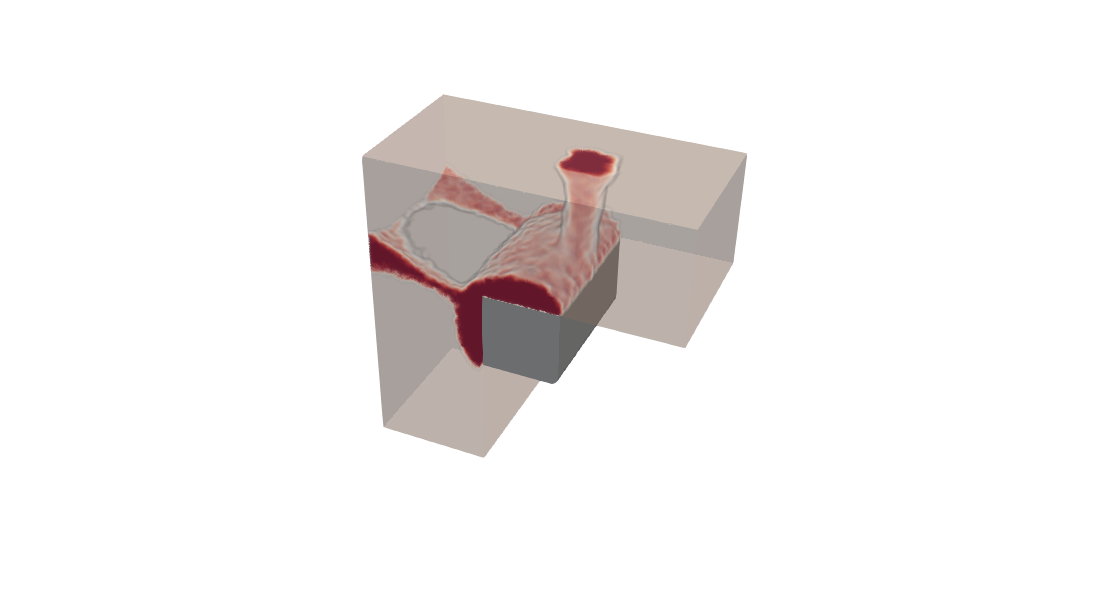} &\includegraphics[trim={4.5in 2.0in 4.5in 1.0in},clip,width=0.4\linewidth]{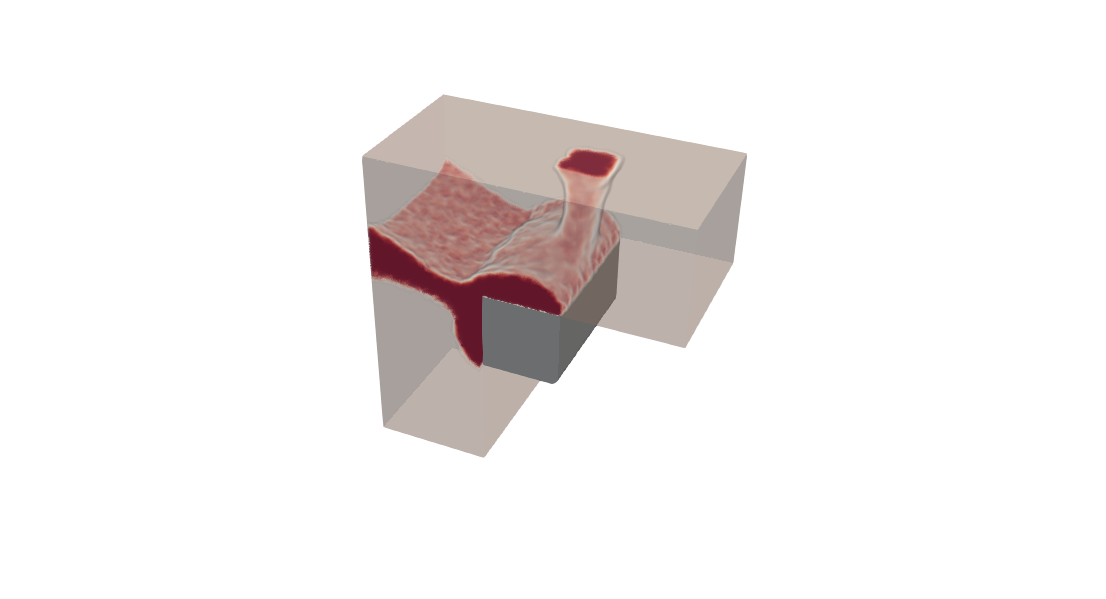} \\
         (e) & (f)\\
    \end{tabular}
    \caption{Evolution of design pattern with continuation scheme. Optimal design obtained at each step of the continuation scheme with increasing $\omega$ as shown in Figure \ref{fig:smooth_function}. Random field as initial condition is shown in (a) and the results obtained for optimal design at mean $\Tilde{m}=0$ for steps in adaptive optimization from step 1 to step 5 shown in (b)-(f). The uncertain parameter $m$ has a correlation length of $L_{CR} = 0.25$ and a variance of $\sigma^2 = {0.1}^2$,the critical chance is $\alpha_c = 0.1$, and the critical stress considered is $T_{cr} = 12.5 \: \text{MPa}$. }
    \label{fig:3D_scenario}
    \vspace{-0.15in}
\end{figure}
We demonstrate the efficiency and scalability of the proposed optimization under uncertainty framework by applying it to a three-dimensional design problem. Specifically, we consider the 3D thermal break scenario shown in Figure \ref{fig:domain}, with a thickness of 0.5. In this case, a uniform circular load is also applied to the middle of the top surface, simulating the influence of additional building envelope assemblies.
The uncertain parameter $m$ is modeled with a horizontal correlation length of $L_{CR} = 0.25$ and a variance of $\sigma^2 = {0.1}^2$. For the chance constraint, we set a critical chance of $\alpha_c = 0.1$ and a critical stress of $T_{cr} = 12.5 \: \text{MPa}$. The finite element discretization of the 3D domain resulted in 919,464 design parameters.
\begin{figure}[h!]
\vspace{-0.15in}
    \centering
    \begin{tabular}{c c}
        \includegraphics[trim={4.5in 1.5in 4.5in 1.0in},clip,width=0.38\linewidth]{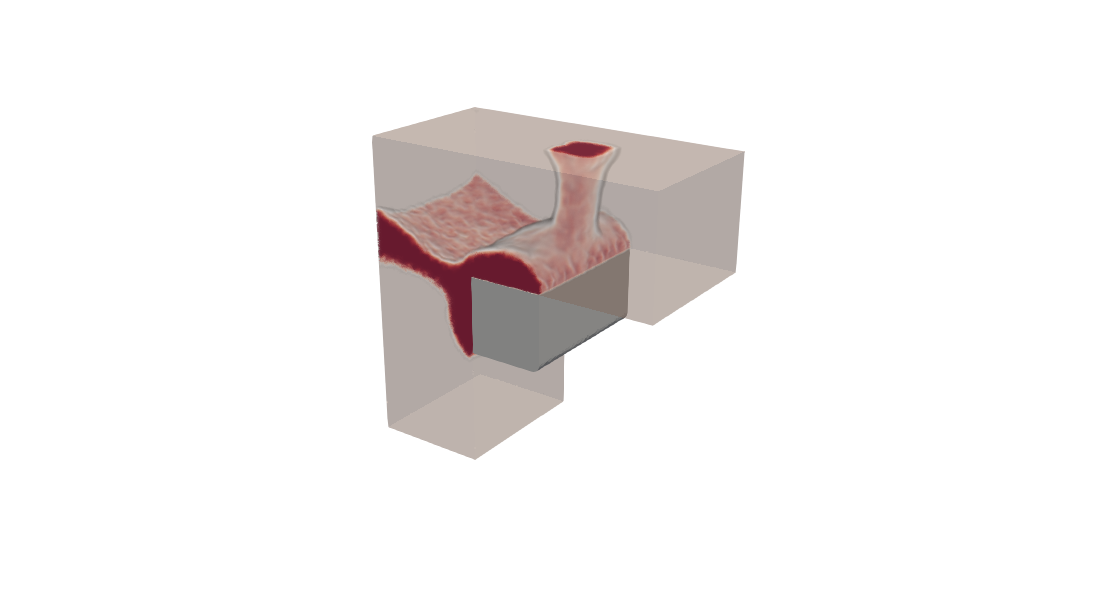} &\includegraphics[trim={4.5in 1.0in 4.5in 1.0in},clip,width=0.38\linewidth]{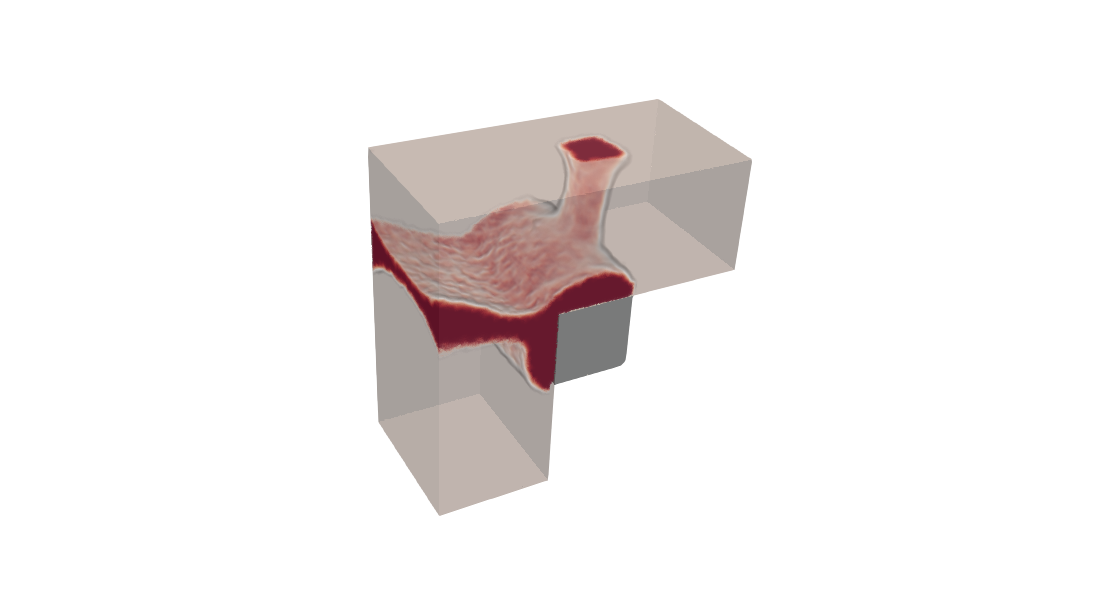}  \\
        (a) & (b)\\
    \end{tabular}
    \caption{
    Optimal design patterns for the 3D scenario. Final design obtained after five steps of the continuation scheme, achieving convergence of the critical chance $\alpha_c$. The 3D domain's finite element discretization, resulted in 919464 design parameters.
    }
    \label{fig:3D_final}
    \vspace{-0.1in}
\end{figure}
Figure \ref{fig:3D_scenario} illustrates the optimal designs obtained at each step of the continuation scheme for a random field as initial condition, with increasing $\omega$ as shown in Figure \ref{fig:smooth_function}.
We tested different initial conditions and arrived at the same final pattern, indicating that for the scenario represented in Figure \ref{fig:3D_scenario}, we achieve global optima. Regions shown in red correspond to material with low porosity ($\phi_f =0.1$).
As the continuation scheme progresses, the red regions start to form, increasing mechanical support in the domain and decreasing the p-norm of von Mises stress. Consequently, the probability in \eqref{eq:inequality_chance} approaches the critical chance threshold of $\alpha_c = 0.1$. By step 5, the final design, shown in Figure \ref{fig:3D_final}, emerges, featuring a column-like structure of low-porosity material $\phi_f =0.1$, providing support aligned with the direction of the applied external load, and also along the left wall to provide support against the stress due to thermal expansion of the beam.

\section{Conclusions}\label{sec:conclusions}
\noindent
This study introduces an efficient and scalable computational framework for chance-constrained optimal design under high-dimensional uncertainty governed by finite element models. The framework optimizes the spatial distribution of a spatially correlated and uncertain porosity field in additively manufactured aerogel thermal breaks to maximize thermal insulation performance while mitigating stress concentrations.
Two key strategies are implemented within the framework. The first utilizes a quadratic Taylor expansion of the design objective with respect to the uncertain parameters to efficiently estimate the mean and variance in the cost function, as well as the chance constraints. The second employs the quadratic approximation as a control variate to enhance the efficiency of Monte Carlo estimations.
The framework incorporates a randomized algorithm to solve the generalized eigenvalue problem, enabling efficient computation of the trace of the covariance-preconditioned Hessian. Scalability with respect of high-dimensional uncertain parameter is achieved through the rapid decay of eigenvalues and the dimension-independent convergence of the optimization iterations.


In future work, we aim to enhance the current design framework under uncertainty by incorporating uncertainties in the forward model parameters. These uncertainties can be rigorously quantified using Bayesian inference methods informed by experimental measurements of aerogel properties. Additionally, our present results demonstrate effective control over interface thickness between materials using Tikhonov regularization. To further improve control for a broader range of design scenarios, we plan to integrate phase-field regularization, as proposed in \cite{tan2024scalable}. This enhancement would be implemented through a continuation scheme, which, while potentially increasing computational demands, offers improved precision in interface control.
Finally, despite the efficiency of the existing framework, its application to large-scale engineering systems remains computationally challenging due to the extensive requirement for repeated PDE solves within the optimization loop. To overcome this limitation, we propose incorporating a Bayesian neural operator surrogate model for the multiphase PDE system, following the construction methodology outlined in \cite{singh2024opal}. This surrogate model would ensure reliable predictions by accurately capturing the relationship between joint uncertainties and design parameters. Furthermore, it would enable efficient approximation of both the PDE state and its sensitivities with respect to the design parameters, facilitating scalable and computationally feasible gradient-based optimization for complex systems.


This study addresses the computational challenges of optimization under high-dimensional uncertainty by leveraging intrinsic dimensionality and incorporating chance constraints to prevent stress concentrations. The framework enhances optimal design reliability, offering a robust and scalable approach while paving the way for future advancements to refine and extend its capabilities.

\section*{Data Availability Statement}
\noindent
Some or all data, models, or code generated or used during the study are available in a repository online.

\section*{Acknowledgments}
\noindent
The authors gratefully acknowledge the financial support received from the U.S. National Science Foundation (NSF) CAREER Award CMMI-2143662.
The authors express their appreciation for Dr.Peng Chen, from Georgia Institute of Technology for valuable discussions on problem formulation and software libraries.
Furthermore, the authors acknowledge the support provided by the Center for Computational Research at the University at Buffalo.



\appendix

\section{Gradient and Hessian action of design objective and chance constraint}\label{sec:gH}
This section outlines the evaluations of the $m$-gradient, $m^f$-gradient, $m$-Hessian, and $m^f$-Hessian at the mean of the uncertain parameters $\Bar{m}$, which are essential for constructing the quadratic Taylor approximations of the design objective and chance constraint, as described in Section \ref{sec:quad}.

$m$-gradient: 
The gradient $Q^{m}(\Bar{m})$ acting
in the direction $\Tilde{m} = m-\Bar{m}$ is expressed as \cite{tan2024scalable},
\begin{equation}
    \langle  \Tilde{m}, \Bar{Q}^m  \rangle = \langle \Tilde{m}, \partial_m\Bar{Q} + \partial_m{\Bar{r}^{Q}} \rangle,
    \label{eq:Qm}
\end{equation}
where $r^Q = r(\mathbf{u}, \mathbf{v}^Q, m, d)$ and the adjoint variable $\mathbf{v}^Q$ is considered as the Lagrange multiplier associated with the state equation with respect to the objective function $Q$.
Here $(\bar{\cdot})$ is the quantities at the mean of $m$, and $(\Tilde{\cdot})$ indicates directional variables referred to the variables that define the direction along which perturbation is considered to compute the change in the functional to obtain the derivative of the functional.
In order to obtain the $m$-gradient, we need to obtain the multipliers $\mathbf{u}$ and $\mathbf{v}^Q$, which can be obtained using the state problem,
\begin{equation}
\textit{Find $\mathbf{u}$, such that} \quad
    \langle  \Tilde{\mathbf{v}}, \partial_\mathbf{v}\Bar{r}  \rangle = 0,
    \label{eq:state_problem}
\end{equation}
and the adjoint problem,
\begin{equation}
\textit{Find $\mathbf{v}^Q$, such that} \quad
    \langle  \Tilde{\mathbf{u}}, \partial_\mathbf{u}{\Bar{r}^{Q}}  \rangle = - \langle  \Tilde{\mathbf{u}}, \partial_\mathbf{u}\Bar{Q}  \rangle.
    \label{eq:adjoint_problem}
\end{equation}
%

$m$-Hessian: 
The Hessian $Q^{mm}(\Bar{m})$ acting on $\hat{m}$ can be expressed as \cite{tan2024scalable},
\begin{equation}
\begin{split}
    \langle\Tilde{m}, \Bar{Q}^{mm}\,\hat{m}^Q\rangle
     = &
    \langle 
    \Tilde{m}, \partial_{m\mathbf{v}}\Bar{r}^{Q}\,\hat{\mathbf{v}}^Q
    +
    \partial_{m\mathbf{u}}\Bar{r}^{Q}\,\hat{\mathbf{u}}^{Q}
    +
    \partial_{m\mathbf{u}}\Bar{Q}\,\hat{\mathbf{u}}^{Q}
    +
    \partial_{mm}\Bar{r}^{Q}\,\hat{m}^Q\\
    +
    \partial_{mm}\Bar{Q}\,\hat{m}^Q
    \rangle.
\end{split}
    \label{eq:Qmm}
\end{equation}
Thus, computing $m$-Hessian requires evaluation of the Lagrange multipliers $\hat{\mathbf{u}}^Q$ and $\hat{\mathbf{v}}^Q$ by solving the incremental state problem:
\begin{equation}
\textit{Find $\hat{\mathbf{u}}^Q$, such that} \quad
    \langle  \Tilde{\mathbf{v}}, \partial_{\mathbf{vu}}\Bar{r}^{Q}\,\hat{\mathbf{u}}^Q  \rangle 
    = - 
    \langle  \Tilde{\mathbf{v}}, \partial_{\mathbf{v}m}\Bar{r}^{Q}\,\hat{m}^Q  \rangle,
    \label{eq:Qmm_incremental_state}
\end{equation}
and the incremental adjoint problem:
\begin{equation} \nonumber
\textit{Find $\hat{\mathbf{v}}^Q$, such that}
\end{equation}
\begin{equation}
    \langle  \Tilde{\mathbf{u}}, \partial_{\mathbf{uv}}\Bar{r}^{Q} \,\hat{\mathbf{v}}^{Q} \rangle 
    = - 
    \langle  
    \Tilde{\mathbf{u}}, \partial_{\mathbf{uu}}\Bar{r}^{Q}\,\hat{\mathbf{u}}
    +
    \partial_{\mathbf{uu}} \Bar{Q}\,\hat{\mathbf{u}}^Q
    +
    \partial_{\mathbf{u}m}\Bar{r}^{Q}\,\hat{m}^Q
    +
    \partial_{\mathbf{u}m}\Bar{Q}\,\hat{m}^Q    
    \rangle.
    \label{eq:Qmm_incremental_adjoint}
\end{equation}
%

$m^f$-gradient: 
The gradient of $f$ at $\Bar{m}$ acting
in the direction $\Tilde{m}$ defined as \cite{chen2021taylor}:
\begin{equation}
    \langle \Tilde{m}, \Bar{f}^{m} \rangle = \langle \Tilde{m}, \partial_{m} \Bar{f} + \partial_{m} \Bar{r}^{f}\rangle
    \label{eq:gradient_chance}
\end{equation}
The computation of $m^f$-gradient requires the lagrange multiplier $\mathbf{v}^f$ which is computed through the adjoint problem as: 
\begin{equation} \nonumber
\textit{Find $\mathbf{v}^f$, such that} 
\end{equation}
\begin{equation}
    \langle \Tilde{\mathbf{u}}, \partial_{\mathbf{u}} \Bar{r}^{f} \rangle = -\langle \Tilde{\mathbf{u}}, \partial_{\mathbf{u}} \Bar{f} \rangle,
    \label{eq:adjoint_problem_chance}
\end{equation}

$m^f$-Hessian: 
The action of Hessian of $f$ can be computed at $\Tilde{m}$ as \cite{chen2021taylor}:
\begin{equation}
\begin{split}
    \langle \Tilde{m}, \Bar{f}^{mm} \hat{m}^f \rangle  = \langle \Tilde{m}, \partial_{m\mathbf{v}} \Bar{r}^{f} \hat{\mathbf{v}}^f + \partial_{m\mathbf{u}} \Bar{r}^{f} \hat{\mathbf{u}}^f + \partial_{m\mathbf{u}} \Bar{f} \hat{\mathbf{u}}^f + \partial_{mm} \Bar{r}^{f} \hat{m}^f
    + \partial_{mm} \Bar{f} \: \hat{m}^f \rangle \quad .
\end{split}
\label{eq:hessian_chance}
\end{equation}
Thus, the solution of $m^f$-Hessian requires the computation of Lagrange multipliers $\hat{\mathbf{u}}^f$ and $\hat{\mathbf{v}}^f$, by solving the incremental state problem,
\begin{equation}
\begin{split}
\textit{Find $\hat{\mathbf{u}}^f$, such that} &\\
    \langle \Tilde{\mathbf{v}}, \partial_{\mathbf{vu}} \Bar{r}^{f} \hat{\mathbf{u}}^f \rangle = - \langle \Tilde{\mathbf{v}}, \partial_{\mathbf{v}m} \Bar{r}^{f} \hat{m}^f \rangle \quad& ,
\end{split}
\label{eq:incremental_state_chance}
\end{equation}
and the incremental adjoint problem,
\begin{equation}
\begin{split}
\textit{Find $\hat{\mathbf{v}}^f$, such that} &\\
    \langle \Tilde{\mathbf{u}}, \partial_{\mathbf{uv}} \Bar{r}^{f} \hat{\mathbf{v}}^f \rangle = -\langle \Tilde{\mathbf{u}}, \partial_{\mathbf{uu}}\Bar{r}^{f}\hat{\mathbf{u}}^f + \partial_{\mathbf{uu}}\Bar{f}\hat{\mathbf{u}}^f + \partial_{\mathbf{u}m}\Bar{r}^{f}\hat{m}^f + &
    \partial_{\mathbf{u}m}\Bar{f}\hat{m}^f \rangle \:.  
\end{split}
\label{eq:incremental_adjoint_chance}
\end{equation}

Also, in order to compute $m^f$-Hessian, we solve a generalized eigenvalue problem in \eqref{eq:eig_chance} for a finite number of eigenvalues $N^f_\mathrm{eig}$.

\section{Gradient of design functional}\label{sec:dCdd}
%
The Lagrangian formalism is employed to derive the $d$-gradient by considering a  meta-Lagrangian functional
\begin{equation}
\footnotesize
    \begin{aligned}
    \mathcal{L}_\mathrm{QUAD} &(d,\mathbf{u},\mathbf{v}^*, \{ {\lambda_{j}^{Q}} \}, \{ {\psi_{j}}^{Q} \}, \{ {\hat{\mathbf{u}}_{j}^{Q}} \}, \{ {\hat{\mathbf{v}}_{j}}^{Q} \}, 
    \\
    & \left(\mathbf{u}^{Q}\right)^{*}, \left(\mathbf{v}^{Q}\right)^{*}, \{ \left(\lambda_{j}^{Q}\right)^{*} \}, \{ \left(\psi_j^{Q}\right)^{*} \}, \{ \left(\hat{\mathbf{u}}_{j}^{Q}\right)^{*} \}, \{ \left(\hat{\mathbf{v}}_{j}^{Q}\right)^{*} \},
    \\
    & \{ {\lambda_{j}^{f}} \}, \{ {\psi_{j}}^{f} \}, \{ {\hat{\mathbf{u}}_{j}}^{f} \}, \{ {\hat{\mathbf{v}}_{j}}^{f} \}, \left(\mathbf{u}^{f}\right)^{*}, \left(\mathbf{v}^{f}\right)^{*}, \{ \left(\lambda_{j}^{f}\right)^{*} \}, \{ \left(\psi_j^{f}\right)^{*} \}, \{ \left(\hat{\mathbf{u}}_{j}^{f}\right)^{*} \}, \{ \left(\hat{\mathbf{v}}_{j}^{f}\right)^{*} \})\\
    &=
    \mathcal{J}_\mathrm{QUAD} (d)
    +
    \langle \mathbf{v}^*, \partial_\mathbf{v} \Bar{r}^Q   \rangle
    \\
    &+ \tau_{\gamma} (\mathbb{E}[l_{\omega}(f_{\text{QUAD}}(m,d))] - \alpha_c) 
    \\
    &+
    \langle \left(\mathbf{u}^{f}\right)^{*}, \partial_\mathbf{u} \Bar{r}_f +\partial_\mathbf{u}\Bar{f}   \rangle
    \\
    &+
    \sum^{N_\mathrm{eig}^f}_{j=1} \langle \left(\psi_j^{f}\right)^{*}, (\Bar{f}^{mm} -{\lambda_{j}^{f}}\mathcal{K}^{-1}) \psi_{j}^{f} \rangle
    \\
    &+
    \sum^{N_\mathrm{eig}^f}_{j=1}  \left(\lambda_{j}^{f}\right)^{*}    ( \langle {\psi_{j}}^{f} ,\mathcal{K}^{-1}{\psi_{j}}^{f} \rangle - 1)
    +
    \sum^{N_\mathrm{eig}^f}_{j=1}  \langle  \left(\hat{\mathbf{v}}_{j}^{f}\right)^{*},  \partial_{\mathbf{vu}}\Bar{r}_f \,{\hat{\mathbf{u}}_{j}}^{f} + \partial_{\mathbf{v}m}\Bar{r}_f\,{\psi_{j}}^{f} \rangle
    \\
    &+
    \sum^{N_\mathrm{eig}^f}_{j=1}   \langle   \left(\hat{\mathbf{u}}_{j}^{f}\right)^{*},     \partial_{\mathbf{uv}}\Bar{r}_f\,{\hat{\mathbf{v}}_{j}}^{f} + \partial_{\mathbf{uu}}\Bar{r}_f\,\left(\mathbf{u}^{f}\right)^{*}  + \partial_{\mathbf{uu}}\Bar{f}\,{\hat{\mathbf{u}}_{j}}^{f}  +\partial_{\mathbf{u}m}\Bar{f} \,{\psi_{j}}^{f} +\partial_{\mathbf{u}m}\Bar{r}_f \,{\psi_{j}}^{f}    \rangle
    \\
    &+
    \langle \left(\mathbf{u}^{Q}\right)^{*}, \partial_\mathbf{u} \Bar{r}^Q +\partial_\mathbf{u}\Bar{Q}   \rangle
    \\
    &+
    \sum^{N_\mathrm{eig}^Q}_{j=1} \langle \left(\psi_j^{Q}\right)^{*}, (\Bar{Q}^{mm} -{\lambda_{j}^{Q}}\mathcal{K}^{-1}) \psi_{j}^{Q} \rangle
    \\
    &+
    \sum^{N_\mathrm{eig}^Q}_{j=1}  \left(\lambda_{j}^{Q}\right)^{*}    ( \langle {\psi_{j}}^{Q} ,\mathcal{K}^{-1}{\psi_{j}}^{Q} \rangle - 1)
    +
    \sum^{N_\mathrm{eig}^Q}_{j=1}  \langle  {\hat{\mathbf{v}}_{j}}^{{Q}^{*}},  \partial_{\mathbf{vu}}\Bar{r}_Q \,{\hat{\mathbf{u}}_{j}}^{Q} + \partial_{\mathbf{v}m}\Bar{r}_Q\,{\psi_{j}}^{Q} \rangle
    \\
    &+
    \sum^{N_\mathrm{eig}^Q}_{j=1}   \langle   \left(\hat{\mathbf{u}}_{j}^{Q}\right)^{*},     \partial_{\mathbf{uv}}\Bar{r}_Q\,{\hat{\mathbf{v}}_{j}}^{Q} + \partial_{\mathbf{uu}}\Bar{r}_Q\,\left(\mathbf{u}^{Q}\right)^{*}  + \partial_{\mathbf{uu}}\Bar{Q}\,{\hat{\mathbf{u}}_{j}}^{Q}  +\partial_{\mathbf{u}m}\Bar{Q} \,{\psi_{j}}^{Q} +\partial_{\mathbf{u}m}\Bar{r}_Q \,{\psi_{j}}^{Q}    \rangle
    \\
    \end{aligned}
    \label{eq:lagrange_quad}
\end{equation}
where 
$(\cdot^*)$ represents the adjoint variables, 
$(\hat{\cdot})$ denotes the incremental variables.
The terms on the right hand side of \eqref{eq:lagrange_quad}, correspond to the following:
the quadratic approximation of the design objective \eqref{eq:cost_quad},
the state and 
adjoint problems for chance constraint $f$ and design objective $Q$,
the generalized eigenvalue problems for chance constraint $f$ and design objective $Q$,
the incremental state and adjoint problems for chance constraint $f$ and design objective $Q$.

For obtaining the design gradient, we need to compute five sets of functional derivatives to solve for the corresponding ten sets of adjoint variables, \\
\{$\left(\mathbf{u}^f\right)^*, \left(\mathbf{u}^Q\right)^*, \left(\mathbf{v}^f\right)^*, \left(\mathbf{v}^Q\right)^*, 
\left(\psi_j^f\right)^* , \left(\psi_j^Q\right)^* ,   
\left(\hat{\mathbf{u}^f_j}\right)^* ,\left(\hat{\mathbf{u}}^Q_j\right)^* ,
\left(\hat{\mathbf{v}^f_j}\right)^*, \left(\hat{\mathbf{v}^Q_j}\right)^* \}  $ in \eqref{eq:lagrange_quad}, while the adjoint variable set for the eigenvalue set $\{ {\lambda_j^f}^*, {\lambda_j^Q}^*  \}$ is not required due to its independency of the design parameter $d$. 

Once we have defined the functional $\mathcal{L}_\mathrm{QUAD}$, 
we need to apply the condition that the first variation of Lagrangian as zero with respect to Lagrange multipliers for both design objective $Q$ and chance constraint function $f$.
We start by equating the variation of \eqref{eq:lagrange_quad} as zero with respect to  $\{\lambda_{j}^{f}\}$,
\begin{equation}
    \{ \left(\psi^f_j\right)^* \} = \frac{1}{2} \bigg(1  + 2 \beta_V\,\lambda^f_j \bigg) \, \{\psi^f_j\}, \quad
    j=1, \cdots, N^f_\mathrm{eig}.
\end{equation}
Next, we equate the first variation of \eqref{eq:lagrange_quad} as zero with respect to $\mathbf{v}_{j}^{f}$ and $\mathbf{u}_{j}^{f}$ which gives the incremental state problem,
\begin{equation}
\textit{Find $\hat{\mathbf{u}}^f$, such that} \quad
    \big< \Tilde{\mathbf{v}} , \partial_\mathbf{vu} \Bar{r}^{f} \, \mathbf{u}^{f*}_j \big>
    =
    - \big<  \Tilde{\mathbf{v}}, \partial_{\mathbf{v}m} \Bar{r}^{f} \,\psi^{{f}^*}_{j}  \big>.
    \label{eq:Jd_incremental_state2}
\end{equation}
%
 and linear adjoint problem:
\begin{equation}
    \begin{aligned}
    \textit{Find $\left(\mathbf{v}^f\right)^*$, such that}\\
            \langle  \Tilde{\mathbf{v}}, \partial_{\mathbf{uv}}\Bar{r}^{f}\,\left(\mathbf{u}^{f}\right)^*  \rangle
            =&
            - \sum_{n=1}^{N_{eig}^f} \langle \Tilde{\mathbf{v}}, \partial_{\mathbf{v}m\mathbf{u}} \Bar{r}^{f} \hat{\mathbf{u}}_{n}^{f} {\left( \psi_{n}^{f} \right)}^{*} +  \partial_{\mathbf{v}mm} \Bar{r}^{f} \psi_{n}^{f} {\left( \psi_{n}^{f} \right)}^{*} \rangle
            \\
            &- \sum_{n=1}^{N_{eig}^f} \langle \Tilde{\mathbf{v}}, \partial_{\mathbf{vuu}} \Bar{r}^{f} \hat{\mathbf{u}}_{n}^{f} {\left( \hat{\mathbf{u}}_{n}^{f} \right)}^{*} +  \partial_{\mathbf{vu}m} \Bar{r}^{f} \psi_{n}^{f} {\left( \hat{\mathbf{u}}_{n}^{f} \right)}^{*} \rangle\\
            & - \langle \Tilde{\mathbf{v}}, \partial_{\mathbf{v}m} \Bar{r}^{f} m^{f} \rangle.
    \end{aligned}
    \label{eq:Jd_linear_adjoint_prob}
\end{equation}
Similarly, we set the first variation of the \eqref{eq:lagrange_quad} to zero for the Lagrange multipliers related to design objective $Q$.
Firstly, we equate the first variation of \eqref{eq:lagrange_quad} w.r.t eigenvalues $\{\lambda^Q_j\}$ to zero, 
\begin{equation}
    \{ {\psi^Q_j}^* \} = \frac{1}{2}\bigg(1 + 2 \beta_V\,\lambda_j^Q \bigg) \, \{\psi_j^Q\}, \quad
    j=1, \cdots, N_\mathrm{eig}^Q.
    \label{eq:adjoint_eigvecs}
\end{equation}
Furthermore, we let the first variation of \eqref{eq:lagrange_quad} as zero w.r.t $\mathbf{v}^Q_j$ and $\mathbf{u}^Q_j$ leading to the incremental state problem: 

\begin{equation}
\textit{Find $\hat{\mathbf{u}}^Q$, such that} \quad
    \big< \Tilde{\mathbf{v}} , \partial_\mathbf{vu} \Bar{r}^Q \, \left(\mathbf{u}^Q_j\right)^* \big>
    =
    - \big<  \Tilde{\mathbf{v}}, \partial_{\mathbf{v}m} \Bar{r}^Q \,\left(\psi^Q_j\right)^*  \big>
    .
    \label{eq:Jd_incremental_state}
\end{equation}
and the incremental adjoint problem:

\begin{equation}
\begin{split}
\textit{Find $\hat{\mathbf{v}}^Q$, such that}& \quad\\
    \big< \Tilde{\mathbf{u}} , \partial_\mathbf{uv} \Bar{r}^Q \, \left(\mathbf{v}^Q_j\right)^* \big>
    =
    - \big< \Tilde{\mathbf{u}}, \partial_{\mathbf{uu}} \bar{r}^Q \, {\left(\hat{\mathbf{u}}^Q_j\right)}^*   + \partial_\mathbf{uu} \bar{Q}\,&{\left(\hat{\mathbf{u}}^Q_j\right)}^* + \partial_{\mathbf{u}m} \bar{r}^Q \,\left(\psi^{Q}_{j}\right)^*  \big>
    ,
\end{split}
    \label{eq:Jd_incremental_adjoint}
\end{equation}
similar to \eqref{eq:Qmm_incremental_state} and \eqref{eq:Qmm_incremental_adjoint}, respectively.
In combination with \eqref{eq:adjoint_eigvecs}, we can express the adjoint variables for both the incremental state and adjoint problems separately for each eigenvector,
\begin{align}
    \{ \left(\hat{\mathbf{u}}\right)^*_j \} &= \frac{1}{2} \bigg(1  + 2 \beta_V\,\lambda^Q_j \bigg) \, \{\hat{\mathbf{u}}^Q_j\}, \quad
    j=1, \cdots, N_\mathrm{eig}^Q,
    \\
    \{ \hat{\mathbf{v}}^*_j \} &= \frac{1}{2} \bigg(1  + 2 \beta_V\,\lambda^Q_j \bigg) \, \{\hat{\mathbf{v}}^Q_j\}, \quad
    j=1, \cdots, N_\mathrm{eig}^Q.
\end{align}
Finally, we equate the first variation of the Lagrangian \eqref{eq:lagrange_quad} to zero with respect to adjoint or state, we obtain the linear state problem:

\textit{Find ${\mathbf{u}^Q}^*$, such that}
\begin{equation}
    \begin{aligned}
            \langle  \Tilde{\mathbf{v}}, \partial_{\mathbf{vu}}\Bar{r}^Q\,\left(\mathbf{u}^Q\right)^*  \rangle
            =&
            -2\beta_V \langle  \Tilde{\mathbf{v}}, \partial_{\mathbf{v}m}\Bar{r}^Q\,(\mathcal{K}\partial_m\Bar{r}^Q)  \rangle
            \\
            &-
            \sum^{N_\mathrm{eig}^Q}_{j=1} \langle  \Tilde{\mathbf{v}}, \partial_{\mathbf{v}mu}\Bar{r}^Q\,\hat{\mathbf{u}}^Q_j\,\left(\psi^Q_j\right)^*  +  \partial_{\mathbf{v}mm} \Bar{r}^Q \, \psi^Q_j\,\left(\psi^Q_j\right)^*  \rangle
            \\
            &-
            \sum^{N_\mathrm{eig}^Q}_{j=1}  \langle  \Tilde{\mathbf{v}}, \partial_{\mathbf{vuu}}\Bar{r}^Q\,\hat{\mathbf{u}}^Q_j\,\left(\hat{\mathbf{u}}^Q_j\right)^*  +  \partial_{\mathbf{vu}m} \Bar{r}^Q \, \psi^Q_j\,\left(\hat{\mathbf{u}}^Q_j\right)^*  \rangle,
    \end{aligned}
    \label{eq:Jd_linear_state_prob}
\end{equation}
 and linear adjoint problem:
 
\textit{Find $\left(\mathbf{v}^Q\right)^*$, such that}
\begin{equation}
    \begin{aligned}
            \langle  \Tilde{\mathbf{u}}, \partial_{\mathbf{uv}}\Bar{r}^Q\,\left(\mathbf{v}^Q\right)^*  \rangle
            =&
            -
            \langle  \Tilde{\mathbf{u}}, \partial_\mathbf{u}\Bar{Q}  \rangle
            -
            2\beta_V \langle  \Tilde{\mathbf{u}}, \partial_{\mathbf{u}m}\Bar{r}^Q\, (\mathcal{K}\partial_m\Bar{r}^Q) \rangle
            \\
            &-
            \langle  \Tilde{\mathbf{u}},   \partial_{\mathbf{uu}}\Bar{r}^Q\,\left(\mathbf{u}^Q\right)^* + \partial_{\mathbf{uu}}\Bar{Q}\,\left(\mathbf{u}^Q\right)^*  \rangle
            \\
            &-
            \sum^{N_\mathrm{eig}^Q}_{j=1} \langle  \Tilde{\mathbf{u}}, \partial_{\mathbf{u}m\mathbf{v}}\Bar{r}^Q\,\hat{\mathbf{v}}_j\,\left(\psi^Q_j\right)^*  
            \\
            &+  \partial_{\mathbf{u}m\mathbf{u}} \Bar{r}^Q \, \hat{\mathbf{u}}_j\,\left(\psi^Q_j\right)^* +  \partial_{\mathbf{u}mm} \Bar{r}^Q \, \psi^Q_j\,\left(\psi^Q_j\right)^*  \rangle
            \\
            &-
            \sum^{N_\mathrm{eig}^Q}_{j=1} \langle  \Tilde{\mathbf{u}}, \partial_{\mathbf{uvu}}\Bar{r}^Q\,\hat{\mathbf{u}}^Q_j\,\left(\hat{\mathbf{v}}^Q_j\right)^*  +  \partial_{\mathbf{uv}m} \Bar{r}^Q \, \psi^Q_j\,\left(\hat{\mathbf{v}}^Q_j\right)^*  \rangle
            \\
            &-
            \sum^{N_\mathrm{eig}^Q}_{j=1}  \langle  \Tilde{\mathbf{u}}, \partial_{\mathbf{uuv}}\Bar{r}^Q\,\hat{\mathbf{v}}^Q_j\,\left(\hat{\mathbf{u}}^Q_j\right)^*  +  \partial_{\mathbf{uuu}} \Bar{r}^Q \, \hat{\mathbf{u}}^Q_j\,\left(\hat{\mathbf{u}}^Q_j\right)^* \\& + \partial_{\mathbf{uuu}}\Bar{Q}\,\hat{\mathbf{u}}^Q_j\,\left(\hat{\mathbf{u}}^Q_j\right)^*    + \partial_{\mathbf{uu}m} \Bar{r}^Q\,\psi^Q_j\,\left(\hat{\mathbf{u}}^Q_j\right)^*   \rangle.
    \end{aligned}
    \label{eq:Jd_linear_adjoint_prob2}
\end{equation}
The computation of $d$-gradient means that we have to compute how the cost functional $\mathcal{J}_\mathrm{QUAD}$ changes with small variations in $d$. This can be done by computing the Fr\'{e}chet derivative of $\mathcal{J}_\mathrm{QUAD}$ w.r.t $d$, which can be expressed as $D_d \mathcal{J}_\mathrm{QUAD}(d)$. However, the computation of this Fr\'{e}chet derivative $D_d \mathcal{J}_\mathrm{QUAD}(d)$ becomes computationally prohibitive when the design parameter $d$ is high-dimensional. Hence, we compute the gradient through the partial derivative of the lagrangian $\mathcal{L}_\mathrm{QUAD}$ with respect to design parameter $d$, which is $\partial_d \mathcal{L}_\mathrm{QUAD}$, as it only requires the solution of lagrange multipliers, which are independent of the dimension of design parameter $d$. 
Also, since the total derivative of $\mathcal{J}_\mathrm{QUAD}$ along a random direction $\Tilde{d}$ is equal to the partial derivative of lagrangian $\mathcal{L}_\mathrm{QUAD}$ w.r.t $d$, as it depends on design parameter only. Therefore, the $d$-gradient can be derived as,
\begin{equation}
    \langle  \Tilde{d},  D_d \mathcal{J}_\mathrm{QUAD}(d)   \rangle  =  \langle  \Tilde{d},  \partial_d \mathcal{L}_\mathrm{QUAD}   \rangle.
    \label{eq:design_gradient}
\end{equation}

\RestyleAlgo{ruled}
\begin{algorithm}[hbt!]
\SetKwInOut{Input}{Input}
\SetKwInOut{Output}{Output}
\caption{Analytical design gradient}\label{algo:Jd}
\setcounter{AlgoLine}{0}

\Input{Design objective \eqref{eq:cost} and weak form of the PDE \eqref{eq:weakform}.}
\Output{Gradient of the quadratic approximation of the design objective $\mathcal{J}_\mathrm{QUAD}(d)$ with respect to the design parameter $d$.}

Compute the $m$-gradient \eqref{eq:Qm}\;

Compute the $m$-Hessian \eqref{eq:Qmm}\;

Compute the $m_f$-gradient \eqref{eq:gradient_chance}\;

Compute the $m_f$-Hessian \eqref{eq:hessian_chance}\; 

Obtain the generalized eigenpairs using the double-pass randomized algorithm\;

Solve for $\mathcal{J}_\mathrm{QUAD}$ by \eqref{eq:Jquad}\;

Find the solution to the linear state problem \eqref{eq:Jd_linear_state_prob}\; 

Find the solution to the linear adjoint problem \eqref{eq:Jd_linear_adjoint_prob}\; 

Solve for the $d$-gradient \eqref{eq:design_gradient}

\end{algorithm}

\pagebreak

\section{Adaptive optimization}
The gradient-based scheme uses a continuation scheme that gradually increases the smoothing parameter $\omega$ and penalty parameter $\gamma$ to achieve convergence \cite{chen2021taylor}. An initial value of the smoothing parameter $\omega_{0}$ and the penalty parameter $\gamma_{0}$ is defined for the continuation scheme. These parameters are updated at the end of each iteration using scaling parameters $\Lambda_{\omega}$ and $\Lambda_{\gamma}$ for smoothing and penalty parameters, respectively. For $k^{th}$ iteration , the updated smoothing parameter is $\omega_{k+1} = \Lambda_{\omega} \omega_{k}$ and the updated penalty parameter is $\gamma_{k+1} = \Lambda_{\gamma} \gamma_{k}$. In an outer loop, we update the parameters $\omega$ and $\gamma$, whereas an Inexact Newton Conjugate-Gradient (INCG) is applied in the inner loop to solve the optimization problem \eqref{eq: optimization_penalty}. The termination criterion for the outer loop is if the maximum number of iterations $k_{\text{M}}$ is reached. For each iteration, the approximated chance is computed through the quadratic approximation $f_{\text{QUAD}}$ of the chance function as given in (\ref{eq:chance_function}). The output is the optimal value of design parameter $d_{\text{opm}}$.

\RestyleAlgo{ruled}
\begin{algorithm}[hbt!]
\SetKwInOut{Input}{Input}
\SetKwInOut{Output}{Output}
\caption{Adaptive Optimization}\label{algo:adaptive_optimization}
\setcounter{AlgoLine}{0}
\Input{Initial design parameter $d_{0}$, smoothing parameter $\omega_{0}$, penalty parameter $\gamma_{0}$, scaling parameters $\Lambda_{\omega}$ and $\Lambda_{\gamma}$, $\hat{f}= f_{\text{QUAD}}(d)$}
\Output{Optimal Design $d_{opm}$}

\textbf{while} $||d_{k} - d_{k-1}|| \geq \mathcal{\epsilon_{\text{out}}}$ or $k<k_{\text{M}}$; 
$d_{k+1} = \text{Inexact Newton Conjugate-Gradient}(d_{k},\mathcal{J}(d), \nabla_{d} \mathcal{J} (d), \epsilon_{\text{in}})$.;
Evaluate approximate chance $\hat{f}_{k+1}$ at $d_{k+1}$.;
Update $\omega_{k+1} = \Lambda_{\omega} \omega_{k}$, $\gamma_{k+1} = \Lambda_{\gamma} \gamma_{k}$;

\textbf{end}

Obtain the optimal design $d_{opm}$;

\end{algorithm}

\pagebreak
 \bibliographystyle{elsarticle-num} 
 \bibliography{references}

\end{document}